\documentclass[twocolumn,prm,preprintnumbers,showpacs,amsmath,amssymb,superscriptaddress]{revtex4-1}
\usepackage[pdftex, colorlinks, citecolor=blue]{hyperref}   
\usepackage{graphicx}
\usepackage{dcolumn}
\usepackage{threeparttable}
\usepackage{multirow}
\usepackage{txfonts}
\usepackage{xcolor}
\usepackage{amssymb}
\usepackage{amsmath}
\usepackage{latexsym}
\usepackage{epsfig}
\usepackage{amsbsy}
\usepackage{array}
\usepackage{tabularx}

\newcommand{\rr}{\mathbf{r}}
\newcommand{\kk}{\mathbf{k}}
\newcommand{\ii}{\mathrm{i}}

\newcommand{\qq}{\mathbf{q}}

\newcommand{\sais}{\ifmmode\mbox{\c{S}}\else\c{S}\fi{}a\ifmmode\mbox{\c{s}}\else\c{s}\fi{}\ifmmode
\imath \else \i \fi{}o\ifmmode \breve{g}\else \u{g}\fi{}lu~}

\begin{document}

\title{Relativistic $GW$+BSE study of the optical properties of Ruddlesden-Popper iridates}

\author{Peitao Liu}
\email{peitao.liu@univie.ac.at}
\affiliation{University of Vienna, Faculty of Physics and Center for
Computational Materials Science, Sensengasse 8, A-1090 Vienna, Austria}
\affiliation{Shenyang National Laboratory for Materials Science, Institute of Metal Research,
Chinese Academy of Sciences, Shenyang, Liaoning 110016, China}

\author{Bongjae Kim}
\affiliation{University of Vienna, Faculty of Physics and Center for
Computational Materials Science, Sensengasse 8, A-1090 Vienna, Austria}
\affiliation{Department of Physics, Kunsan National University, Gunsan 54150, Korea}

\author{Xing-Qiu Chen}
\affiliation{Shenyang National Laboratory for Materials Science, Institute of Metal Research,
Chinese Academy of Sciences, Shenyang, Liaoning 110016, China}

\author{D.D. Sarma}
\affiliation{Solid State and Structural Chemistry Unit, Indian Institute of Science, Bangalore-560012, India}

\author{Georg Kresse}
\affiliation{University of Vienna, Faculty of Physics and Center for
Computational Materials Science, Sensengasse 8, A-1090 Vienna, Austria}

\author{Cesare Franchini}
\affiliation{University of Vienna, Faculty of Physics and Center for
Computational Materials Science, Sensengasse 8, A-1090 Vienna, Austria}

\begin{abstract}

We study the optical properties of the Ruddlesden-Popper series of iridates Sr$_{n+1}$Ir$_n$O$_{3n+1}$  ($n$=1, 2 and $\infty$) by solving the Bethe-Salpeter equation (BSE), where the quasiparticle (QP) energies and screened interactions $W$ are obtained by the $GW$ approximation including spin-orbit coupling.
The computed optical conductivity spectra show strong excitonic effects and reproduce very well the experimentally observed double-peak structure, in particular for the spin-orbital Mott insulators Sr$_2$IrO$_4$ and Sr$_3$Ir$_2$O$_7$. However, $GW$ does not account well for the correlated metallic state of SrIrO$_3$
owing to a much too small band renormalization, and this affects the overall quality of the optical conductivity.
Our analysis describes well the progressive redshift of the main optical peaks as a function of dimensionality ($n$), which is correlated with the gradual decrease of the electronic correlation (quantified by the constrained random phase approximation) towards the metallic $n=\infty$ limit.
We have also assessed the quality of a computationally cheaper BSE approach that is based on a model dielectric function and conducted on top of DFT+$U$ one-electron energies.
Unfortunately, this model BSE approach does not accurately reproduce the outcome of the full $GW$+BSE method and leads to larger deviations to the measured spectra.
\end{abstract}

\maketitle

\section{Introduction}

5$d$ Ir-based transition metal oxides (TMOs) have stimulated a lot of  interest
due to the anticipation of novel phases and exotic properties
resulting from the cooperative interplay among the
crystalline electric field, spin-orbit coupling (SOC), Coulomb
repulsion ($U$), and different spin-exchange interactions
\cite{Kim2008_PRL,Moon_2008_MIT, Kim2009_science, PhysRevLett.102.017205,
Shitade2009,Chaloupka2010, Pesin2010, Khaliullin2012,PhysRevLett.110.117207,
Okada2013,Kim2014_natcom, PhysRevLett.112.056402,Zhao2015,Chun2015,He2015,Battisti2016}.
Of particular interest is the Ruddlesden-Popper (RP) series of iridates
Sr$_{n+1}$Ir$_n$O$_{3n+1}$  ($n$=1, 2 and $\infty$), which has been the subject of numerous works ~\cite{Kim2008_PRL,Moon_2008_MIT,William2014_review,Jeffrey2016_review,Yamasaki2016}.
It is found that as $n$ increases from 1 to $\infty$,
a dimensionality-controlled insulator-metal transition (IMT) occurs~\cite{Moon_2008_MIT,Yamasaki2016}.
In particular, the first member of the series, Sr$_2$IrO$_4$, provides a prototypical model system to
investigate the entanglement of the spin and orbital degrees of
freedom due to the strong SOC, which triggers a novel relativistic $J_\text{eff}$=1/2 Mott-like insulating state in an otherwise metallic compound~\cite{Kim2008_PRL,PhysRevLett.105.216410,Liu2015},
and an unusual in-plane canted antiferromagnetism (AFM)
with a weak net ferromagnetic component and Dzyaloshinskii-Moriya interaction~\cite{Liu2015,PhysRevLett.108.247212,PhysRevB.87.144405,PhysRevB.87.140406}.
In addition, Sr$_2$IrO$_4$ exhibits striking structural and magnetic similarities to high-$T_\text{c}$ cuprate
superconductors~\cite{PhysRevLett.108.177003,PhysRevLett.108.247212}. This stimulates the search for a new family of superconductors by doping~\cite{PhysRevLett.106.136402,PhysRevLett.110.027002,Kim2014_doping,Kim2015_doping,PhysRevB.94.195145}
or strain engineering~\cite{PhysRevLett.112.147201}.
The $n=2$ compound, Sr$_3$Ir$_2$O$_7$, exhibits structural and electronic properties similar to
its sister $n=1$ counterpart Sr$_2$IrO$_4$~\cite{Wojek2012, Wang2013};
however, unlike Sr$_2$IrO$_4$, it shows a $c$-collinear AFM state, owing to the
stronger interlayer coupling, and shows a smaller insulating gap
(0.13 eV~\cite{Moon2009_optics} vs. 0.30 eV~\cite{Okada2013})~\cite{PhysRevB.86.174414,Park2014,Boseggia2012, Khaliullin2012}.
The end member of the RP series is the perovskite-like SrIrO$_3$ compound. It has a three-dimensional crystal structure and exhibits nonmagnetic correlated
and topological crystalline semimetal character~\cite{Moon_2008_MIT, Heung-Sik2015,Chen2015,Fujioka2017,Lunyong2017_review113}, associated with  surface states protected by the lattice symmetry~\cite{Heung-Sik2015,Chen2015}, a large quasiparticle mass enhancement~\cite{Moon_2008_MIT},
and an unusual positive magnetoresistance~\cite{Fujioka2017}.

To identify the electronic properties of these correlated iridates,
optical spectroscopies have been widely used~\cite{Kim2008_PRL,Moon_2008_MIT,Moon2009_optics,Lee2012_optics,
Kim2012_PRLoptics,Nichols2013_APL,Nichols2013_optics,Liu2013,Kim2016_SL,Park2014,Sohn2014_optics,
Propper2016_optics,Fujioka2017,Souri2017_optics}, since they can provide important information
on the low-energy excitations, charge dynamics, and degree of electron correlations~\cite{RevModPhys.83.471}.
Moreover, due to the sensitivity of the optical conductivity upon
the variation of the electric properties,  optical spectroscopy
often serves as a direct probe to inspect the evolution of electronic structure across an IMT upon the application of internal and external stimuli such as dimensionality~\cite{Moon_2008_MIT},
electron/hole doping or SOC strength~\cite{Lee2012_optics}.
Also, optical spectroscopy is used to probe the optical excitations assisted by phonons~\cite{Park2014}
and the tuning of the spin-orbit coupled $J_\text{eff}$=1/2 state
by pressure~\cite{Haskel2012} and epitaxial strain~\cite{Nichols2013_optics,Liu2013,Kim2016_SL}.

For Sr$_2$IrO$_4$ and Sr$_3$Ir$_2$O$_7$, the experimental conductivity spectra near the band gap region
show a typical two-peak structure, named $\alpha$ and $\beta$~\cite{Kim2008_PRL,Moon_2008_MIT,Moon2009_optics,Park2014}.
These peaks are interpreted as interband $d$-$d$ transitions from the
$J_\text{eff}$=1/2 lower Hubbard band (LHB) to the $J_\text{eff}$=1/2 upper Hubbard band (UHB) ($\alpha$ peak),
and from the occupied $J_\text{eff}$=3/2 manifold to the $J_\text{eff}$=1/2 UHB ($\beta$ peak)~\cite{Kim2008_PRL,Moon_2008_MIT,Moon2009_optics,Park2014}.
As the dimensionality increases from Sr$_2$IrO$_4$ ($n=1$) to Sr$_3$Ir$_2$O$_7$ ($n=2$),
the $\alpha$ and $\beta$ peaks shift down to lower energies, in accordance with the decrease of the band gaps~\cite{Moon_2008_MIT}. For $n=\infty$ the system approaches the topological semimetal state:
The $\alpha$ peak loses intensity and only the $\beta$ peak along with the metallic Drude peak were observed in SrIrO$_3$~\cite{Moon_2008_MIT}.
Although the full disappearance of the $\alpha$ transition was confirmed by dynamical mean field theory (DMFT)~\cite{Zhang2013_DMFT}, very recent optical spectroscopy have identified the persistence of the
$\alpha$ peak even in SrIrO$_3$~\cite{Kim2016_SL,Fujioka2017}, raising doubts on the evolution of the optical transition across the RP series and their relative intensities.

In this context, it would be interesting and desirable to assess and describe the optical spectra from advanced first-principles calculations, at a level of theory capable to treat SOC, strong on-site Coulomb interactions, noncollinear spins, and lattice distortions simultaneously and precisely.
Up to now, only a few theoretical studies have focused on the optical properties of RP iridates,
either by microscopic model Hamiltonians~\cite{Kim2012_PRLoptics,Kim2016_optics, Souri2017_optics},
density functional theory (DFT) with an additional Hubbard $U$ correction (DFT+$U$)~\cite{Propper2016_optics,Kim2016_SL},
hybrid functionals~\cite{Kim2016_optics}, or DMFT~\cite{Zhang2013_DMFT}.
Microscopic model Hamiltonians are generally superior to DFT-like approaches in capturing the dominant physics and offer a transparent interpretation of the main interactions, but restrictions on the cluster size and the dependence on adjustable interacting parameters could limit the accuracy of the results. Within these limits, the obtained optical conductivity, only available for Sr$_2$IrO$_4$, are in good agreement with experiments~\cite{Kim2012_PRLoptics,Kim2016_optics, Souri2017_optics}.
DFT-based schemes can provide an atomistic interpretation by solving a simplified Schr\"odinger equation and the DFT+$U$ variant (with suitable values of the on-site parameter $U$) reproduces relatively well the value of the gap and the two-peak structure in Sr$_2$IrO$_4$~\cite{Liu2015,PhysRevB.94.195145}. However, the resulting optical conductivity computed within the independent-particle approximation is in less good agreement with experiments~\cite{Propper2016_optics,Kim2016_SL}.
The inclusion of a fraction of non-local exchange within the hybrid functional technique
yields good optical spectra in iridates~\cite{Kim2016_optics}, but also in this case the results are strongly dependent on the specific fraction, which is not easy to determine, in particular for complex TMOs~\cite{PhysRevB.86.235117}.
On the other side, the incorporation of local dynamical correlations in DMFT leads to a successful account of band renormalization effects as well as transfer of spectral weights
in strongly correlated materials~\cite{RevModPhys.68.13}. In combination with DFT, DMFT
yields the correct trends for the optical conductivity as a function of the dimensionality in the RP iridates,
consistent with experiments~\cite{Zhang2013_DMFT}, except for
the absence of the $\alpha$ peak in SrIrO$_3$, as mentioned above~\cite{Fujioka2017}.
Nevertheless, the predicted energy positions of the peaks are blueshifted compared to experiments mostly likely due to the large on-site $U$ used in the calculations~\cite{Zhang2013_DMFT}.

In principle, from an \emph{ab initio} perspective the established proper way
 of calculating the optical properties is to solve the Bethe-Salpeter equation, where the excitonic effects resulting from the electron-hole interactions are explicitly accounted for~\cite{Hanke1980, RevModPhys.74.601}.
In order to obtain precise positions of the peaks in the spectra, accurate evaluations of quasiparticle (QP) energies are important, in particular, the size of the band gap.
However, DFT is a single-particle ground state theory and does not provide an accurate account of the
excited state properties, e.g., electron addition/removal energies as well as electron-hole interactions.
This leads to the well-known underestimation of the band gaps~\cite{Schilfgaarde2006}.
A successful approach for the calculation of QP energies is the $GW$ approximation~\cite{Hedin1965,Strinati1982,Louie1985}. It provides a good approximation for the evaluation of the  self-energy of a many-body system of electrons by including the screening effects in the electron-electron Coulomb interactions.
The $GW$ methods has been applied to a wide variety of systems ranging from elemental semiconductors to TMO perovskites~\cite{Schilfgaarde2006, Klime2014,PhysRevB.94.165109,Zeynep2018}, delivering band gaps in rather good agreement with experiments.
The optimal procedure to calculate the optical spectra consists in
the solution of the BSE using the $GW$ QP energies, adopting static screening for the electron-hole interactions within the random phase approximation (RPA)~\cite{RevModPhys.74.601}.
This is referred to as $GW$+BSE. Although $GW$+BSE has been widely used to predict the optical spectra
of various systems, e.g., molecules, clusters, semiconductors and
insulators~\cite{Rohlfing2000,Tiago2006,Sabine2014,Jiangang2014,PhysRevMaterials.2.034603}, its applications to TMO perovskites are
very rare due to the technical challenges and the huge computational demand~\cite{PhysRevB.81.085213,PhysRevB.87.235102,PhysRevB.89.045104,PhysRevB.91.195137}.

In this work, the electronic and optical properties of the three RP iridates $n$=1, 2, and $\infty$ are  investigated by fully relativistic (spin-orbit coupling included) $GW$ and many-body electron-hole interactions through the solution of the BSE.
The work we present here is intended to contribute to a comprehensive understanding
on how well the $GW$ approximation describes the electronic structures of these 5$d$ relativistic iridates.
It gives some insights for the interpretation of the experimental optical spectra from the perspective of state-of-the-art \emph{ab initio} calculations based on the explicit calculations of the oscillator strength and characters of the optical transitions.
The $GW$ band structure and density of states (DOS) are compared to the DFT+$U$ calculations with a Hubbard $U$ calculated fully \emph{ab initio} within the constrained RPA (cRPA). Both sets of calculations give satisfactory results in terms of band gap and band topology. The main difference is that $GW$ significantly pushes down in energy the O-2$p$ states, leading to a decreased hybridization between Ir-$t_\text{2g}$ and O-2$p$ states.
This results in QP band gaps of Sr$_2$IrO$_4$ and Sr$_3$Ir$_2$O$_7$ that are in good agreement with experiments.
However, the $GW$ approximation is not adequate to describe the correlated metallic state in SrIrO$_3$
characteristic of strong QP peaks around the Fermi energy in the spectral function. This implies that
the partial correlations included in the $GW$ self-energy are not sufficient and a theory
beyond the $GW$ approximation is needed. The computed optical conductivity spectra
show strong excitonic effects and reproduce well the experimentally observed double-peak structures
in all three RP iridates. As the dimensionality increases, the $\alpha$ and $\beta$ peaks shift towards
lower energies, in line with the experiments.
Furthermore, we have used a less expensive model Bethe-Salpeter scheme (mBSE),
which avoids calculating the screened Coulomb interactions but instead uses an analytic model. It is found that
the calculated spectra from mBSE agree qualitatively with the ones from full $GW$+BSE.

\begin{figure*}[t]
\begin{center}
\includegraphics*[width=0.80\textwidth]{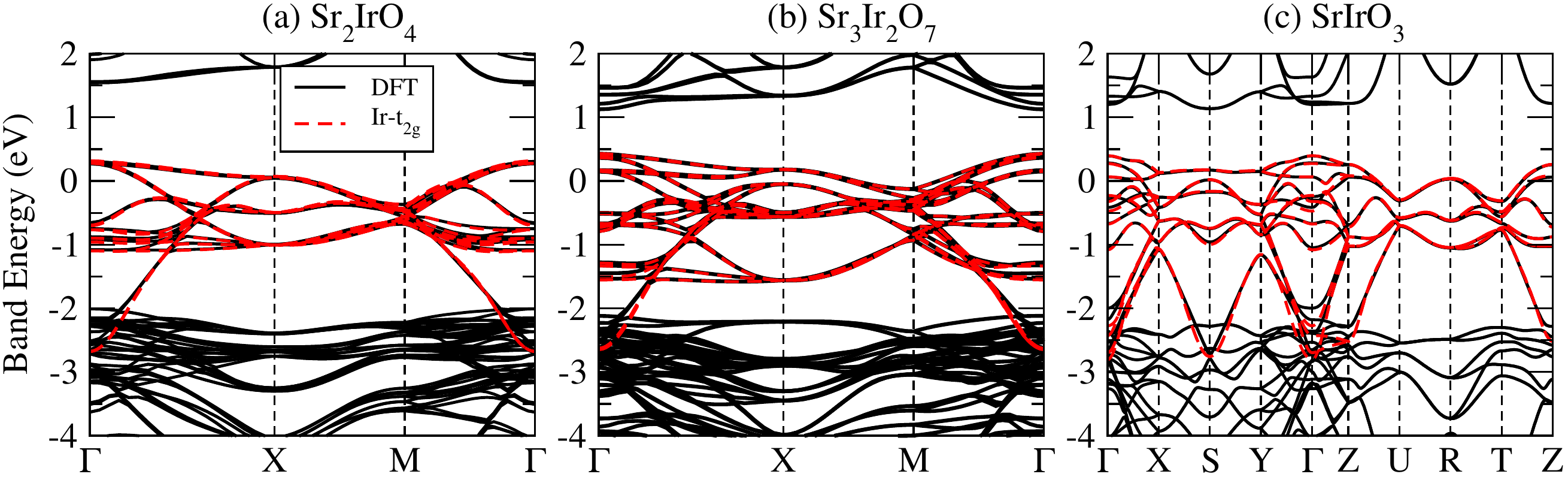}
\end{center}
\caption{The nonmagnetic DFT bands (black line) superposed with Wannier interpolated bands (red dashed line).}
\label{fig:CRPA}
\end{figure*}

\section{Computational details}

Our first-principles calculations were performed using the
projector augmented wave method (PAW)~\cite{PhysRevB.50.17953} as implemented in the Vienna
\emph{Ab initio} Simulation Package (VASP)~\cite{PhysRevB.47.558, PhysRevB.54.11169}.
The ultrasoft PAW potentials with an appendix ({\tt $\_$GW})
released with VASP.5.2 were used. The plane-wave cutoff for the orbitals was chosen to be the maximum
of all elements in the considered material. The energy cutoff for the response function
was chosen to be half of the plane-wave cutoff. To  sample the Brillouin zone,
6$\times$6$\times$1 $k$-point grids shifted off $\Gamma$ generated by the Monkhorst-Pack (MP) scheme were used for $n$=1 and 2
and the grids were increased to 6$\times$6$\times$4  for $n$=$\infty$, unless explicitly stated otherwise.
The atomic positions were optimized with the lattice parameters fixed at the
experimental values~\cite{PhysRevB.49.9198,SUBRAMANIAN1994645,Zhao2008,Puggioni2016}
(see Table~\ref{Table_structure} in the Appendix). All calculations were performed using a fully relativistic setup with the inclusion
of SOC~\cite{Hobbs2000_ncl_VASP, Biermann2008_GW_SOC, PhysRevB.93.224425}, at all levels of theory DFT+$U$, $GW$ and BSE.

\subsection{$G_0W_0$@DFT+$U$}

Due to the large computational cost of $GW$ calculations on such large systems, we adopted the
single-shot $G_0W_0$ variant of the $GW$ approximation. The $G_0W_0$ method is known to
predict relatively satisfactory band gaps~\cite{Fuchs2007, Zeynep2018}.
The justification of the good performance of $G_0W_0$ arises from the error cancellation
stemming from the lack of self-consistency and the absence of vertex
corrections~\cite{Shishkin2007_PRL}. However, $G_0W_0$ results are clearly dependent on the
starting one-electron energies and orbitals~\cite{Fuchs2007, PhysRevB.89.045104}.
For systems with localized $d$ or $f$ states, DFT+$U$ obtained orbitals are much closer to the ground state, and hence
DFT+$U$ are shown to be a better starting point than DFT~\cite{JiangHong_2009PRL,JiangHong_2010PRB}.
To this end, a small effective Hubbard $U_\text{eff}$=$U-J$=1 eV was introduced for the Ir-5$d$ states in all three considered iridates,
using the scheme introduced by Dudarev \emph{et al.}~\cite{PhysRevB.57.1505}.
It was also found that $G_0W_0$@DFT+$U$ exhibits only a weak dependence on $U$ in a physically meaningful range of $U$ values~\cite{JiangHong_2009PRL}. In the case of iridates, there is an another important reason to start from DFT+$U$ orbitals: Without including $U$, the initial DFT band structure is metallic and $G_0W_0$ is not capable to open the gap.

For the calculation of the response function at the $G_0W_0$ level, 128 frequency points
and about 600 virtual orbitals were used. Though important~\cite{Klime2014,Zeynep2018},
the basis-set converged limit is not considered here because it is beyond the scope of the present work.
However, based on our previous systematic analysis of the convergence of $G_0W_0$ results for a representative dataset of 3$d$, 4$d$ and 5$d$ TMO perovskites, we found that 600 virtual orbitals are sufficient to obtained well converged results with error of the order of $\sim$50 meV~\cite{Zeynep2018}.

\subsection{Constrained random phase approximation}

\begin{table}[h!]
\footnotesize
\caption {On-site Coulomb and exchange interactions (in eV)
calculated by cRPA for three RP iridates.
$U_{ij}=U_{ijij}$ and $J_{ij}=U_{ijji}$ with $i$ and $j$ representing $t_\text{2g}$-like Wannier orbitals.
See Eq.~(\ref{eq:Uijkl_2}) for the notations used.
}
\begin{ruledtabular}
\begin{tabular}{lcccccccc}
                                   & \multicolumn{3}{c}{$U_{ij}$}  & & \multicolumn{3}{c}{$J_{ij}$}  \\
                                   \hline
 Sr$_2$IrO$_4$       & $d_\text{yz}$ & $d_\text{zx}$ & $d_\text{xy}$  && $d_\text{yz}$ & $d_\text{zx}$ & $d_\text{xy}$  \\
$d_\text{yz}$       & 2.30 & 1.72 & 1.57  && --      & 0.23 & 0.22  \\
$d_\text{zx}$       & 1.72 & 2.30 & 1.57  && 0.23 & --      & 0.22  \\
$d_\text{xy}$       & 1.57 & 1.57 & 2.03  && 0.22 & 0.22 & --  \\
\hline
Sr$_3$Ir$_2$O$_7$      & $d_\text{yz}$ & $d_\text{zx}$ & $d_\text{xy}$  && $d_\text{yz}$ & $d_\text{zx}$ & $d_\text{xy}$  \\
$d_\text{yz}$       & 2.16 & 1.58 & 1.43  && --      & 0.23 & 0.22  \\
$d_\text{zx}$       & 1.58 & 2.16 & 1.43  && 0.23 & --      & 0.22  \\
$d_\text{xy}$       & 1.43 & 1.43 & 1.85  && 0.22 & 0.22 & --  \\
\hline
 SrIrO$_3$                  & $d_\text{yz}$ & $d_\text{zx}$ & $d_\text{xy}$  && $d_\text{yz}$ & $d_\text{zx}$ & $d_\text{xy}$   \\
$d_\text{yz}$       & 1.78 & 1.21 & 1.19  && --      & 0.22 & 0.22  \\
$d_\text{zx}$       & 1.21 & 1.73 & 1.22  && 0.22 & --      & 0.22  \\
$d_\text{xy}$       & 1.19 & 1.22 & 1.74  && 0.22 & 0.22 & --
\end{tabular}
\end{ruledtabular}
\label{Table_cRPA}
\end{table}

For a comparison, we have also performed  DFT+$U$+SOC electronic structure calculations using $U$ values calculated fully \emph{ab initio} by the cRPA~\cite{PhysRevB.74.125106}.
The central idea of cRPA is to remove from the total polarizability $\chi$ the contribution in the
target correlated Ir-$t_\text{2g}$ states $\chi^c$
\begin{equation}
\chi^r=\chi-\chi^c.
\label{eq:Uijkl_1}
\end{equation}
We follow the Kubo formalism derived by Kaltak~\emph{et al.} \cite{Merzuk2015}.
To this end, the maximally localized Ir-$t_\text{2g}$ Wannier functions $w(\rr)$
obtained by the Wannier90 suite~\cite{PhysRevB.56.12847, MOSTOFI2008685}
with an interface to VASP~\cite{Franchini2012} are used as the local basis, and
the matrix elements of $U$ are evaluated by
\begin{equation}
U_{ijkl}=\lim_{\omega \to 0}\iint d\rr d\rr' w^*_i(\rr) w^*_j(\rr') \mathcal{U}(\rr,\rr',\omega) w_k(\rr) w_l(\rr').
\label{eq:Uijkl_2}
\end{equation}
Here, $\mathcal{U}$ is the partially screened interaction kernel,
which is calculated by solving the equation
 \begin{equation}
\mathcal{U}^{-1}=\mathcal{V}^{-1}-\chi^r,
\label{eq:Uijkl_3}
\end{equation}
and $\mathcal{V}$ is the bare (unscreened) interaction kernel.
For more computational details about cRPA, please refer to Ref.~\cite{Merzuk2015}.

For the description of the three iridates we choose the Ir-$t_\text{2g}$ states as target correlated subspace.
To prove the reliability of this choice we show in Fig.~\ref{fig:CRPA} the comparison between the nonmagnetic DFT bands
and the corresponding one for the Ir-$t_\text{2g}$  manifold obtained by Wannier interpolation.

The obtained values of $U_{ij}$ and $J_{ij}$ are listed in Table~\ref{Table_cRPA}. For the DFT+$U$ calculations we have used the average values of $U_{ij}$ and $J_{ij}$, specifically:
(i) for Sr$_2$IrO$_4$: $U=1.82$ eV and $J$=0.22 eV ;
(ii) for Sr$_3$Ir$_2$O$_7$: $U=1.67$ eV and $J$=0.22 eV ;
(iii) for SrIrO$_3$: $U=1.37$ eV and $J$=0.22 eV.
We found that $U$ decreases with increasing $n$ as a consequence of the gradual increase of the bandwidth which leads to enhanced screening. We will discuss this issue in more details in Sec.~\ref{res:el}.
Note that the $U$ and $J$ values of Sr$_2$IrO$_4$ were also calculated by Arita~\emph{et al.}~\cite{PhysRevLett.108.086403},
yielding $U$=1.93 eV  and $J$=0.16 eV, in very good agreement with our estimation. The small deviation might
arise from the neglect of in-plane octahedral rotations in the crystal structure used by Arita~\emph{et al.}~\cite{PhysRevLett.108.086403}.

\subsection{BSE and optical conductivity}

The optical conductivity was calculated through the solution of the BSE within the Tamm-Dancoff approximation (TDA) using $G_0W_0$ as a starting point for the construction of the screening properties and QP energies using 6$\times$6$\times$1 ($n$=1 and $n$=2) and 6$\times$6$\times$4 ($n$=$\infty$) $k$-meshes.
It is important to note that the TDA shows tiny differences in spectra as compared to the full solution of the BSE and going beyond TDA is technically intractable if SOC is considered~\cite{Tobias2015}.
The specific procedure for calculating the optical conductivity involves four steps:

($i$) Standard self-consistent DFT+$U$ (small $U$=1.0 eV) calculations.

($ii$) Additional DFT+$U$ step, in which the one-electron wave functions and eigenenergies of all virtual
orbitals spanned by the plane wave basis set are evaluated by an exact diagonalization of the previously
determined self-consistent DFT+$U$ Hamiltonian.

($iii$)  $G_0W_0$ runs to compute the QP energies and RPA screened interactions $W$.

($iv$) Finally, the BSE in the TDA is solved, yielding the frequency-dependent macroscopic dielectric function~\cite{Tobias2015}
\begin{equation}\label{eq:DF}
\begin{split}
\varepsilon(\omega)
 = 1 - &
 \lim_{\qq \to 0}\,  V(\qq)
 \sum_{\Lambda} \left(\frac{1}{\omega-\Omega_{\Lambda}+\ii\eta  } - \frac{1}{\omega +\Omega_{\Lambda}-\ii\eta} \right) \\
&\times
 \Bigg\lbrace
\sum_\kk  w_\kk \sum_{v,c}
\langle \psi_{c\kk} |e^{\rm{i}\qq\cdot \rr}| {\psi_{v\kk}}\rangle
X^\Lambda_{cv\kk} \Bigg\rbrace\times \Bigg\lbrace c.c.\Bigg\rbrace,
\end{split}
\end{equation}
with the oscillator strengths $S_{\Lambda}$ associated with the optical transitions defined by
\begin{equation} \label{eq:oscillator_strength}
\begin{split}
S_{\Lambda} &= {\rm \textbf{Tr}}
\Bigg[
 \bigg\lbrace
\sum_\kk  w_\kk \sum_{v,c}
\langle \psi_{c\kk} |e^{\rm{i}\qq\cdot \rr}| {\psi_{v\kk}}\rangle
X^\Lambda_{cv\kk} \bigg\rbrace\times \bigg\lbrace c.c.\bigg\rbrace
 \Bigg].
\end{split}
\end{equation}
Here, $\Omega_{\Lambda}$ and $X^\Lambda$  are BSE eigenvalues and eigenvectors, respectively.
$V$ is the bare interaction, $\eta$ is a positive infinitesimal, and $w_\kk$ are the $k$-point weights.
$\psi_{v\kk}$ and $\psi_{c\kk}$ refer to occupied and unoccupied DFT+$U$ wave functions, respectively.
From $\varepsilon(\omega)$, the real part of the optical conductivity is then derived by
\begin{eqnarray}
\label{conduct1}
{\rm Re}[\sigma(\omega)] &=&\frac{\omega}{4\pi} {\rm Im}[\varepsilon(\omega)].
\end{eqnarray}
We have also compared the full BSE spectra with the RPA one, obtained by neglecting $W$ in the calculation of the full polarizability~\cite{Tobias2015}.

For the calculation of the optical conductivity of metallic SrIrO$_3$, we have also considered an intraband contribution  by means of the  Drude-like model~\cite{Kim2016_optics,Ambrosch-Draxl2004}
\begin{eqnarray}
\label{Drude}
\sigma_D(\omega) = \frac{ \Gamma \omega^2_p }{4\pi(\omega^2+\Gamma^2)},
\end{eqnarray}
where $\Gamma$ is lifetime broadening, which is set to 0.1 eV in our study according to Ref.~\cite{Kim2016_optics}
and $\omega_p$  is the plasma frequency. The converged $\omega^2_p$ is calculated to be about 2.26 eV.

Considering that the $GW$-based calculations of the RPA screened interaction $W$ and of the QP energies are rather expensive and do not scale favorably with the number of $k$ points, we have also tested an analytic model for the treatment of the static screening required as input in the BSE~\cite{BECHSTEDT1992765,PhysRevB.78.085103,Bokdam2016}:
\begin{equation}
 {\varepsilon}_{\mathbf{G},\mathbf{G}}^{-1}(\qq)={{\varepsilon}_{\infty}^{-1}}+(1-{{\varepsilon}_{\infty}^{-1}})[1-\text{exp}(-\frac{|\mathbf{q+G}|^2}{4{\lambda}^2})],
\label{eq:mBSE}
\end{equation}
where ${\varepsilon}_{\infty}$ is the static ion-clamped dielectric function in the long-wave limit
and the screening length parameter $\lambda$ is derived by fitting the screening ${\varepsilon}^{-1}$ at small wave vectors
with respect to $ |\mathbf{q+G}|$ with $\mathbf{q}$ and $\mathbf{G}$ being the wave vector and lattice vector of the reciprocal cell, respectively.
This approach is typically referred to as model-BSE (mBSE).
Within this model, one can test the convergence of the optical spectra as a function of the number and distribution of $k$ points without the need to perform the demanding preliminary $GW$ calculations~\cite{Tobias2015}. Here, we have used as input for the BSE calculations the DFT+$U$ one-electron energies.
However, instead of progressively increasing the $k$-point grid, a procedure that does not significantly improve the quality of the spectra, we have used a set of different suitably shifted $k$-meshes
and averaged over the obtained individual mBSE spectra~\cite{Tobias2015,PhysRevB.78.121201}.
Specifically, we have created eight different $k$-meshes by centering the standard 6$\times$6$\times$1 and 6$\times$6$\times$4 grids on the eight irreducible $k$ points compatible with a 4$\times$4$\times$1 $k$-mesh.
The final spectra were obtained by averaging the spectra obtained from the eight independent mBSE calculations
with predetermined weights according to the symmetry of the 4$\times$4$\times$1 $k$-mesh~\cite{Tobias2015,PhysRevB.78.121201}.

\subsection{Crystal structures and magnetic orderings}

\begin{figure}[h!]
\begin{center}
 \includegraphics*[width=0.47\textwidth]{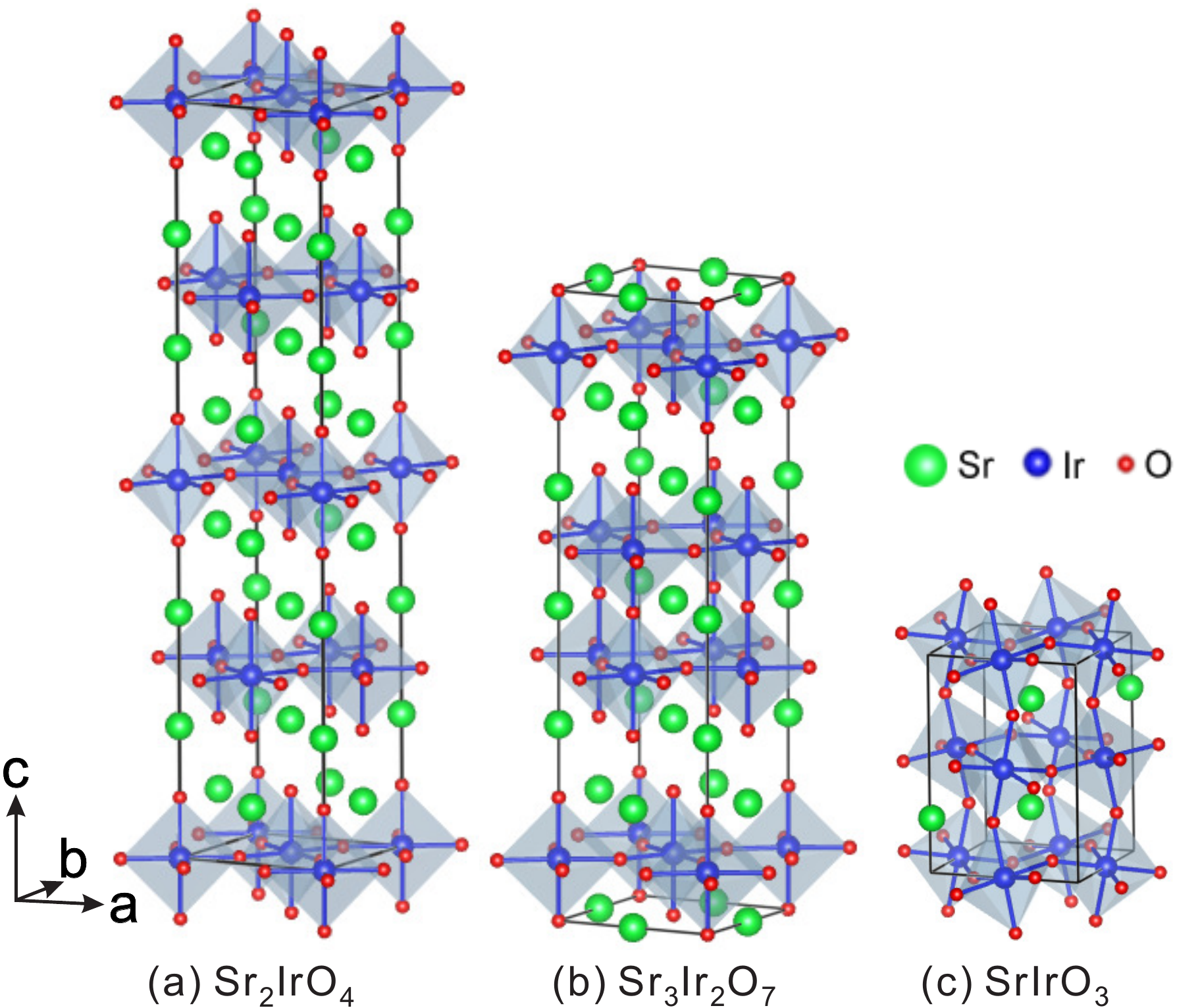}
\end{center}
\caption{Crystal structures of RP iridates
Sr$_{n+1}$Ir$_n$O$_{3n+1}$  ($n$=1, 2, and $\infty$). Sr, Ir and O atoms are shown in green,
blue, and red, respectively.}
\label{crystal}
\end{figure}

The RP series of perovskite-like iridates Sr$_{n+1}$Ir$_n$O$_{3n+1}$ is a family of materials
with $n$ being the number of SrIrO$_3$ perovskite layers sandwiched between SrO layers~\cite{Beznosikov2000},
as shown in Fig.~\ref{crystal}.
We have adopted the experimental lattice constants and fully relaxed all internal atomic positions
using standard convergence criteria (forces smaller than 0.01~eV/\AA) by DFT+$U$+SOC method.
The resulting optimized structural data are collected in Table~\ref{Table_structure} in the Appendix.
The calculated data are in good agreement with available measured data, with a relative deviation of about $1 \%$.

Similarly, starting from the experimentally measured magnetic orderings ($n=1$: in-plane canted AFM~\cite{PhysRevLett.108.247212,PhysRevB.87.140406};
$n=2$: $c$-collinear AFM~\cite{PhysRevB.86.174414,Park2014,Boseggia2012, Khaliullin2012} and $n=\infty$: nonmagnetic~\cite{Zhao2008}),
all spin-orbital degrees of freedom have been fully relaxed within noncollinear and relativistic DFT+$U$+SOC and $G_0W_0$+SOC.
The optimized and experimental magnetic data for the magnetic $n=1$ and $n=2$ compounds are summarized in Table~\ref{Table_mag_moment} in the Appendix. The calculated total moments are generally larger than the measured one, in particular for Sr$_{3}$Ir$_2$O$_{7}$: 0.32 $\mu_{\rm{B}}$ ($G_0W_0$) vs. 0.21 $\mu_{\rm{B}}$ for $n=1$ and 0.45  $\mu_{\rm{B}}$ ($G_0W_0$) vs. 0.1 $\mu_{\rm{B}}$ for $n=2$.

\section{Results and discussions}

This section focuses on the presentation of the results. Since the orbital properties and the band topology are essential ingredients for the calculation of the optical excitations, we start by discussing the electronic band structures in Sec.~\ref{res:el}. The subsequent section is dedicated to the BSE results.

\subsection{Electronic structures}
\label{res:el}

\begin{table}
\footnotesize
\caption{A summary of
cRPA estimated $U$ and $J$
(the average of matrix elements $U_{ij}$ and $J_{ij}$ shown in Table~\ref{Table_cRPA}),
DFT+$U_\text{eff}$  and $GW$ predicted band gaps as well as experimental gaps
for three iridates. The energies are given in eV.
}
\begin{ruledtabular}
\begin{tabular}{cccc}
            & Sr$_2$IrO$_4$   & Sr$_3$Ir$_2$O$_7$  & SrIrO$_3$  \\
\hline
 $U$   & 1.82 & 1.67 & 1.39 \\
 $J$    & 0.22 & 0.22 & 0.22  \\
       &      &      &      \\
DFT+$U_\text{eff}$ gap  & 0.23 & 0.14 & metal \\
 $GW$ gap  & 0.25 & 0.16 & metal \\
 Expt. gap & 0.30\cite{Moon2009_optics} & 0.13\cite{Okada2013} & metal\cite{Zhao2008,LONGO1971174}
\end{tabular}
\end{ruledtabular}
\label{Table_sum}
\end{table}

\begin{table}
\footnotesize
\caption{Relative energy difference (in meV)
between the valence band maximum (VBM) at $X$ and $\Gamma$ points, i.e., $X_\text{VBM}-\Gamma_\text{VBM}$,
for Sr$_2$IrO$_4$ and  Sr$_3$Ir$_2$O$_7$ predicted by DFT+$U_\text{eff}$+SOC, $GW$+SOC, and DMFT methods.
The estimated experimental values are also shown for comparison.}
\begin{ruledtabular}
\begin{tabular}{ccccc}
   & DFT+$U_\text{eff}$+SOC  & $GW$+SOC & DMFT~\cite{Zhang2013_DMFT}  &  Expt.~\cite{Wang2013} \\
\hline
 Sr$_2$IrO$_4$           & $-$20  & 10   & 70 & 250 \\
 Sr$_3$Ir$_2$O$_7$  & $-$70  & $-$40 &30  & 200 \\
\end{tabular}
\end{ruledtabular}
\label{Table_QP_relative_positions}
\end{table}

\begin{figure*}
\begin{center}
\includegraphics*[width=0.94 \textwidth]{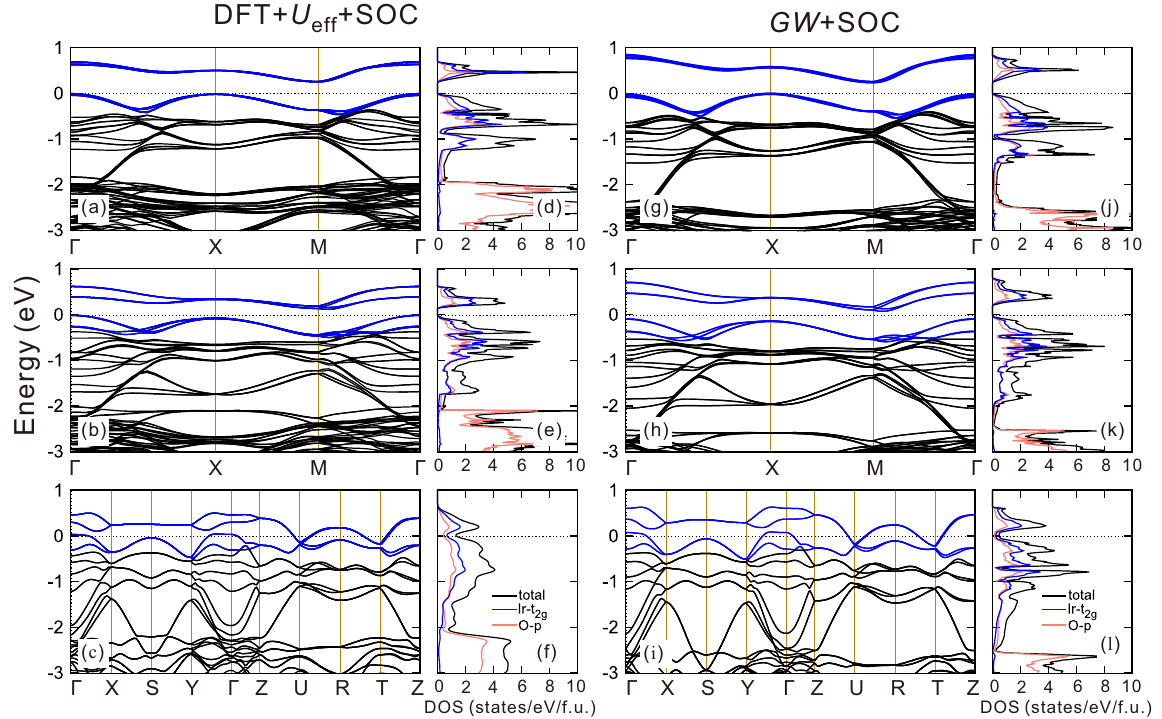}
\end{center}
\caption{Electronic band structures and density of states (DOS) obtained from
(a)-(f) DFT+$U_\text{eff}$+SOC  and (g)-(l) $GW$+SOC calculations for three iridates
(upper panel for Sr$_2$IrO$_4$, middle panel for Sr$_3$Ir$_2$O$_7$ and bottom panel for SrIrO$_3$).
The Fermi energy has been aligned to zero. Due to the large crystal field, the
high-energy Ir-$e_\text{g}$ states are not shown.}
\label{GW_bands}
\end{figure*}

The DFT+$U_\text{eff}$+SOC and $GW$+SOC band structures are compared in Fig.~\ref{GW_bands}.
We remind that the DFT+$U$ calculations were performed using the cRPA value of $U_\text{eff}$=$U-J$ reported in Table~\ref{Table_sum}, whereas the $GW$ runs are done starting from DFT+$U$ one-electron energies and orbitals using a smaller  $U_\text{eff}$=1 eV.

Both approaches deliver results consistent with experiments, and there are only residual differences between the two sets of electronic dispersions and density of states.
The most noticeable difference is that $GW$ pushes down the O-2$p$ states by about 0.5 eV, which in turn
decreases the hybridizations between Ir-$d$ states and  O-2$p$ states.

The calculations reproduce relatively well the $J_\text{eff}$=1/2 spin-orbital Mott insulating
state of Sr$_2$IrO$_4$ and  Sr$_3$Ir$_2$O$_7$ originating from the cooperative action of $U_\text{eff}$ and SOC,
and correctly describe SrIrO$_3$ as a semimetal. The obtained (indirect) fundamental gaps, calculated from the difference between the (quasi)particles energies at the bottom of the UHB and at the top of the LHB agrees very well with measurements (see Table~\ref{Table_sum}).

Sr$_2$IrO$_4$ and Sr$_3$Ir$_2$O$_7$ exhibit very similar band structures owing to
their similar crystal structures. The most noticeable difference is the splitting at the top of the valence band at $\Gamma$ induced by the bilayer structure in the $n=2$ system.
The splitting computed for Sr$_3$Ir$_2$O$_7$, 0.23 eV, is consistent with the one obtained by angle-resolved photoemission spectroscopy (ARPES)~\cite{Wang2013}.
It is important to note that for both $n=1$ and $n$=2 compounds the $J_\text{eff}$=1/2 LHB and $J_\text{eff}$=3/2 states are not separated by a well-defined gap, in contrast to the ideal $J_\text{eff}$=1/2 picture.

A feature of the electronic dispersions that is not well accounted for by our theoretical calculations is the relative energy order between the top of the valence band at $\Gamma$ and $X$. ARPES indicates that the maximum of the LHB at $\Gamma$ lays 150-250 meV lower compared to $X$ for both
Sr$_2$IrO$_4$ and  Sr$_3$Ir$_2$O$_7$
~\cite{Kim2008_PRL,Kim2015_doping,Wang2013,PhysRevLett.114.016401,PhysRevLett.115.176402}.
However, DFT+$U_\text{eff}$+SOC predicts that the $\Gamma$ point is 20 and 70 meV higher in energy than the $X$ point for Sr$_2$IrO$_4$ and  Sr$_3$Ir$_2$O$_7$, respectively.
$GW$ improves the description only marginally for Sr$_2$IrO$_4$, where the top of the valence band at $X$ is found 10 meV higher than at $\Gamma$, but fails in
reproducing the correct order for Sr$_3$Ir$_2$O$_7$, even though the QP difference $\Gamma - X$ is reduced to 40 meV.
The relative energy difference between  $X$ and $\Gamma$ points is compiled in Table~\ref{Table_QP_relative_positions}.
This clearly implies that the correlations included in
the $GW$ self-energies are not adequate enough to reproduce accurately the local band topology, likely due to the neglect of dynamical correlation effects. Indeed, it has been shown that DMFT is capable to cure this problem and delivers better relative energies at $\Gamma$ and $X$~\cite{Zhang2013_DMFT} (see Table~\ref{Table_QP_relative_positions}).

The band structure of SrIrO$_3$ is metallic and clearly different from the other two cases. The most important characteristic is the Dirac node at the $U$ point, which is protected by the lattice symmetry as discussed in previous publications~\cite{Kee2012,Carter2012}. The (multiple) Dirac cone is the only crossing between the conduction and valence band in the entire Brillouin zone, and is associated with a pseudogap with a small DOS at the Fermi energy [see Fig.~\ref{GW_bands}(f) and \ref{GW_bands}(l)], consistent with the experimentally measured small charge carrier density~\cite{Liu2005}, proving the semimetallic character of SrIrO$_3$.
A second drawback of the employed level of theory is the reduced degree of electronic correlation. In fact,
$GW$ gives a renormalization factor $Z$ for the $J_\text{eff}=1/2$ bands close to the Fermi level of 0.61, yielding a mass enhancement of 1.64, far lower than the experimental one, $\approx$ 6~\cite{Moon_2008_MIT}.
This is also reflected by the absence of the characteristic QP peak close to the Fermi level detected by
ARPES~\cite{PhysRevLett.114.016401} and confirmed by DMFT~\cite{Zhang2013_DMFT}. This clearly indicates that
the type and degree of correlations included in the $GW$ self-energy diagrams are
insufficient to describe the correlated metallic state of SrIrO$_3$ and going beyond the $GW$ approximation is needed.
These limitations influence the quality of the BSE spectrum of SrIrO$_3$, as discussed in the next section.

\subsection{Optical spectra}

\subsubsection{BSE}

\begin{figure*}
\begin{center}
\includegraphics*[width=0.94\textwidth]{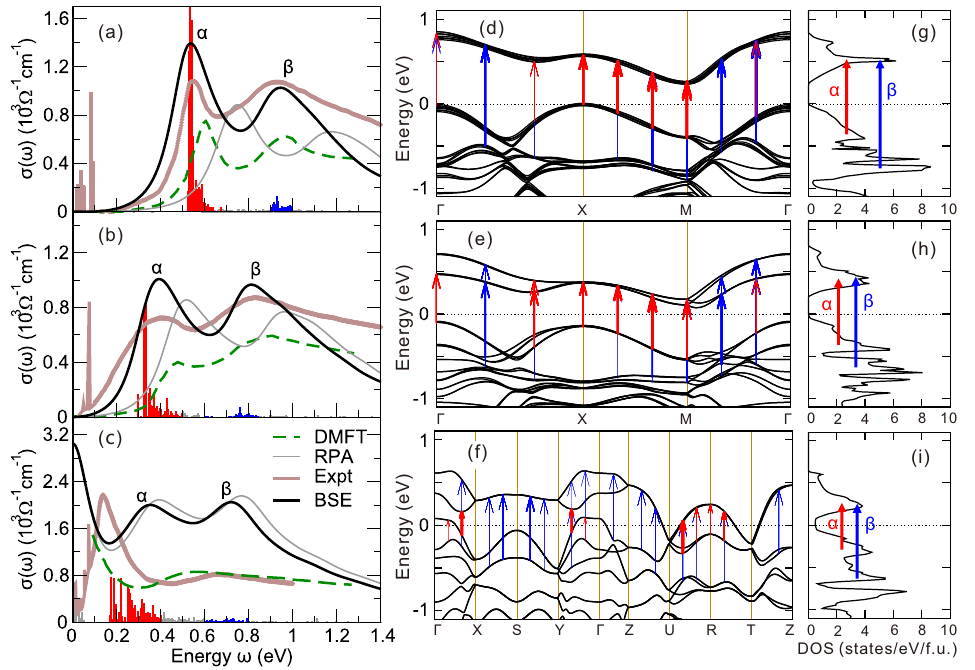}
\end{center}
\caption{The experimental and calculated optical conductivity spectra
${\sigma}$($\omega$) of (a) Sr$_2$IrO$_4$, (b) Sr$_3$Ir$_2$O$_7$ and (c) SrIrO$_3$.
The experimental data at 10 K are adapted from Ref.~\cite{Moon2009_optics} for Sr$_2$IrO$_4$ single crystals,
Ref.~\cite{Park2014} for Sr$_3$Ir$_2$O$_7$  single crystals and Ref.~\cite{Fujioka2017} for SrIrO$_3$ polycrystalline samples.
Note that the sharp peaks below 0.1 eV in the experimental spectra arise from optical phonon modes.
The DMFT simulated spectra are taken from Ref.~\cite{Zhang2013_DMFT}.
The gray vertical lines represent the oscillator strength (divided by 10$^4$ here) whose contributions
to $\alpha$ and $\beta$ peaks are highlighted in red and blue colors, respectively.
(d)-(f) $GW$ band structure.  (g)-(i) $GW$ total density of states.
The red and blue arrows in (d)-(f) represent
the dominant interband transitions for the $\alpha$ and $\beta$ peaks, respectively.
The width of the arrows denotes the normalized amplitude of BSE eigenvectors $|X^{\Lambda }_{cv\kk}|$.
Arrows in (g)-(i) show the involved optical transitions schematically.
}
\label{optics}
\end{figure*}

\begin{figure}
\begin{center}
\includegraphics*[width=0.40\textwidth]{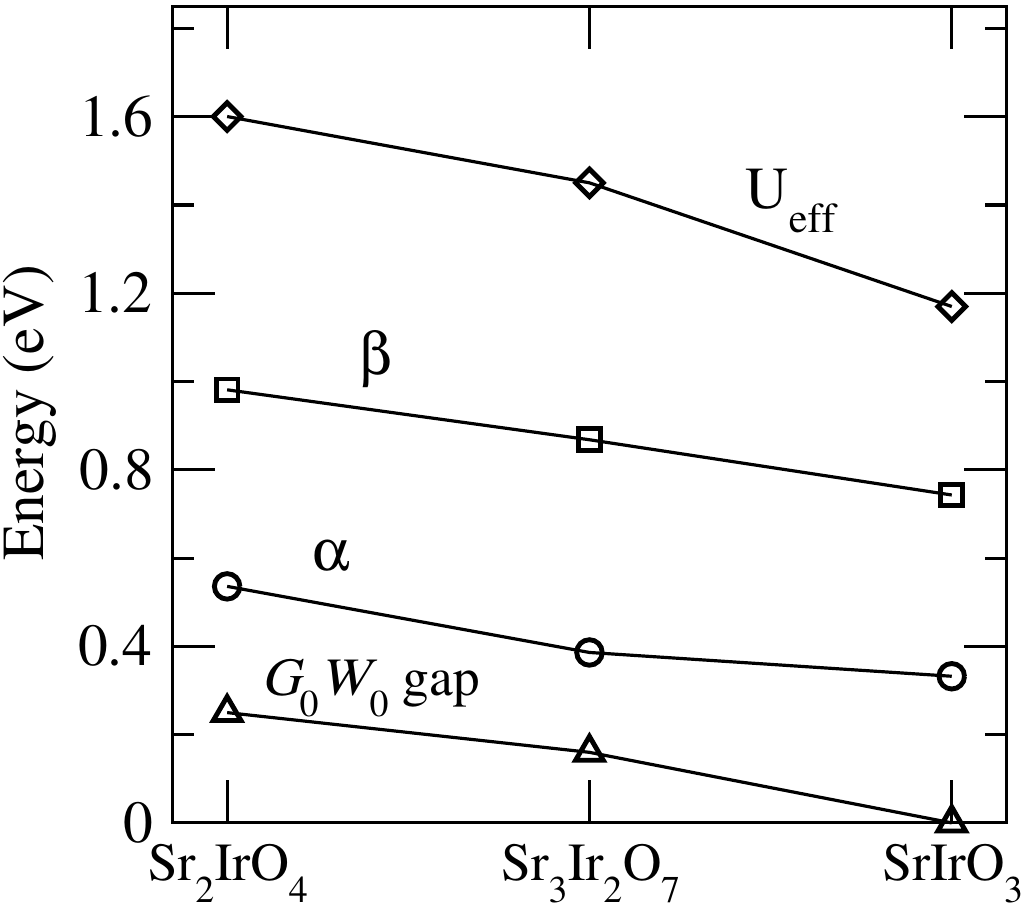}
\end{center}
\caption{Positions of $\alpha$ and $\beta$ peaks extracted from the BSE spectra
(Lorentz oscillator model analysis), and calculated band gaps ($G_0W_0$) and  $U_{\rm eff}$ (cRPA) for the three iridates under scrutiny. }
\label{fig:U_alpha_beta}
\end{figure}

With the QP energies and RPA screened interactions $W$ derived from
the $GW$ calculations presented in the previous section, we computed the optical conductivity
$\sigma$($\omega$) through the solution of the BSE. The results are graphically summarized in Fig.~\ref{optics}, which shows a comparison between experimental and calculated (BSE \& RPA) spectra [Fig.~\ref{optics}(a)-\ref{optics}(c)].
We found that all compounds are characterized by a double-peak structure ($\alpha$ and $\beta$), consistent with experiments~\cite{Moon2009_optics,Park2014,Fujioka2017}, but the agreement with the measured data depends on the level of theory and is also materials dependent.
Even though both, BSE and RPA, predict two dominant peaks, a good quantitative agreement with experiments is only obtained at the BSE level
for Sr$_2$IrO$_4$ and Sr$_3$Ir$_2$O$_7$,
for which the calculated and measured $\alpha$ and $\beta$ transitions are centered almost at the same excitation energies.
The dramatic redshift of the $\alpha$ and $\beta$ peaks at the BSE level as compared to RPA indicates
strong excitonic effects, but no bound exciton is discerned in our BSE calculations.
Also SrIrO$_3$ was found to exhibit a double-peak structure, qualitatively consistent with the most recent
experimental findings~\cite{Fujioka2017}. However, the agreement between theory and experiment is much less satisfactory compared with the $n$=1 and $n$=2 cases. In fact, the calculated $\alpha$ and $\beta$ peaks are centered at higher energies than the experimental ones~\cite{Fujioka2017}, and the Drude peak is broader and more intense. We note that available data based on DMFT calculations,
which include dynamical correlations not incorporated in the $GW$ framework, provide only a marginally improved description [see
Fig.~\ref{optics}(c) and also Fig.~\ref{fig:DMFT_optics} in the Appendix]. In particular, DMFT reproduces relatively well the $\beta$ peak, but the $\alpha$ peak is not detected.

To identify the character of the optical transitions, we have calculated the BSE oscillator strengths
 $S_{\Lambda}$ associated with the optical transitions [see Eq.~(\ref{eq:oscillator_strength})], shown as histograms in Fig.~\ref{optics}(a)-\ref{optics}(c) (red and blue colors are used to distinguish the contribution to the $\alpha$ and $\beta$ peaks, respectively). The  oscillator strengths are associated with the
dominant $k$-point dependent interband transitions represented by arrows in the band structure plots of Fig.~\ref{optics}(d)-\ref{optics}(f). The width of the arrows is proportional to the corresponding amplitude of BSE eigenvectors. This analysis clearly shows that the $\alpha$ peak arises from transition from the
$J_\text{eff}$=1/2 LHB to the $J_\text{eff}$=1/2 UHB, whereas the $\beta$ peak comes from $J_\text{eff}$=3/2 to
$J_\text{eff}$=1/2 UHB excitations. This is also schematically shown in the DOS given in Fig.~\ref{optics}(g)-\ref{optics}(i). A closer look to the interband transitions shows that the $\alpha$-type excitations are particularly strong for Sr$_2$IrO$_4$ and, to a lesser extent, Sr$_3$Ir$_2$O$_7$ along the $X$-$M$ direction, owing to the fact that the $J_\text{eff}$=1/2 LHB and UHB are rather flat (small band width) and parallel. This peculiar band topology leads to a significant enhancement of the joint DOS (not shown) and favors intense transitions localized in a relatively small energy windows. Indeed in Sr$_2$IrO$_4$, the $\alpha$ peak is narrower than the $\beta$ peak.
Moving to  Sr$_3$Ir$_2$O$_7$, the LHB and UHB split and are less parallel than those in
Sr$_2$IrO$_4$. As a result, the $\alpha$ peak is less intense and broader.
The $J_\text{eff}$=1/2 band topology is strongly perturbed in SrIrO$_3$ due to the substantial hybridization between  Ir-5$d$ and O-2$p$ orbitals originating from the underlying three-dimensional crystal structure with distorted orthorhombic symmetry. Ultimately, this leads to an admixing of the $J_\text{eff}$=3/2 states with the lower $J_\text{eff}$=1/2 band. As a result, the characteristic $\alpha$ and $\beta$ peaks are much broader. However, the agreement with experiment is not satisfactory. The reasons are twofold: First, DFT and $GW$ do not properly account for the strong bandwidth renormalization observed experimentally~\cite{PhysRevLett.114.016401} which drastically changes the band topology near the Fermi level and thus affects the optical excitations. Moreover, the experimental temperature evolution
in the optical spectra of Sr$_2$IrO$_4$ showed a large electron-phonon interaction~\cite{Sohn2014_optics},
which is completely neglected in our calculations.
Second, it appears that there are experimental complications
(difficulties in synthesizing stoichiometric crystals, degradation in ambient conditions, sensitivity to lithographic processing, presence of oxygen vacancies) that make it difficult to perform systematic and reproducible measurements of transport properties~\cite{doi:10.1063/1.4960101, Kazunori2016,doi:10.1080/14786435.2015.1134835}.
This clearly hinders a direct comparison with theory. In fact, depending on the specific type of sample (polycrystalline~\cite{Fujioka2017} or SrIrO$_3$ films grown on MgO~\cite{Moon_2008_MIT} or SrTiO$_3$~\cite{Kim2016_SL}) different optical conductivity spectra have been reported in literatures which differ even in fundamental aspects
such as the presence or absence of the $\alpha$ peak.

As a final note on the evolution of the electron and optical properties as a function of dimensionality, we found that by going from $n=1$ to $n=\infty$, the $\alpha$ and $\beta$  peaks are progressively shifted towards lower energies, in agreement with observations. This trend is correlated with the progressive decrease of the effective interaction $U_{\rm eff}$ and with the gradual closing of the gap, as summarized in Fig.~\ref{fig:U_alpha_beta} and in line with experimental observations~\cite{Moon_2008_MIT}.

\subsubsection{Model BSE}

\begin{figure*}
\begin{center}
\includegraphics*[width=0.85\textwidth]{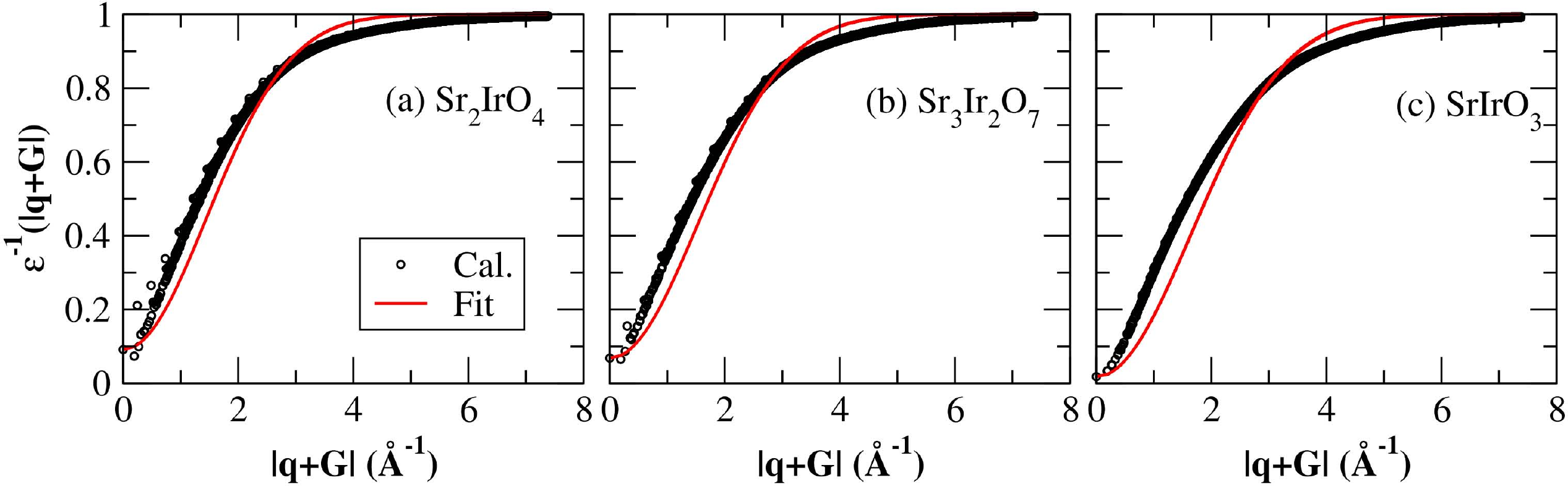}
\end{center}
\caption{The computed inverse of the dielectric function ${\varepsilon}^{-1}$ with respect to $\mid$\textbf{q+G}$\mid$  for
three iridates. The red cure is obtained by fitting based on Eq.~(\ref{eq:mBSE}). }
\label{screen_G}
\end{figure*}

\begin{table}
\footnotesize
\caption{The calculated static ion-clamped dielectric function ${\varepsilon}_{\infty}$
and  the screening length parameter $\lambda$ ($\AA^{-1}$) used in mBSE [Eq.~(\ref{eq:mBSE})]
for three iridates.
}
\begin{ruledtabular}
\begin{tabular}{cccc}
            & Sr$_2$IrO$_4$   & Sr$_3$Ir$_2$O$_7$  & SrIrO$_3$  \\
\hline
  ${\varepsilon}_{\infty}$  & 10.989 & 14.706 &  55.556 \\
 $\lambda$    & 1.026 & 1.090 & 1.165
\end{tabular}
\end{ruledtabular}
\label{Table_model}
\end{table}

The $GW$+BSE approach used in the above section to compute the optical spectra is a reliable and predictive scheme that generally delivers high-quality results provided that the input band structure and screening properties are well described. However, the calculations are computationally very demanding owing to the slow-convergence of the BSE spectrum with respect to the density of $k$ points. This makes the $GW$+BSE calculations on large systems prohibitive.
To overcome this issue, a less expensive but robust mBSE approach  was proposed
~\cite{BECHSTEDT1992765,PhysRevB.78.085103,Bokdam2016}, which is based on two (generally valid) approximations:
\begin{enumerate}
 \item[(i)] The RPA static screening $W$ is replaced by a simple analytical model, given in
 Eq.~(\ref{eq:mBSE}). The parameters are determined through a fitting of the RPA static screening computed
 for a standard $k$-point grid. The parameters obtained for the RP iridates are collected in Table~\ref{Table_model}. To demonstrate the quality of the fitting we provide in Fig.~\ref{screen_G} the comparison between the RPA  and model inverse of the dielectric function ${\varepsilon}^{-1}$ as a function of $\mid$\textbf{q+G}$\mid$ using the standard 6$\times$6$\times$1 ($n$=1 and $n$=2) and 6$\times$6$\times$4 ($n=\infty$) $k$-point grids. In passing, we note here that using a hybrid PBE0 approach would be equivalent to adopting a constant inverse of the dielectric function (${\varepsilon}^{-1}$=0.25), resulting in a much worse description of the screening.
 \item[(ii)] The $GW$+SOC QP energies are replaced by the corresponding DFT+$U_\text{eff}$+SOC one-electron energies.
 For the RP iridates family this is a valid approximation, as shown by the comparison between the $GW$ and DFT+$U_\text{eff}$ band structure given in Fig.~\ref{GW_bands}.
\end{enumerate}

\begin{figure*}
\begin{center}
\includegraphics*[width=0.90\textwidth]{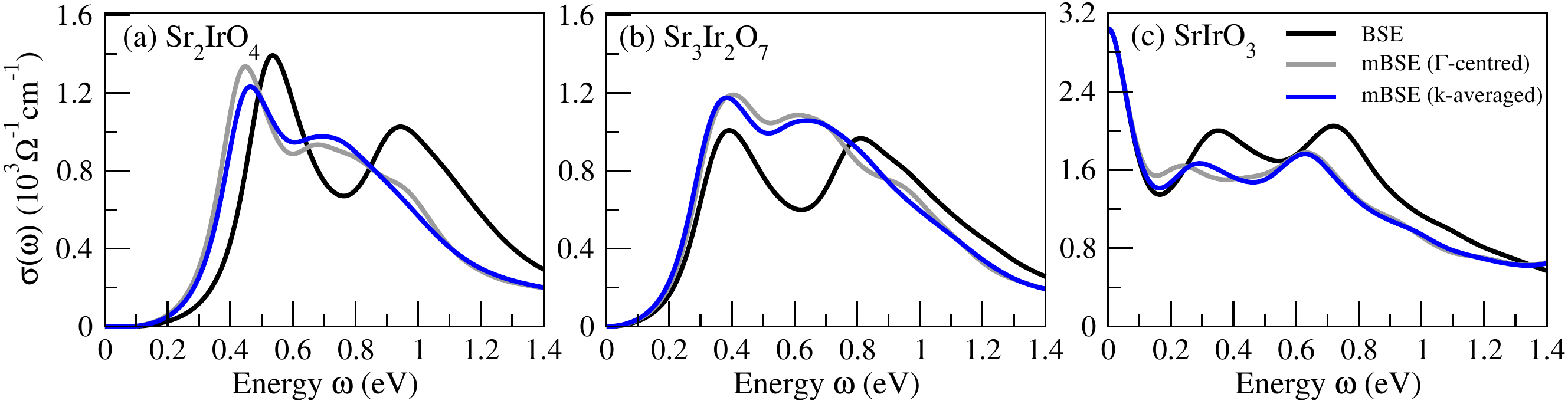}
\end{center}
\caption{The mBSE calculated spectra ($k$-averaged and $\Gamma$-centered) as compared to the spectra from BSE (MP shifted $k$-mesh). }
\label{fig:mBSE_BSE}
\end{figure*}

Fig.~\ref{fig:mBSE_BSE} shows the optical conductivity spectra predicted by mBSE (both the $k$-averaged and the $\Gamma$-centered one) as compared to BSE. For the full BSE calculations (also those shown in the previous section) we have adopted the shifted $k$-mesh that best reproduced the averaged mBSE spectra.
It is important to remark that using a $\Gamma$-center grid leads to the appearance of a slight spurious peak for the $n$=1 and $n$=2 systems located at around 1~eV, that is not seen in experiments. The $k$-averaging washes out this peak and improves the agreement with the experimental and BSE results.
In general, the agreement between BSE and the $k$-averaged mBSE spectra is not particularly good. Even though the two-peak structure is correctly reproduced,  the $\beta$ peak is systematically redshifted within the mBSE approach. This is
primarily due to
the upward shift of valence bands predicted by DFT+$U_\text{eff}$+SOC calculations, in particular the $J_\text{eff}$=3/2 states,
and to a lesser extent, due to the increased modeled dielectric function $\varepsilon$
in the long-wave range compared to the RPA (see Fig.~\ref{screen_G}).
Both facts reduce the separation between valence and conduction bands, resulting in a decrease of the excitation energies (redshift).

Summing up, mBSE is not capable to fully reproduce the outcome of a full BSE calculation, at least in the present case, but represents a viable compromise for extracting the main optical characteristics for large systems (e.g., the two-peak structure).

\section{Conclusion}

In conclusion, we have studied the electronic and optical properties
of the RP iridates of Sr$_{n+1}$Ir$_n$O$_{3n+1}$  ($n$=1, 2 and $\infty$)
by \emph{ ab initio} $GW$+BSE calculations including SOC.
The computed optical conductivity spectra reproduce well the experimentally observed double-peak structure
and describe well the progressive redshift of the main optical peaks as a function of dimensionality.
Though no bound exciton state is observed,  the computed spectra demonstrate strong excitonic effects.
In addition, we calculated the Coulomb repulsions $U$ and exchange interactions $J$ via cRPA,
showing that the correlation is reduced as $n$ increases, consistent with the gradual redshift of the
$\alpha$ and $\beta$ peaks in the optical spectra.
The comparative study between DFT+$U_\text{eff}$+SOC ($U_\text{eff}$ from cRPA)
and $GW$+SOC shows that both methods yield similar band structures
(except that $GW$ pushes down the O-2$p$ states by about 0.5 eV) and band gaps
for  Sr$_2$IrO$_4$ and Sr$_3$Ir$_2$O$_7$ in excellent agreement with measurements.
However, as expected both methods fail to describe the correlated metallic state of SrIrO$_3$,
incorrectly predicting a pseudogap at the Fermi level, and $GW$ finds a mass enhancement of only 1.64, largely underestimated compared to experimental estimations.
This deficiency  clearly influences the overall quality of the optical conductivity spectra of SrIrO$_3$
and implies that going beyond the $GW$ approximation is required in order to achieve a satisfactory account of the correlated metallic state.
Finally, we have assessed the performance of a model BSE approach which uses as input a
model dielectric function and  DFT+$U$ one-particle energies. The advantage of this scheme is the low
computational cost. It hence allows for an inspection of the convergence of the spectra with respect to the $k$-point sampling,
in particular for large systems where $GW$+BSE calculations are prohibitive.
We found that mBSE qualitatively captures the chracteristic two-peak structure but the overall spectra deviates substantially from those
obtained from the full $GW$+BSE procedure.

\section*{Acknowledgements}

P. Liu thanks M. Bokdam for helpful discussions.
This work was supported by the Austrian Science Fund (FWF) within the SFB ViCoM (Grant No. F41),
and by the joint DST (Indian Department of Science and Technology)-FWF project INDOX (I1490-N19).
X.-Q. Chen was supported by the National Science Fund for Distinguished Young Scholars (No. 51725103).
Supercomputing time on the Vienna Scientific cluster (VSC) is gratefully acknowledged.

\appendix*

\section{Structural, magnetic data and BSE/DMFT vs. experiments.}\label{appendix}

\begin{figure}[h!]
\begin{center}
\includegraphics[width=0.4\textwidth]{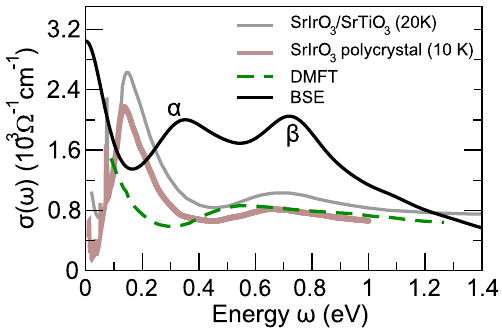}
\end{center}
\caption{Same as Fig.~\ref{optics}(c), but also showing
the experimental spectra for SrIrO$_3$ film grown on SrTiO$_3$
substrate at 20 K adapted from Kim \emph{et al.}~\cite{Kim2016_SL}.
}
\label{fig:DMFT_optics}
\end{figure}

In this appendix, we provide detailed information of
the structural data (experimental lattice constants and Wyckoff positions of relaxed and experimental crystal structures) in Table~\ref{Table_structure},
the magnetic data (DFT+$U_\text{eff}$+SOC and $GW$+SOC predicted ordered magnetic moments of Ir atoms including the orbital and spin contributions
for Sr$_2$IrO$_4$ and Sr$_3$Ir$_2$O$_7$ along with the experimental data)  in Table~\ref{Table_mag_moment}
and comparison between BSE, DMFT, and available low-temperature experimental optical conductively spectra
of SrIrO$_3$ in Fig.~\ref{fig:DMFT_optics}.

\begin{table*}[ht!]
\footnotesize
\caption{
Experimental lattice constants and Wyckoff positions of relaxed and experimental crystal structures
for three RP iridates.
}
\begin{ruledtabular}
\begin{tabular}{ccccccccc}
\multirow{7}{2.8cm}{Sr$_2$IrO$_4$ [$I4_1/acd$ (142)]} &
\multicolumn{2}{c}{ } &  \multicolumn{3}{c}{Relaxed} & \multicolumn{3}{c}{Experiments~\cite{PhysRevB.49.9198}} \\
 \cline{4-6}  \cline{7-9}
& \multicolumn{2}{c}{ Lattice constant ($\AA$)} &  \multicolumn{3}{c}{  } & $a$=5.4846 & $b$=5.4846 & $c$=25.804 \\
 \cline{2-9}
& Atom & Wyckoff  & $x$ & $y$ & $z$ & $x$ & $y$ & $z$ \\
& Ir & 8$a$  & 0.00000 & 0.25000 & 0.37500 & 0.00000 & 0.25000 & 0.37500 \\
& Sr & 16$d$  & 0.00000 & 0.25000 & 0.54976 & 0.00000 & 0.25000 & 0.55053 \\
& O1 & 16$d$ & 0.00000 & 0.25000 & 0.45526 & 0.00000 & 0.25000 & 0.45473 \\
& O2 & 16$f$  &  0.19127 &  0.44127  & 0.12500 & 0.19910 & 0.44910  & 0.12500 \\
\hline
\multirow{7}{2.8cm}{Sr$_3$Ir$_2$O$_7$ [$Ccce$ (68)]} &
\multicolumn{2}{c}{ } &  \multicolumn{3}{c}{Relaxed} & \multicolumn{3}{c}{Experiments~\cite{SUBRAMANIAN1994645}} \\
 \cline{4-6}  \cline{7-9}
& \multicolumn{2}{c}{ Lattice constant ($\AA$)} &  \multicolumn{3}{c}{  } & $a$=5.5098 & $b$=5.5098 & $c$=20.879 \\
 \cline{2-9}
& Atom & Wyckoff  & $x$ & $y$ & $z$ & $x$ & $y$ & $z$ \\
& Ir & 8$f$  & 0.00000 & 0.00000 & 0.09790 & 0.00000 & 0.00000 & 0.09743 \\
& Sr1 & 4$b$  & 0.00000 & 0.50000 & 0.00000 & 0.00000 & 0.50000 & 0.00000 \\
& Sr2 & 8$f$  & 0.00000 & 0.50000 & 0.18824 & 0.00000 & 0.50000 & 0.18720 \\
& O1 & 4$a$ & 0.00000 & 0.00000 & 0.00000 & 0.00000 & 0.00000 & 0.00000 \\
& O2 & 8$f$  & 0.00000 & 0.00000 & 0.19530 & 0.00000 & 0.00000 & 0.19390 \\
& O3 & 16$i$  &  0.31034 &  0.18965  & 0.09668 &  0.30215 &  0.19785   & 0.09600 \\
\hline
\multirow{7}{2.8cm}{SrIrO$_3$ [$Pbnm$ (62)]} &
\multicolumn{2}{c}{ } &  \multicolumn{3}{c}{Relaxed} & \multicolumn{3}{c}{Experiments~\cite{Zhao2008,Puggioni2016}} \\
 \cline{4-6}  \cline{7-9}
& \multicolumn{2}{c}{ Lattice constant ($\AA$)} &  \multicolumn{3}{c}{  } & $a$=5.5617 & $b$=5.5909 & $c$=7.8821 \\
 \cline{2-9}
& Atom & Wyckoff  & $x$ & $y$ & $z$ & $x$ & $y$ & $z$ \\
& Ir & 4$b$  & 0.00000 & 0.50000 & 0.00000 & 0.00000 & 0.50000 & 0.00000 \\
& Sr & 4$c$  & 0.50832 & 0.54140 & 0.25000 & 0.49010 & 0.50850 & 0.25000 \\
& O1 & 4$c$ & 0.07998 & 0.48449 & 0.25000 & 0.07300 & 0.50600 & 0.25000 \\
& O2 & 8$d$  & 0.70665 & 0.29265 & 0.04117 & 0.71400 & 0.29200 & 0.04400
\end{tabular}
\end{ruledtabular}
\label{Table_structure}
\end{table*}

\begin{table*}[ht!]
\footnotesize
\caption {Predicted ordered magnetic moments (in ${\mu}_B$) of Ir atoms including the orbital and spin contributions
for Sr$_2$IrO$_4$ and Sr$_3$Ir$_2$O$_7$ by DFT+$U_\text{eff}$+SOC and $GW$+SOC calculations.
The experimental results are also given for comparison.
}
\begin{ruledtabular}
\begin{tabular}{ccccccc}
\multirow{8}{1.8cm}{Sr$_2$IrO$_4$ DFT+$U_\text{eff}$+SOC ($U_\text{eff}$=1.6 eV)} &
 Ir-site   & Positions & Orbital  & Spin   & Total & Total (Expt.~\cite{PhysRevB.87.140406}) \\
\cline{2-7}
& 1 & (0.0, 0.0, 0.00) & ($~0.05$,$~0.26$, 0) & ($~0.03$,$~0.12$, 0) & ($~0.08$,$~0.38$, 0) & ($~0.049$,$~0.202$, 0) \\
& 2 & (0.5, 0.5, 0.00) & ($~0.05$,$-0.26$, 0) & ($~0.03$,$-0.12$, 0) & ($~0.08$,$-0.38$, 0) & ($~0.049$,$-0.202$, 0) \\
& 3 & (0.5, 0.0, 0.25) & ($-0.05$,$~0.26$, 0) & ($-0.03$,$~0.12$, 0) & ($-0.08$,$~0.38$, 0) & ($-0.049$,$~0.202$, 0) \\
& 4 & (0.0, 0.5, 0.25) & ($-0.05$,$-0.26$, 0) & ($-0.03$,$-0.12$, 0) & ($-0.08$,$-0.38$, 0) & ($-0.049$,$-0.202$, 0) \\
& 5 & (0.0, 0.0, 0.50) & ($-0.05$,$~0.26$, 0) & ($-0.03$,$~0.12$, 0) & ($-0.08$,$~0.38$, 0) & ($-0.049$,$~0.202$, 0) \\
& 6 & (0.5, 0.5, 0.50) & ($-0.05$,$-0.26$, 0) & ($-0.03$,$-0.12$, 0) & ($-0.08$,$-0.38$, 0) & ($-0.049$,$-0.202$, 0) \\
& 7 & (0.5, 0.0, 0.75) & ($~0.05$,$~0.26$, 0) & ($~0.03$,$~0.12$, 0) & ($~0.08$,$~0.38$, 0) & ($~0.049$,$~0.202$, 0) \\
& 8 & (0.0, 0.5, 0.75) & ($~0.05$,$-0.26$, 0) & ($~0.03$,$-0.12$, 0) & ($~0.08$,$-0.38$, 0) & ($~0.049$,$-0.202$, 0) \\
\hline
\multirow{8}{1.8cm}{Sr$_2$IrO$_4$ $GW$+SOC} &
 Ir-site   & Positions & Orbital  & Spin   & Total & Total (Expt.~\cite{PhysRevB.87.140406}) \\
\cline{2-7}
& 1 & (0.0, 0.0, 0.00) & ($~0.05$,$~0.21$, 0) & ($~0.04$,$~0.10$, 0) & ($~0.09$,$~0.31$, 0)  & ($~0.049$,$~0.202$, 0) \\
& 2 & (0.5, 0.5, 0.00) & ($~0.05$,$-0.21$, 0) & ($~0.04$,$-0.10$, 0) & ($~0.09$,$-0.31$, 0)  & ($~0.049$,$-0.202$, 0) \\
& 3 & (0.5, 0.0, 0.25) & ($-0.05$,$~0.21$, 0) & ($-0.04$,$~0.10$, 0) & ($-0.09$,$~0.31$, 0)  & ($-0.049$,$~0.202$, 0) \\
& 4 & (0.0, 0.5, 0.25) & ($-0.05$,$-0.21$, 0) & ($-0.04$,$-0.10$, 0) & ($-0.09$,$-0.31$, 0)  & ($-0.049$,$-0.202$, 0) \\
& 5 & (0.0, 0.0, 0.50) & ($-0.05$,$~0.21$, 0) & ($-0.04$,$~0.10$, 0) & ($-0.09$,$~0.31$, 0)  & ($-0.049$,$~0.202$, 0) \\
& 6 & (0.5, 0.5, 0.50) & ($-0.05$,$-0.21$, 0) & ($-0.04$,$-0.10$, 0) & ($-0.09$,$-0.31$, 0)  & ($-0.049$,$-0.202$, 0) \\
& 7 & (0.5, 0.0, 0.75) & ($~0.05$,$~0.21$, 0) & ($~0.04$,$~0.10$, 0) & ($~0.09$,$~0.31$, 0)  & ($~0.049$,$~0.202$, 0) \\
& 8 & (0.0, 0.5, 0.75) & ($~0.05$,$-0.21$, 0) & ($~0.04$,$-0.10$, 0) & ($~0.09$,$-0.31$, 0)  & ($~0.049$,$-0.202$, 0) \\
\hline
\multirow{8}{1.8cm}{Sr$_3$Ir$_2$O$_7$ DFT+$U_\text{eff}$+SOC ($U_\text{eff}$=1.45 eV)} &
 Ir-site   & Positions & Orbital  & Spin   & Total & Total (Expt.~\cite{PhysRevB.86.174414}) \\
\cline{2-7}
& 1 & (0.0, 0.0, 0.0979) & (0, 0,$~0.29$) & (0, 0,$~0.23$) & (0, 0,$~0.52$) & (0, 0,$~0.1$) \\
& 2 & (0.5, 0.5, 0.0979) & (0, 0,$-0.29$) & (0, 0,$-0.23$) & (0, 0,$-0.52$) & (0, 0,$-0.1$) \\
& 3 & (0.0, 0.5, 0.4021) & (0, 0,$~0.29$) & (0, 0,$~0.23$) & (0, 0,$~0.52$) & (0, 0,$~0.1$) \\
& 4 & (0.5, 0.0, 0.4021) & (0, 0,$-0.29$) & (0, 0,$-0.23$) & (0, 0,$-0.52$) & (0, 0,$-0.1$) \\
& 5 & (0.0, 0.5, 0.5979) & (0, 0,$-0.29$) & (0, 0,$-0.23$) & (0, 0,$-0.52$) & (0, 0,$-0.1$) \\
& 6 & (0.5, 0.0, 0.5979) & (0, 0,$~0.29$) & (0, 0,$~0.23$) & (0, 0,$~0.52$) & (0, 0,$~0.1$) \\
& 7 & (0.0, 0.0, 0.9021) & (0, 0,$-0.29$) & (0, 0,$-0.23$) & (0, 0,$-0.52$) & (0, 0,$-0.1$) \\
& 8 & (0.5, 0.5, 0.9021) & (0, 0,$~0.29$) & (0, 0,$~0.23$) & (0, 0,$~0.52$) & (0, 0,$~0.1$) \\
\hline
\multirow{8}{1.8cm}{Sr$_3$Ir$_2$O$_7$ $GW$+SOC} &
 Ir-site   & Positions & Orbital  & Spin   & Total & Total (Expt.~\cite{PhysRevB.86.174414}) \\
\cline{2-7}
& 1 & (0.0, 0.0, 0.0979) & (0, 0,$~0.25$) & (0, 0,$~0.20$) & (0, 0,$~0.45$)   & (0, 0,$~0.1$) \\
& 2 & (0.5, 0.5, 0.0979) & (0, 0,$-0.25$) & (0, 0,$-0.20$) & (0, 0,$-0.45$)   & (0, 0,$-0.1$) \\
& 3 & (0.0, 0.5, 0.4021) & (0, 0,$~0.25$) & (0, 0,$~0.20$) & (0, 0,$~0.45$)   & (0, 0,$~0.1$) \\
& 4 & (0.5, 0.0, 0.4021) & (0, 0,$-0.25$) & (0, 0,$-0.20$) & (0, 0,$-0.45$)   & (0, 0,$-0.1$) \\
& 5 & (0.0, 0.5, 0.5979) & (0, 0,$-0.25$) & (0, 0,$-0.20$) & (0, 0,$-0.45$)   & (0, 0,$-0.1$) \\
& 6 & (0.5, 0.0, 0.5979) & (0, 0,$~0.25$) & (0, 0,$~0.20$) & (0, 0,$~0.45$)   & (0, 0,$~0.1$) \\
& 7 & (0.0, 0.0, 0.9021) & (0, 0,$-0.25$) & (0, 0,$-0.20$) & (0, 0,$-0.45$)   & (0, 0,$-0.1$) \\
& 8 & (0.5, 0.5, 0.9021) & (0, 0,$~0.25$) & (0, 0,$~0.20$) & (0, 0,$~0.45$)   & (0, 0,$~0.1$)
\end{tabular}
\end{ruledtabular}
\label{Table_mag_moment}
\end{table*}

\bibliographystyle{apsrev4-1}
\bibliography{reference} 

\begin{thebibliography}{111}%
\makeatletter
\providecommand \@ifxundefined [1]{%
 \@ifx{#1\undefined}
}%
\providecommand \@ifnum [1]{%
 \ifnum #1\expandafter \@firstoftwo
 \else \expandafter \@secondoftwo
 \fi
}%
\providecommand \@ifx [1]{%
 \ifx #1\expandafter \@firstoftwo
 \else \expandafter \@secondoftwo
 \fi
}%
\providecommand \natexlab [1]{#1}%
\providecommand \enquote  [1]{``#1''}%
\providecommand \bibnamefont  [1]{#1}%
\providecommand \bibfnamefont [1]{#1}%
\providecommand \citenamefont [1]{#1}%
\providecommand \href@noop [0]{\@secondoftwo}%
\providecommand \href [0]{\begingroup \@sanitize@url \@href}%
\providecommand \@href[1]{\@@startlink{#1}\@@href}%
\providecommand \@@href[1]{\endgroup#1\@@endlink}%
\providecommand \@sanitize@url [0]{\catcode `\\12\catcode `\$12\catcode
  `\&12\catcode `\#12\catcode `\^12\catcode `\_12\catcode `\%12\relax}%
\providecommand \@@startlink[1]{}%
\providecommand \@@endlink[0]{}%
\providecommand \url  [0]{\begingroup\@sanitize@url \@url }%
\providecommand \@url [1]{\endgroup\@href {#1}{\urlprefix }}%
\providecommand \urlprefix  [0]{URL }%
\providecommand \Eprint [0]{\href }%
\providecommand \doibase [0]{http://dx.doi.org/}%
\providecommand \selectlanguage [0]{\@gobble}%
\providecommand \bibinfo  [0]{\@secondoftwo}%
\providecommand \bibfield  [0]{\@secondoftwo}%
\providecommand \translation [1]{[#1]}%
\providecommand \BibitemOpen [0]{}%
\providecommand \bibitemStop [0]{}%
\providecommand \bibitemNoStop [0]{.\EOS\space}%
\providecommand \EOS [0]{\spacefactor3000\relax}%
\providecommand \BibitemShut  [1]{\csname bibitem#1\endcsname}%
\let\auto@bib@innerbib\@empty
\bibitem [{\citenamefont {Kim}\ \emph {et~al.}(2008)\citenamefont {Kim},
  \citenamefont {Jin}, \citenamefont {Moon}, \citenamefont {Kim}, \citenamefont
  {Park}, \citenamefont {Leem}, \citenamefont {Yu}, \citenamefont {Noh},
  \citenamefont {Kim}, \citenamefont {Oh}, \citenamefont {Park}, \citenamefont
  {Durairaj}, \citenamefont {Cao},\ and\ \citenamefont
  {Rotenberg}}]{Kim2008_PRL}%
  \BibitemOpen
  \bibfield  {author} {\bibinfo {author} {\bibfnamefont {B.~J.}\ \bibnamefont
  {Kim}}, \bibinfo {author} {\bibfnamefont {H.}~\bibnamefont {Jin}}, \bibinfo
  {author} {\bibfnamefont {S.~J.}\ \bibnamefont {Moon}}, \bibinfo {author}
  {\bibfnamefont {J.-Y.}\ \bibnamefont {Kim}}, \bibinfo {author} {\bibfnamefont
  {B.-G.}\ \bibnamefont {Park}}, \bibinfo {author} {\bibfnamefont {C.~S.}\
  \bibnamefont {Leem}}, \bibinfo {author} {\bibfnamefont {J.}~\bibnamefont
  {Yu}}, \bibinfo {author} {\bibfnamefont {T.~W.}\ \bibnamefont {Noh}},
  \bibinfo {author} {\bibfnamefont {C.}~\bibnamefont {Kim}}, \bibinfo {author}
  {\bibfnamefont {S.-J.}\ \bibnamefont {Oh}}, \bibinfo {author} {\bibfnamefont
  {J.-H.}\ \bibnamefont {Park}}, \bibinfo {author} {\bibfnamefont
  {V.}~\bibnamefont {Durairaj}}, \bibinfo {author} {\bibfnamefont
  {G.}~\bibnamefont {Cao}}, \ and\ \bibinfo {author} {\bibfnamefont
  {E.}~\bibnamefont {Rotenberg}},\ }\href {\doibase
  10.1103/PhysRevLett.101.076402} {\bibfield  {journal} {\bibinfo  {journal}
  {Phys. Rev. Lett.}\ }\textbf {\bibinfo {volume} {101}},\ \bibinfo {pages}
  {076402} (\bibinfo {year} {2008})}\BibitemShut {NoStop}%
\bibitem [{\citenamefont {Moon}\ \emph {et~al.}(2008)\citenamefont {Moon},
  \citenamefont {Jin}, \citenamefont {Kim}, \citenamefont {Choi}, \citenamefont
  {Lee}, \citenamefont {Yu}, \citenamefont {Cao}, \citenamefont {Sumi},
  \citenamefont {Funakubo}, \citenamefont {Bernhard},\ and\ \citenamefont
  {Noh}}]{Moon_2008_MIT}%
  \BibitemOpen
  \bibfield  {author} {\bibinfo {author} {\bibfnamefont {S.~J.}\ \bibnamefont
  {Moon}}, \bibinfo {author} {\bibfnamefont {H.}~\bibnamefont {Jin}}, \bibinfo
  {author} {\bibfnamefont {K.~W.}\ \bibnamefont {Kim}}, \bibinfo {author}
  {\bibfnamefont {W.~S.}\ \bibnamefont {Choi}}, \bibinfo {author}
  {\bibfnamefont {Y.~S.}\ \bibnamefont {Lee}}, \bibinfo {author} {\bibfnamefont
  {J.}~\bibnamefont {Yu}}, \bibinfo {author} {\bibfnamefont {G.}~\bibnamefont
  {Cao}}, \bibinfo {author} {\bibfnamefont {A.}~\bibnamefont {Sumi}}, \bibinfo
  {author} {\bibfnamefont {H.}~\bibnamefont {Funakubo}}, \bibinfo {author}
  {\bibfnamefont {C.}~\bibnamefont {Bernhard}}, \ and\ \bibinfo {author}
  {\bibfnamefont {T.~W.}\ \bibnamefont {Noh}},\ }\href {\doibase
  10.1103/PhysRevLett.101.226402} {\bibfield  {journal} {\bibinfo  {journal}
  {Phys. Rev. Lett.}\ }\textbf {\bibinfo {volume} {101}},\ \bibinfo {pages}
  {226402} (\bibinfo {year} {2008})}\BibitemShut {NoStop}%
\bibitem [{\citenamefont {Kim}\ \emph {et~al.}(2009)\citenamefont {Kim},
  \citenamefont {Ohsumi}, \citenamefont {Komesu}, \citenamefont {Sakai},
  \citenamefont {Morita}, \citenamefont {Takagi},\ and\ \citenamefont
  {Arima}}]{Kim2009_science}%
  \BibitemOpen
  \bibfield  {author} {\bibinfo {author} {\bibfnamefont {B.~J.}\ \bibnamefont
  {Kim}}, \bibinfo {author} {\bibfnamefont {H.}~\bibnamefont {Ohsumi}},
  \bibinfo {author} {\bibfnamefont {T.}~\bibnamefont {Komesu}}, \bibinfo
  {author} {\bibfnamefont {S.}~\bibnamefont {Sakai}}, \bibinfo {author}
  {\bibfnamefont {T.}~\bibnamefont {Morita}}, \bibinfo {author} {\bibfnamefont
  {H.}~\bibnamefont {Takagi}}, \ and\ \bibinfo {author} {\bibfnamefont
  {T.}~\bibnamefont {Arima}},\ }\href {\doibase 10.1126/science.1167106}
  {\bibfield  {journal} {\bibinfo  {journal} {Science}\ }\textbf {\bibinfo
  {volume} {323}},\ \bibinfo {pages} {1329} (\bibinfo {year}
  {2009})}\BibitemShut {NoStop}%
\bibitem [{\citenamefont {Jackeli}\ and\ \citenamefont
  {Khaliullin}(2009)}]{PhysRevLett.102.017205}%
  \BibitemOpen
  \bibfield  {author} {\bibinfo {author} {\bibfnamefont {G.}~\bibnamefont
  {Jackeli}}\ and\ \bibinfo {author} {\bibfnamefont {G.}~\bibnamefont
  {Khaliullin}},\ }\href {\doibase 10.1103/PhysRevLett.102.017205} {\bibfield
  {journal} {\bibinfo  {journal} {Phys. Rev. Lett.}\ }\textbf {\bibinfo
  {volume} {102}},\ \bibinfo {pages} {017205} (\bibinfo {year}
  {2009})}\BibitemShut {NoStop}%
\bibitem [{\citenamefont {Shitade}\ \emph {et~al.}(2009)\citenamefont
  {Shitade}, \citenamefont {Katsura}, \citenamefont
  {Kune\ifmmode~\check{s}\else \v{s}\fi{}}, \citenamefont {Qi}, \citenamefont
  {Zhang},\ and\ \citenamefont {Nagaosa}}]{Shitade2009}%
  \BibitemOpen
  \bibfield  {author} {\bibinfo {author} {\bibfnamefont {A.}~\bibnamefont
  {Shitade}}, \bibinfo {author} {\bibfnamefont {H.}~\bibnamefont {Katsura}},
  \bibinfo {author} {\bibfnamefont {J.}~\bibnamefont
  {Kune\ifmmode~\check{s}\else \v{s}\fi{}}}, \bibinfo {author} {\bibfnamefont
  {X.-L.}\ \bibnamefont {Qi}}, \bibinfo {author} {\bibfnamefont {S.-C.}\
  \bibnamefont {Zhang}}, \ and\ \bibinfo {author} {\bibfnamefont
  {N.}~\bibnamefont {Nagaosa}},\ }\href {\doibase
  10.1103/PhysRevLett.102.256403} {\bibfield  {journal} {\bibinfo  {journal}
  {Phys. Rev. Lett.}\ }\textbf {\bibinfo {volume} {102}},\ \bibinfo {pages}
  {256403} (\bibinfo {year} {2009})}\BibitemShut {NoStop}%
\bibitem [{\citenamefont {Chaloupka}\ \emph {et~al.}(2010)\citenamefont
  {Chaloupka}, \citenamefont {Jackeli},\ and\ \citenamefont
  {Khaliullin}}]{Chaloupka2010}%
  \BibitemOpen
  \bibfield  {author} {\bibinfo {author} {\bibfnamefont {J.}~\bibnamefont
  {Chaloupka}}, \bibinfo {author} {\bibfnamefont {G.}~\bibnamefont {Jackeli}},
  \ and\ \bibinfo {author} {\bibfnamefont {G.}~\bibnamefont {Khaliullin}},\
  }\href {\doibase 10.1103/PhysRevLett.105.027204} {\bibfield  {journal}
  {\bibinfo  {journal} {Phys. Rev. Lett.}\ }\textbf {\bibinfo {volume} {105}},\
  \bibinfo {pages} {027204} (\bibinfo {year} {2010})}\BibitemShut {NoStop}%
\bibitem [{\citenamefont {Pesin}\ and\ \citenamefont
  {Balents}(2010)}]{Pesin2010}%
  \BibitemOpen
  \bibfield  {author} {\bibinfo {author} {\bibfnamefont {D.}~\bibnamefont
  {Pesin}}\ and\ \bibinfo {author} {\bibfnamefont {L.}~\bibnamefont
  {Balents}},\ }\href {https://www.nature.com/articles/nphys1606} {\bibfield
  {journal} {\bibinfo  {journal} {Nature Physics}\ }\textbf {\bibinfo {volume}
  {6}},\ \bibinfo {pages} {376} (\bibinfo {year} {2010})}\BibitemShut {NoStop}%
\bibitem [{\citenamefont {Kim}\ \emph {et~al.}(2012{\natexlab{a}})\citenamefont
  {Kim}, \citenamefont {Choi}, \citenamefont {Kim}, \citenamefont {Mitchell},
  \citenamefont {Jackeli}, \citenamefont {Daghofer}, \citenamefont {van~den
  Brink}, \citenamefont {Khaliullin},\ and\ \citenamefont
  {Kim}}]{Khaliullin2012}%
  \BibitemOpen
  \bibfield  {author} {\bibinfo {author} {\bibfnamefont {J.~W.}\ \bibnamefont
  {Kim}}, \bibinfo {author} {\bibfnamefont {Y.}~\bibnamefont {Choi}}, \bibinfo
  {author} {\bibfnamefont {J.}~\bibnamefont {Kim}}, \bibinfo {author}
  {\bibfnamefont {J.~F.}\ \bibnamefont {Mitchell}}, \bibinfo {author}
  {\bibfnamefont {G.}~\bibnamefont {Jackeli}}, \bibinfo {author} {\bibfnamefont
  {M.}~\bibnamefont {Daghofer}}, \bibinfo {author} {\bibfnamefont
  {J.}~\bibnamefont {van~den Brink}}, \bibinfo {author} {\bibfnamefont
  {G.}~\bibnamefont {Khaliullin}}, \ and\ \bibinfo {author} {\bibfnamefont
  {B.~J.}\ \bibnamefont {Kim}},\ }\href {\doibase
  10.1103/PhysRevLett.109.037204} {\bibfield  {journal} {\bibinfo  {journal}
  {Phys. Rev. Lett.}\ }\textbf {\bibinfo {volume} {109}},\ \bibinfo {pages}
  {037204} (\bibinfo {year} {2012}{\natexlab{a}})}\BibitemShut {NoStop}%
\bibitem [{\citenamefont {Boseggia}\ \emph {et~al.}(2013)\citenamefont
  {Boseggia}, \citenamefont {Springell}, \citenamefont {Walker}, \citenamefont
  {R\o{}nnow}, \citenamefont {R\"uegg}, \citenamefont {Okabe}, \citenamefont
  {Isobe}, \citenamefont {Perry}, \citenamefont {Collins},\ and\ \citenamefont
  {McMorrow}}]{PhysRevLett.110.117207}%
  \BibitemOpen
  \bibfield  {author} {\bibinfo {author} {\bibfnamefont {S.}~\bibnamefont
  {Boseggia}}, \bibinfo {author} {\bibfnamefont {R.}~\bibnamefont {Springell}},
  \bibinfo {author} {\bibfnamefont {H.~C.}\ \bibnamefont {Walker}}, \bibinfo
  {author} {\bibfnamefont {H.~M.}\ \bibnamefont {R\o{}nnow}}, \bibinfo {author}
  {\bibfnamefont {C.}~\bibnamefont {R\"uegg}}, \bibinfo {author} {\bibfnamefont
  {H.}~\bibnamefont {Okabe}}, \bibinfo {author} {\bibfnamefont
  {M.}~\bibnamefont {Isobe}}, \bibinfo {author} {\bibfnamefont {R.~S.}\
  \bibnamefont {Perry}}, \bibinfo {author} {\bibfnamefont {S.~P.}\ \bibnamefont
  {Collins}}, \ and\ \bibinfo {author} {\bibfnamefont {D.~F.}\ \bibnamefont
  {McMorrow}},\ }\href {\doibase 10.1103/PhysRevLett.110.117207} {\bibfield
  {journal} {\bibinfo  {journal} {Phys. Rev. Lett.}\ }\textbf {\bibinfo
  {volume} {110}},\ \bibinfo {pages} {117207} (\bibinfo {year}
  {2013})}\BibitemShut {NoStop}%
\bibitem [{\citenamefont {Okada}\ \emph {et~al.}(2013)\citenamefont {Okada},
  \citenamefont {Walkup}, \citenamefont {Lin}, \citenamefont {Dhital},
  \citenamefont {Chang}, \citenamefont {Khadka}, \citenamefont {Zhou},
  \citenamefont {Jeng}, \citenamefont {Paranjape}, \citenamefont {Bansil},
  \citenamefont {Wang}, \citenamefont {Wilson},\ and\ \citenamefont
  {Madhavan}}]{Okada2013}%
  \BibitemOpen
  \bibfield  {author} {\bibinfo {author} {\bibfnamefont {Y.}~\bibnamefont
  {Okada}}, \bibinfo {author} {\bibfnamefont {D.}~\bibnamefont {Walkup}},
  \bibinfo {author} {\bibfnamefont {H.}~\bibnamefont {Lin}}, \bibinfo {author}
  {\bibfnamefont {C.}~\bibnamefont {Dhital}}, \bibinfo {author} {\bibfnamefont
  {T.-R.}\ \bibnamefont {Chang}}, \bibinfo {author} {\bibfnamefont
  {S.}~\bibnamefont {Khadka}}, \bibinfo {author} {\bibfnamefont
  {W.}~\bibnamefont {Zhou}}, \bibinfo {author} {\bibfnamefont {H.-T.}\
  \bibnamefont {Jeng}}, \bibinfo {author} {\bibfnamefont {M.}~\bibnamefont
  {Paranjape}}, \bibinfo {author} {\bibfnamefont {A.}~\bibnamefont {Bansil}},
  \bibinfo {author} {\bibfnamefont {Z.}~\bibnamefont {Wang}}, \bibinfo {author}
  {\bibfnamefont {S.~D.}\ \bibnamefont {Wilson}}, \ and\ \bibinfo {author}
  {\bibfnamefont {V.}~\bibnamefont {Madhavan}},\ }\href
  {https://www.nature.com/articles/nmat3653} {\bibfield  {journal} {\bibinfo
  {journal} {Nature materials}\ }\textbf {\bibinfo {volume} {12}},\ \bibinfo
  {pages} {707} (\bibinfo {year} {2013})}\BibitemShut {NoStop}%
\bibitem [{\citenamefont {Kim}\ \emph {et~al.}(2014{\natexlab{a}})\citenamefont
  {Kim}, \citenamefont {Daghofer}, \citenamefont {Said}, \citenamefont {Gog},
  \citenamefont {van~den Brink}, \citenamefont {Khaliullin},\ and\
  \citenamefont {Kim}}]{Kim2014_natcom}%
  \BibitemOpen
  \bibfield  {author} {\bibinfo {author} {\bibfnamefont {J.}~\bibnamefont
  {Kim}}, \bibinfo {author} {\bibfnamefont {M.}~\bibnamefont {Daghofer}},
  \bibinfo {author} {\bibfnamefont {A.~H.}\ \bibnamefont {Said}}, \bibinfo
  {author} {\bibfnamefont {T.}~\bibnamefont {Gog}}, \bibinfo {author}
  {\bibfnamefont {J.}~\bibnamefont {van~den Brink}}, \bibinfo {author}
  {\bibfnamefont {G.}~\bibnamefont {Khaliullin}}, \ and\ \bibinfo {author}
  {\bibfnamefont {B.~J.}\ \bibnamefont {Kim}},\ }\href
  {https://www.nature.com/articles/ncomms5453} {\bibfield  {journal} {\bibinfo
  {journal} {Nat Commun}\ }\textbf {\bibinfo {volume} {5}},\ \bibinfo {pages}
  {4453} (\bibinfo {year} {2014}{\natexlab{a}})}\BibitemShut {NoStop}%
\bibitem [{\citenamefont {Cao}\ \emph {et~al.}(2014)\citenamefont {Cao},
  \citenamefont {Qi}, \citenamefont {Li}, \citenamefont {Terzic}, \citenamefont
  {Yuan}, \citenamefont {DeLong}, \citenamefont {Murthy},\ and\ \citenamefont
  {Kaul}}]{PhysRevLett.112.056402}%
  \BibitemOpen
  \bibfield  {author} {\bibinfo {author} {\bibfnamefont {G.}~\bibnamefont
  {Cao}}, \bibinfo {author} {\bibfnamefont {T.~F.}\ \bibnamefont {Qi}},
  \bibinfo {author} {\bibfnamefont {L.}~\bibnamefont {Li}}, \bibinfo {author}
  {\bibfnamefont {J.}~\bibnamefont {Terzic}}, \bibinfo {author} {\bibfnamefont
  {S.~J.}\ \bibnamefont {Yuan}}, \bibinfo {author} {\bibfnamefont {L.~E.}\
  \bibnamefont {DeLong}}, \bibinfo {author} {\bibfnamefont {G.}~\bibnamefont
  {Murthy}}, \ and\ \bibinfo {author} {\bibfnamefont {R.~K.}\ \bibnamefont
  {Kaul}},\ }\href {\doibase 10.1103/PhysRevLett.112.056402} {\bibfield
  {journal} {\bibinfo  {journal} {Phys. Rev. Lett.}\ }\textbf {\bibinfo
  {volume} {112}},\ \bibinfo {pages} {056402} (\bibinfo {year}
  {2014})}\BibitemShut {NoStop}%
\bibitem [{\citenamefont {Zhao}\ \emph {et~al.}(2015)\citenamefont {Zhao},
  \citenamefont {Torchinsky}, \citenamefont {Chu}, \citenamefont {Ivanov},
  \citenamefont {Lifshitz}, \citenamefont {Flint}, \citenamefont {Qi},
  \citenamefont {Cao},\ and\ \citenamefont {Hsieh}}]{Zhao2015}%
  \BibitemOpen
  \bibfield  {author} {\bibinfo {author} {\bibfnamefont {L.}~\bibnamefont
  {Zhao}}, \bibinfo {author} {\bibfnamefont {D.~H.}\ \bibnamefont
  {Torchinsky}}, \bibinfo {author} {\bibfnamefont {H.}~\bibnamefont {Chu}},
  \bibinfo {author} {\bibfnamefont {V.}~\bibnamefont {Ivanov}}, \bibinfo
  {author} {\bibfnamefont {R.}~\bibnamefont {Lifshitz}}, \bibinfo {author}
  {\bibfnamefont {R.}~\bibnamefont {Flint}}, \bibinfo {author} {\bibfnamefont
  {T.}~\bibnamefont {Qi}}, \bibinfo {author} {\bibfnamefont {G.}~\bibnamefont
  {Cao}}, \ and\ \bibinfo {author} {\bibfnamefont {D.}~\bibnamefont {Hsieh}},\
  }\href {https://www.nature.com/articles/nphys3517} {\bibfield  {journal}
  {\bibinfo  {journal} {Nature Physics}\ }\textbf {\bibinfo {volume} {12}},\
  \bibinfo {pages} {32} (\bibinfo {year} {2015})}\BibitemShut {NoStop}%
\bibitem [{\citenamefont {Hwan~Chun}\ \emph {et~al.}(2015)\citenamefont
  {Hwan~Chun}, \citenamefont {Kim}, \citenamefont {Kim}, \citenamefont {Zheng},
  \citenamefont {Stoumpos}, \citenamefont {Malliakas}, \citenamefont
  {Mitchell}, \citenamefont {Mehlawat}, \citenamefont {Singh}, \citenamefont
  {Choi}, \citenamefont {Gog}, \citenamefont {Al-Zein}, \citenamefont {Sala},
  \citenamefont {Krisch}, \citenamefont {Chaloupka}, \citenamefont {Jackeli},
  \citenamefont {Khaliullin},\ and\ \citenamefont {Kim}}]{Chun2015}%
  \BibitemOpen
  \bibfield  {author} {\bibinfo {author} {\bibfnamefont {S.}~\bibnamefont
  {Hwan~Chun}}, \bibinfo {author} {\bibfnamefont {J.-W.}\ \bibnamefont {Kim}},
  \bibinfo {author} {\bibfnamefont {J.}~\bibnamefont {Kim}}, \bibinfo {author}
  {\bibfnamefont {H.}~\bibnamefont {Zheng}}, \bibinfo {author} {\bibfnamefont
  {C.}~\bibnamefont {Stoumpos}}, \bibinfo {author} {\bibfnamefont {C.~D.}\
  \bibnamefont {Malliakas}}, \bibinfo {author} {\bibfnamefont {J.~F.}\
  \bibnamefont {Mitchell}}, \bibinfo {author} {\bibfnamefont {K.}~\bibnamefont
  {Mehlawat}}, \bibinfo {author} {\bibfnamefont {Y.}~\bibnamefont {Singh}},
  \bibinfo {author} {\bibfnamefont {Y.}~\bibnamefont {Choi}}, \bibinfo {author}
  {\bibfnamefont {T.}~\bibnamefont {Gog}}, \bibinfo {author} {\bibfnamefont
  {A.}~\bibnamefont {Al-Zein}}, \bibinfo {author} {\bibfnamefont {M.~M.}\
  \bibnamefont {Sala}}, \bibinfo {author} {\bibfnamefont {M.}~\bibnamefont
  {Krisch}}, \bibinfo {author} {\bibfnamefont {J.}~\bibnamefont {Chaloupka}},
  \bibinfo {author} {\bibfnamefont {G.}~\bibnamefont {Jackeli}}, \bibinfo
  {author} {\bibfnamefont {G.}~\bibnamefont {Khaliullin}}, \ and\ \bibinfo
  {author} {\bibfnamefont {B.~J.}\ \bibnamefont {Kim}},\ }\href
  {https://doi.org/10.1038/nphys3322} {\bibfield  {journal} {\bibinfo
  {journal} {Nature Physics}\ }\textbf {\bibinfo {volume} {11}},\ \bibinfo
  {pages} {462} (\bibinfo {year} {2015})}\BibitemShut {NoStop}%
\bibitem [{\citenamefont {He}\ \emph {et~al.}(2015)\citenamefont {He},
  \citenamefont {Hogan}, \citenamefont {Mion}, \citenamefont {Hafiz},
  \citenamefont {He}, \citenamefont {Denlinger}, \citenamefont {Mo},
  \citenamefont {Dhital}, \citenamefont {Chen}, \citenamefont {Lin},
  \citenamefont {Zhang}, \citenamefont {Hashimoto}, \citenamefont {Pan},
  \citenamefont {Lu}, \citenamefont {Arita}, \citenamefont {Shimada},
  \citenamefont {Markiewicz}, \citenamefont {Wang}, \citenamefont {Kempa},
  \citenamefont {Naughton}, \citenamefont {Bansil}, \citenamefont {Wilson},\
  and\ \citenamefont {He}}]{He2015}%
  \BibitemOpen
  \bibfield  {author} {\bibinfo {author} {\bibfnamefont {J.}~\bibnamefont
  {He}}, \bibinfo {author} {\bibfnamefont {T.}~\bibnamefont {Hogan}}, \bibinfo
  {author} {\bibfnamefont {T.~R.}\ \bibnamefont {Mion}}, \bibinfo {author}
  {\bibfnamefont {H.}~\bibnamefont {Hafiz}}, \bibinfo {author} {\bibfnamefont
  {Y.}~\bibnamefont {He}}, \bibinfo {author} {\bibfnamefont {J.~D.}\
  \bibnamefont {Denlinger}}, \bibinfo {author} {\bibfnamefont {S.~K.}\
  \bibnamefont {Mo}}, \bibinfo {author} {\bibfnamefont {C.}~\bibnamefont
  {Dhital}}, \bibinfo {author} {\bibfnamefont {X.}~\bibnamefont {Chen}},
  \bibinfo {author} {\bibfnamefont {Q.}~\bibnamefont {Lin}}, \bibinfo {author}
  {\bibfnamefont {Y.}~\bibnamefont {Zhang}}, \bibinfo {author} {\bibfnamefont
  {M.}~\bibnamefont {Hashimoto}}, \bibinfo {author} {\bibfnamefont
  {H.}~\bibnamefont {Pan}}, \bibinfo {author} {\bibfnamefont {D.~H.}\
  \bibnamefont {Lu}}, \bibinfo {author} {\bibfnamefont {M.}~\bibnamefont
  {Arita}}, \bibinfo {author} {\bibfnamefont {K.}~\bibnamefont {Shimada}},
  \bibinfo {author} {\bibfnamefont {R.~S.}\ \bibnamefont {Markiewicz}},
  \bibinfo {author} {\bibfnamefont {Z.}~\bibnamefont {Wang}}, \bibinfo {author}
  {\bibfnamefont {K.}~\bibnamefont {Kempa}}, \bibinfo {author} {\bibfnamefont
  {M.~J.}\ \bibnamefont {Naughton}}, \bibinfo {author} {\bibfnamefont
  {A.}~\bibnamefont {Bansil}}, \bibinfo {author} {\bibfnamefont {S.~D.}\
  \bibnamefont {Wilson}}, \ and\ \bibinfo {author} {\bibfnamefont {R.-H.}\
  \bibnamefont {He}},\ }\href {https://www.nature.com/articles/nmat4273}
  {\bibfield  {journal} {\bibinfo  {journal} {Nature materials}\ }\textbf
  {\bibinfo {volume} {14}},\ \bibinfo {pages} {577} (\bibinfo {year}
  {2015})}\BibitemShut {NoStop}%
\bibitem [{\citenamefont {Battisti}\ \emph {et~al.}(2016)\citenamefont
  {Battisti}, \citenamefont {Bastiaans}, \citenamefont {Fedoseev},
  \citenamefont {de~la Torre}, \citenamefont {Iliopoulos}, \citenamefont
  {Tamai}, \citenamefont {Hunter}, \citenamefont {Perry}, \citenamefont
  {Zaanen}, \citenamefont {Baumberger},\ and\ \citenamefont
  {Allan}}]{Battisti2016}%
  \BibitemOpen
  \bibfield  {author} {\bibinfo {author} {\bibfnamefont {I.}~\bibnamefont
  {Battisti}}, \bibinfo {author} {\bibfnamefont {K.~M.}\ \bibnamefont
  {Bastiaans}}, \bibinfo {author} {\bibfnamefont {V.}~\bibnamefont {Fedoseev}},
  \bibinfo {author} {\bibfnamefont {A.}~\bibnamefont {de~la Torre}}, \bibinfo
  {author} {\bibfnamefont {N.}~\bibnamefont {Iliopoulos}}, \bibinfo {author}
  {\bibfnamefont {A.}~\bibnamefont {Tamai}}, \bibinfo {author} {\bibfnamefont
  {E.~C.}\ \bibnamefont {Hunter}}, \bibinfo {author} {\bibfnamefont {R.~S.}\
  \bibnamefont {Perry}}, \bibinfo {author} {\bibfnamefont {J.}~\bibnamefont
  {Zaanen}}, \bibinfo {author} {\bibfnamefont {F.}~\bibnamefont {Baumberger}},
  \ and\ \bibinfo {author} {\bibfnamefont {M.~P.}\ \bibnamefont {Allan}},\
  }\href {https://www.nature.com/articles/nphys3894} {\bibfield  {journal}
  {\bibinfo  {journal} {Nature Physics}\ }\textbf {\bibinfo {volume} {13}},\
  \bibinfo {pages} {21} (\bibinfo {year} {2016})}\BibitemShut {NoStop}%
\bibitem [{\citenamefont {Witczak-Krempa}\ \emph {et~al.}(2014)\citenamefont
  {Witczak-Krempa}, \citenamefont {Chen}, \citenamefont {Kim},\ and\
  \citenamefont {Balents}}]{William2014_review}%
  \BibitemOpen
  \bibfield  {author} {\bibinfo {author} {\bibfnamefont {W.}~\bibnamefont
  {Witczak-Krempa}}, \bibinfo {author} {\bibfnamefont {G.}~\bibnamefont
  {Chen}}, \bibinfo {author} {\bibfnamefont {Y.~B.}\ \bibnamefont {Kim}}, \
  and\ \bibinfo {author} {\bibfnamefont {L.}~\bibnamefont {Balents}},\ }\href
  {\doibase 10.1146/annurev-conmatphys-020911-125138} {\bibfield  {journal}
  {\bibinfo  {journal} {Annual Review of Condensed Matter Physics}\ }\textbf
  {\bibinfo {volume} {5}},\ \bibinfo {pages} {57} (\bibinfo {year}
  {2014})}\BibitemShut {NoStop}%
\bibitem [{\citenamefont {Rau}\ \emph {et~al.}(2016)\citenamefont {Rau},
  \citenamefont {Lee},\ and\ \citenamefont {Kee}}]{Jeffrey2016_review}%
  \BibitemOpen
  \bibfield  {author} {\bibinfo {author} {\bibfnamefont {J.~G.}\ \bibnamefont
  {Rau}}, \bibinfo {author} {\bibfnamefont {E.~K.-H.}\ \bibnamefont {Lee}}, \
  and\ \bibinfo {author} {\bibfnamefont {H.-Y.}\ \bibnamefont {Kee}},\ }\href
  {\doibase 10.1146/annurev-conmatphys-031115-011319} {\bibfield  {journal}
  {\bibinfo  {journal} {Annual Review of Condensed Matter Physics}\ }\textbf
  {\bibinfo {volume} {7}},\ \bibinfo {pages} {195} (\bibinfo {year}
  {2016})}\BibitemShut {NoStop}%
\bibitem [{\citenamefont {Yamasaki}\ \emph {et~al.}(2016)\citenamefont
  {Yamasaki}, \citenamefont {Fujiwara}, \citenamefont {Tachibana},
  \citenamefont {Iwasaki}, \citenamefont {Higashino}, \citenamefont {Yoshimi},
  \citenamefont {Nakagawa}, \citenamefont {Nakatani}, \citenamefont {Yamagami},
  \citenamefont {Aratani}, \citenamefont {Kirilmaz}, \citenamefont {Sing},
  \citenamefont {Claessen}, \citenamefont {Watanabe}, \citenamefont
  {Shirakawa}, \citenamefont {Yunoki}, \citenamefont {Naitoh}, \citenamefont
  {Takase}, \citenamefont {Matsuno}, \citenamefont {Takagi}, \citenamefont
  {Sekiyama},\ and\ \citenamefont {Saitoh}}]{Yamasaki2016}%
  \BibitemOpen
  \bibfield  {author} {\bibinfo {author} {\bibfnamefont {A.}~\bibnamefont
  {Yamasaki}}, \bibinfo {author} {\bibfnamefont {H.}~\bibnamefont {Fujiwara}},
  \bibinfo {author} {\bibfnamefont {S.}~\bibnamefont {Tachibana}}, \bibinfo
  {author} {\bibfnamefont {D.}~\bibnamefont {Iwasaki}}, \bibinfo {author}
  {\bibfnamefont {Y.}~\bibnamefont {Higashino}}, \bibinfo {author}
  {\bibfnamefont {C.}~\bibnamefont {Yoshimi}}, \bibinfo {author} {\bibfnamefont
  {K.}~\bibnamefont {Nakagawa}}, \bibinfo {author} {\bibfnamefont
  {Y.}~\bibnamefont {Nakatani}}, \bibinfo {author} {\bibfnamefont
  {K.}~\bibnamefont {Yamagami}}, \bibinfo {author} {\bibfnamefont
  {H.}~\bibnamefont {Aratani}}, \bibinfo {author} {\bibfnamefont
  {O.}~\bibnamefont {Kirilmaz}}, \bibinfo {author} {\bibfnamefont
  {M.}~\bibnamefont {Sing}}, \bibinfo {author} {\bibfnamefont {R.}~\bibnamefont
  {Claessen}}, \bibinfo {author} {\bibfnamefont {H.}~\bibnamefont {Watanabe}},
  \bibinfo {author} {\bibfnamefont {T.}~\bibnamefont {Shirakawa}}, \bibinfo
  {author} {\bibfnamefont {S.}~\bibnamefont {Yunoki}}, \bibinfo {author}
  {\bibfnamefont {A.}~\bibnamefont {Naitoh}}, \bibinfo {author} {\bibfnamefont
  {K.}~\bibnamefont {Takase}}, \bibinfo {author} {\bibfnamefont
  {J.}~\bibnamefont {Matsuno}}, \bibinfo {author} {\bibfnamefont
  {H.}~\bibnamefont {Takagi}}, \bibinfo {author} {\bibfnamefont
  {A.}~\bibnamefont {Sekiyama}}, \ and\ \bibinfo {author} {\bibfnamefont
  {Y.}~\bibnamefont {Saitoh}},\ }\href {\doibase 10.1103/PhysRevB.94.115103}
  {\bibfield  {journal} {\bibinfo  {journal} {Phys. Rev. B}\ }\textbf {\bibinfo
  {volume} {94}},\ \bibinfo {pages} {115103} (\bibinfo {year}
  {2016})}\BibitemShut {NoStop}%
\bibitem [{\citenamefont {Watanabe}\ \emph {et~al.}(2010)\citenamefont
  {Watanabe}, \citenamefont {Shirakawa},\ and\ \citenamefont
  {Yunoki}}]{PhysRevLett.105.216410}%
  \BibitemOpen
  \bibfield  {author} {\bibinfo {author} {\bibfnamefont {H.}~\bibnamefont
  {Watanabe}}, \bibinfo {author} {\bibfnamefont {T.}~\bibnamefont {Shirakawa}},
  \ and\ \bibinfo {author} {\bibfnamefont {S.}~\bibnamefont {Yunoki}},\ }\href
  {\doibase 10.1103/PhysRevLett.105.216410} {\bibfield  {journal} {\bibinfo
  {journal} {Phys. Rev. Lett.}\ }\textbf {\bibinfo {volume} {105}},\ \bibinfo
  {pages} {216410} (\bibinfo {year} {2010})}\BibitemShut {NoStop}%
\bibitem [{\citenamefont {Liu}\ \emph {et~al.}(2015)\citenamefont {Liu},
  \citenamefont {Khmelevskyi}, \citenamefont {Kim}, \citenamefont {Marsman},
  \citenamefont {Li}, \citenamefont {Chen}, \citenamefont {Sarma},
  \citenamefont {Kresse},\ and\ \citenamefont {Franchini}}]{Liu2015}%
  \BibitemOpen
  \bibfield  {author} {\bibinfo {author} {\bibfnamefont {P.}~\bibnamefont
  {Liu}}, \bibinfo {author} {\bibfnamefont {S.}~\bibnamefont {Khmelevskyi}},
  \bibinfo {author} {\bibfnamefont {B.}~\bibnamefont {Kim}}, \bibinfo {author}
  {\bibfnamefont {M.}~\bibnamefont {Marsman}}, \bibinfo {author} {\bibfnamefont
  {D.}~\bibnamefont {Li}}, \bibinfo {author} {\bibfnamefont {X.-Q.}\
  \bibnamefont {Chen}}, \bibinfo {author} {\bibfnamefont {D.~D.}\ \bibnamefont
  {Sarma}}, \bibinfo {author} {\bibfnamefont {G.}~\bibnamefont {Kresse}}, \
  and\ \bibinfo {author} {\bibfnamefont {C.}~\bibnamefont {Franchini}},\ }\href
  {\doibase 10.1103/PhysRevB.92.054428} {\bibfield  {journal} {\bibinfo
  {journal} {Phys. Rev. B}\ }\textbf {\bibinfo {volume} {92}},\ \bibinfo
  {pages} {054428} (\bibinfo {year} {2015})}\BibitemShut {NoStop}%
\bibitem [{\citenamefont {Fujiyama}\ \emph
  {et~al.}(2012{\natexlab{a}})\citenamefont {Fujiyama}, \citenamefont {Ohsumi},
  \citenamefont {Komesu}, \citenamefont {Matsuno}, \citenamefont {Kim},
  \citenamefont {Takata}, \citenamefont {Arima},\ and\ \citenamefont
  {Takagi}}]{PhysRevLett.108.247212}%
  \BibitemOpen
  \bibfield  {author} {\bibinfo {author} {\bibfnamefont {S.}~\bibnamefont
  {Fujiyama}}, \bibinfo {author} {\bibfnamefont {H.}~\bibnamefont {Ohsumi}},
  \bibinfo {author} {\bibfnamefont {T.}~\bibnamefont {Komesu}}, \bibinfo
  {author} {\bibfnamefont {J.}~\bibnamefont {Matsuno}}, \bibinfo {author}
  {\bibfnamefont {B.~J.}\ \bibnamefont {Kim}}, \bibinfo {author} {\bibfnamefont
  {M.}~\bibnamefont {Takata}}, \bibinfo {author} {\bibfnamefont
  {T.}~\bibnamefont {Arima}}, \ and\ \bibinfo {author} {\bibfnamefont
  {H.}~\bibnamefont {Takagi}},\ }\href {\doibase
  10.1103/PhysRevLett.108.247212} {\bibfield  {journal} {\bibinfo  {journal}
  {Phys. Rev. Lett.}\ }\textbf {\bibinfo {volume} {108}},\ \bibinfo {pages}
  {247212} (\bibinfo {year} {2012}{\natexlab{a}})}\BibitemShut {NoStop}%
\bibitem [{\citenamefont {Dhital}\ \emph {et~al.}(2013)\citenamefont {Dhital},
  \citenamefont {Hogan}, \citenamefont {Yamani}, \citenamefont {de~la Cruz},
  \citenamefont {Chen}, \citenamefont {Khadka}, \citenamefont {Ren},\ and\
  \citenamefont {Wilson}}]{PhysRevB.87.144405}%
  \BibitemOpen
  \bibfield  {author} {\bibinfo {author} {\bibfnamefont {C.}~\bibnamefont
  {Dhital}}, \bibinfo {author} {\bibfnamefont {T.}~\bibnamefont {Hogan}},
  \bibinfo {author} {\bibfnamefont {Z.}~\bibnamefont {Yamani}}, \bibinfo
  {author} {\bibfnamefont {C.}~\bibnamefont {de~la Cruz}}, \bibinfo {author}
  {\bibfnamefont {X.}~\bibnamefont {Chen}}, \bibinfo {author} {\bibfnamefont
  {S.}~\bibnamefont {Khadka}}, \bibinfo {author} {\bibfnamefont
  {Z.}~\bibnamefont {Ren}}, \ and\ \bibinfo {author} {\bibfnamefont {S.~D.}\
  \bibnamefont {Wilson}},\ }\href {\doibase 10.1103/PhysRevB.87.144405}
  {\bibfield  {journal} {\bibinfo  {journal} {Phys. Rev. B}\ }\textbf {\bibinfo
  {volume} {87}},\ \bibinfo {pages} {144405} (\bibinfo {year}
  {2013})}\BibitemShut {NoStop}%
\bibitem [{\citenamefont {Ye}\ \emph {et~al.}(2013)\citenamefont {Ye},
  \citenamefont {Chi}, \citenamefont {Chakoumakos}, \citenamefont
  {Fernandez-Baca}, \citenamefont {Qi},\ and\ \citenamefont
  {Cao}}]{PhysRevB.87.140406}%
  \BibitemOpen
  \bibfield  {author} {\bibinfo {author} {\bibfnamefont {F.}~\bibnamefont
  {Ye}}, \bibinfo {author} {\bibfnamefont {S.}~\bibnamefont {Chi}}, \bibinfo
  {author} {\bibfnamefont {B.~C.}\ \bibnamefont {Chakoumakos}}, \bibinfo
  {author} {\bibfnamefont {J.~A.}\ \bibnamefont {Fernandez-Baca}}, \bibinfo
  {author} {\bibfnamefont {T.}~\bibnamefont {Qi}}, \ and\ \bibinfo {author}
  {\bibfnamefont {G.}~\bibnamefont {Cao}},\ }\href {\doibase
  10.1103/PhysRevB.87.140406} {\bibfield  {journal} {\bibinfo  {journal} {Phys.
  Rev. B}\ }\textbf {\bibinfo {volume} {87}},\ \bibinfo {pages} {140406}
  (\bibinfo {year} {2013})}\BibitemShut {NoStop}%
\bibitem [{\citenamefont {Kim}\ \emph {et~al.}(2012{\natexlab{b}})\citenamefont
  {Kim}, \citenamefont {Casa}, \citenamefont {Upton}, \citenamefont {Gog},
  \citenamefont {Kim}, \citenamefont {Mitchell}, \citenamefont {van
  Veenendaal}, \citenamefont {Daghofer}, \citenamefont {van~den Brink},
  \citenamefont {Khaliullin},\ and\ \citenamefont
  {Kim}}]{PhysRevLett.108.177003}%
  \BibitemOpen
  \bibfield  {author} {\bibinfo {author} {\bibfnamefont {J.}~\bibnamefont
  {Kim}}, \bibinfo {author} {\bibfnamefont {D.}~\bibnamefont {Casa}}, \bibinfo
  {author} {\bibfnamefont {M.~H.}\ \bibnamefont {Upton}}, \bibinfo {author}
  {\bibfnamefont {T.}~\bibnamefont {Gog}}, \bibinfo {author} {\bibfnamefont
  {Y.-J.}\ \bibnamefont {Kim}}, \bibinfo {author} {\bibfnamefont {J.~F.}\
  \bibnamefont {Mitchell}}, \bibinfo {author} {\bibfnamefont {M.}~\bibnamefont
  {van Veenendaal}}, \bibinfo {author} {\bibfnamefont {M.}~\bibnamefont
  {Daghofer}}, \bibinfo {author} {\bibfnamefont {J.}~\bibnamefont {van~den
  Brink}}, \bibinfo {author} {\bibfnamefont {G.}~\bibnamefont {Khaliullin}}, \
  and\ \bibinfo {author} {\bibfnamefont {B.~J.}\ \bibnamefont {Kim}},\ }\href
  {\doibase 10.1103/PhysRevLett.108.177003} {\bibfield  {journal} {\bibinfo
  {journal} {Phys. Rev. Lett.}\ }\textbf {\bibinfo {volume} {108}},\ \bibinfo
  {pages} {177003} (\bibinfo {year} {2012}{\natexlab{b}})}\BibitemShut
  {NoStop}%
\bibitem [{\citenamefont {Wang}\ and\ \citenamefont
  {Senthil}(2011)}]{PhysRevLett.106.136402}%
  \BibitemOpen
  \bibfield  {author} {\bibinfo {author} {\bibfnamefont {F.}~\bibnamefont
  {Wang}}\ and\ \bibinfo {author} {\bibfnamefont {T.}~\bibnamefont {Senthil}},\
  }\href {\doibase 10.1103/PhysRevLett.106.136402} {\bibfield  {journal}
  {\bibinfo  {journal} {Phys. Rev. Lett.}\ }\textbf {\bibinfo {volume} {106}},\
  \bibinfo {pages} {136402} (\bibinfo {year} {2011})}\BibitemShut {NoStop}%
\bibitem [{\citenamefont {Watanabe}\ \emph {et~al.}(2013)\citenamefont
  {Watanabe}, \citenamefont {Shirakawa},\ and\ \citenamefont
  {Yunoki}}]{PhysRevLett.110.027002}%
  \BibitemOpen
  \bibfield  {author} {\bibinfo {author} {\bibfnamefont {H.}~\bibnamefont
  {Watanabe}}, \bibinfo {author} {\bibfnamefont {T.}~\bibnamefont {Shirakawa}},
  \ and\ \bibinfo {author} {\bibfnamefont {S.}~\bibnamefont {Yunoki}},\ }\href
  {\doibase 10.1103/PhysRevLett.110.027002} {\bibfield  {journal} {\bibinfo
  {journal} {Phys. Rev. Lett.}\ }\textbf {\bibinfo {volume} {110}},\ \bibinfo
  {pages} {027002} (\bibinfo {year} {2013})}\BibitemShut {NoStop}%
\bibitem [{\citenamefont {Kim}\ \emph {et~al.}(2014{\natexlab{b}})\citenamefont
  {Kim}, \citenamefont {Krupin}, \citenamefont {Denlinger}, \citenamefont
  {Bostwick}, \citenamefont {Rotenberg}, \citenamefont {Zhao}, \citenamefont
  {Mitchell}, \citenamefont {Allen},\ and\ \citenamefont
  {Kim}}]{Kim2014_doping}%
  \BibitemOpen
  \bibfield  {author} {\bibinfo {author} {\bibfnamefont {Y.~K.}\ \bibnamefont
  {Kim}}, \bibinfo {author} {\bibfnamefont {O.}~\bibnamefont {Krupin}},
  \bibinfo {author} {\bibfnamefont {J.~D.}\ \bibnamefont {Denlinger}}, \bibinfo
  {author} {\bibfnamefont {A.}~\bibnamefont {Bostwick}}, \bibinfo {author}
  {\bibfnamefont {E.}~\bibnamefont {Rotenberg}}, \bibinfo {author}
  {\bibfnamefont {Q.}~\bibnamefont {Zhao}}, \bibinfo {author} {\bibfnamefont
  {J.~F.}\ \bibnamefont {Mitchell}}, \bibinfo {author} {\bibfnamefont {J.~W.}\
  \bibnamefont {Allen}}, \ and\ \bibinfo {author} {\bibfnamefont {B.~J.}\
  \bibnamefont {Kim}},\ }\href {\doibase 10.1126/science.1251151} {\bibfield
  {journal} {\bibinfo  {journal} {Science}\ }\textbf {\bibinfo {volume}
  {345}},\ \bibinfo {pages} {187} (\bibinfo {year}
  {2014}{\natexlab{b}})}\BibitemShut {NoStop}%
\bibitem [{\citenamefont {Kim}\ \emph {et~al.}(2015{\natexlab{a}})\citenamefont
  {Kim}, \citenamefont {Sung}, \citenamefont {Denlinger},\ and\ \citenamefont
  {Kim}}]{Kim2015_doping}%
  \BibitemOpen
  \bibfield  {author} {\bibinfo {author} {\bibfnamefont {Y.~K.}\ \bibnamefont
  {Kim}}, \bibinfo {author} {\bibfnamefont {N.~H.}\ \bibnamefont {Sung}},
  \bibinfo {author} {\bibfnamefont {J.~D.}\ \bibnamefont {Denlinger}}, \ and\
  \bibinfo {author} {\bibfnamefont {B.~J.}\ \bibnamefont {Kim}},\ }\href
  {https://doi.org/10.1038/nphys3503} {\bibfield  {journal} {\bibinfo
  {journal} {Nature Physics}\ }\textbf {\bibinfo {volume} {12}},\ \bibinfo
  {pages} {37} (\bibinfo {year} {2015}{\natexlab{a}})}\BibitemShut {NoStop}%
\bibitem [{\citenamefont {Liu}\ \emph {et~al.}(2016{\natexlab{a}})\citenamefont
  {Liu}, \citenamefont {Reticcioli}, \citenamefont {Kim}, \citenamefont
  {Continenza}, \citenamefont {Kresse}, \citenamefont {Sarma}, \citenamefont
  {Chen},\ and\ \citenamefont {Franchini}}]{PhysRevB.94.195145}%
  \BibitemOpen
  \bibfield  {author} {\bibinfo {author} {\bibfnamefont {P.}~\bibnamefont
  {Liu}}, \bibinfo {author} {\bibfnamefont {M.}~\bibnamefont {Reticcioli}},
  \bibinfo {author} {\bibfnamefont {B.}~\bibnamefont {Kim}}, \bibinfo {author}
  {\bibfnamefont {A.}~\bibnamefont {Continenza}}, \bibinfo {author}
  {\bibfnamefont {G.}~\bibnamefont {Kresse}}, \bibinfo {author} {\bibfnamefont
  {D.~D.}\ \bibnamefont {Sarma}}, \bibinfo {author} {\bibfnamefont {X.-Q.}\
  \bibnamefont {Chen}}, \ and\ \bibinfo {author} {\bibfnamefont
  {C.}~\bibnamefont {Franchini}},\ }\href {\doibase 10.1103/PhysRevB.94.195145}
  {\bibfield  {journal} {\bibinfo  {journal} {Phys. Rev. B}\ }\textbf {\bibinfo
  {volume} {94}},\ \bibinfo {pages} {195145} (\bibinfo {year}
  {2016}{\natexlab{a}})}\BibitemShut {NoStop}%
\bibitem [{\citenamefont {Lupascu}\ \emph {et~al.}(2014)\citenamefont
  {Lupascu}, \citenamefont {Clancy}, \citenamefont {Gretarsson}, \citenamefont
  {Nie}, \citenamefont {Nichols}, \citenamefont {Terzic}, \citenamefont {Cao},
  \citenamefont {Seo}, \citenamefont {Islam}, \citenamefont {Upton},
  \citenamefont {Kim}, \citenamefont {Casa}, \citenamefont {Gog}, \citenamefont
  {Said}, \citenamefont {Katukuri}, \citenamefont {Stoll}, \citenamefont
  {Hozoi}, \citenamefont {van~den Brink},\ and\ \citenamefont
  {Kim}}]{PhysRevLett.112.147201}%
  \BibitemOpen
  \bibfield  {author} {\bibinfo {author} {\bibfnamefont {A.}~\bibnamefont
  {Lupascu}}, \bibinfo {author} {\bibfnamefont {J.~P.}\ \bibnamefont {Clancy}},
  \bibinfo {author} {\bibfnamefont {H.}~\bibnamefont {Gretarsson}}, \bibinfo
  {author} {\bibfnamefont {Z.}~\bibnamefont {Nie}}, \bibinfo {author}
  {\bibfnamefont {J.}~\bibnamefont {Nichols}}, \bibinfo {author} {\bibfnamefont
  {J.}~\bibnamefont {Terzic}}, \bibinfo {author} {\bibfnamefont
  {G.}~\bibnamefont {Cao}}, \bibinfo {author} {\bibfnamefont {S.~S.~A.}\
  \bibnamefont {Seo}}, \bibinfo {author} {\bibfnamefont {Z.}~\bibnamefont
  {Islam}}, \bibinfo {author} {\bibfnamefont {M.~H.}\ \bibnamefont {Upton}},
  \bibinfo {author} {\bibfnamefont {J.}~\bibnamefont {Kim}}, \bibinfo {author}
  {\bibfnamefont {D.}~\bibnamefont {Casa}}, \bibinfo {author} {\bibfnamefont
  {T.}~\bibnamefont {Gog}}, \bibinfo {author} {\bibfnamefont {A.~H.}\
  \bibnamefont {Said}}, \bibinfo {author} {\bibfnamefont {V.~M.}\ \bibnamefont
  {Katukuri}}, \bibinfo {author} {\bibfnamefont {H.}~\bibnamefont {Stoll}},
  \bibinfo {author} {\bibfnamefont {L.}~\bibnamefont {Hozoi}}, \bibinfo
  {author} {\bibfnamefont {J.}~\bibnamefont {van~den Brink}}, \ and\ \bibinfo
  {author} {\bibfnamefont {Y.-J.}\ \bibnamefont {Kim}},\ }\href {\doibase
  10.1103/PhysRevLett.112.147201} {\bibfield  {journal} {\bibinfo  {journal}
  {Phys. Rev. Lett.}\ }\textbf {\bibinfo {volume} {112}},\ \bibinfo {pages}
  {147201} (\bibinfo {year} {2014})}\BibitemShut {NoStop}%
\bibitem [{\citenamefont {Wojek}\ \emph {et~al.}(2012)\citenamefont {Wojek},
  \citenamefont {Berntsen}, \citenamefont {Boseggia}, \citenamefont
  {Boothroyd}, \citenamefont {Prabhakaran}, \citenamefont {McMorrow},
  \citenamefont {Ronnow}, \citenamefont {Chang},\ and\ \citenamefont
  {Tjernberg}}]{Wojek2012}%
  \BibitemOpen
  \bibfield  {author} {\bibinfo {author} {\bibfnamefont {B.~M.}\ \bibnamefont
  {Wojek}}, \bibinfo {author} {\bibfnamefont {M.~H.}\ \bibnamefont {Berntsen}},
  \bibinfo {author} {\bibfnamefont {S.}~\bibnamefont {Boseggia}}, \bibinfo
  {author} {\bibfnamefont {A.~T.}\ \bibnamefont {Boothroyd}}, \bibinfo {author}
  {\bibfnamefont {D.}~\bibnamefont {Prabhakaran}}, \bibinfo {author}
  {\bibfnamefont {D.~F.}\ \bibnamefont {McMorrow}}, \bibinfo {author}
  {\bibfnamefont {H.~M.}\ \bibnamefont {Ronnow}}, \bibinfo {author}
  {\bibfnamefont {J.}~\bibnamefont {Chang}}, \ and\ \bibinfo {author}
  {\bibfnamefont {O.}~\bibnamefont {Tjernberg}},\ }\href
  {http://stacks.iop.org/0953-8984/24/i=41/a=415602} {\bibfield  {journal}
  {\bibinfo  {journal} {J. Phys.: Condens. Matter}\ }\textbf {\bibinfo {volume}
  {24}},\ \bibinfo {pages} {415602} (\bibinfo {year} {2012})}\BibitemShut
  {NoStop}%
\bibitem [{\citenamefont {Wang}\ \emph {et~al.}(2013)\citenamefont {Wang},
  \citenamefont {Cao}, \citenamefont {Waugh}, \citenamefont {Park},
  \citenamefont {Qi}, \citenamefont {Korneta}, \citenamefont {Cao},\ and\
  \citenamefont {Dessau}}]{Wang2013}%
  \BibitemOpen
  \bibfield  {author} {\bibinfo {author} {\bibfnamefont {Q.}~\bibnamefont
  {Wang}}, \bibinfo {author} {\bibfnamefont {Y.}~\bibnamefont {Cao}}, \bibinfo
  {author} {\bibfnamefont {J.~A.}\ \bibnamefont {Waugh}}, \bibinfo {author}
  {\bibfnamefont {S.~R.}\ \bibnamefont {Park}}, \bibinfo {author}
  {\bibfnamefont {T.~F.}\ \bibnamefont {Qi}}, \bibinfo {author} {\bibfnamefont
  {O.~B.}\ \bibnamefont {Korneta}}, \bibinfo {author} {\bibfnamefont
  {G.}~\bibnamefont {Cao}}, \ and\ \bibinfo {author} {\bibfnamefont {D.~S.}\
  \bibnamefont {Dessau}},\ }\href {\doibase 10.1103/PhysRevB.87.245109}
  {\bibfield  {journal} {\bibinfo  {journal} {Phys. Rev. B}\ }\textbf {\bibinfo
  {volume} {87}},\ \bibinfo {pages} {245109} (\bibinfo {year}
  {2013})}\BibitemShut {NoStop}%
\bibitem [{\citenamefont {Moon}\ \emph {et~al.}(2009)\citenamefont {Moon},
  \citenamefont {Jin}, \citenamefont {Choi}, \citenamefont {Lee}, \citenamefont
  {Seo}, \citenamefont {Yu}, \citenamefont {Cao}, \citenamefont {Noh},\ and\
  \citenamefont {Lee}}]{Moon2009_optics}%
  \BibitemOpen
  \bibfield  {author} {\bibinfo {author} {\bibfnamefont {S.~J.}\ \bibnamefont
  {Moon}}, \bibinfo {author} {\bibfnamefont {H.}~\bibnamefont {Jin}}, \bibinfo
  {author} {\bibfnamefont {W.~S.}\ \bibnamefont {Choi}}, \bibinfo {author}
  {\bibfnamefont {J.~S.}\ \bibnamefont {Lee}}, \bibinfo {author} {\bibfnamefont
  {S.~S.~A.}\ \bibnamefont {Seo}}, \bibinfo {author} {\bibfnamefont
  {J.}~\bibnamefont {Yu}}, \bibinfo {author} {\bibfnamefont {G.}~\bibnamefont
  {Cao}}, \bibinfo {author} {\bibfnamefont {T.~W.}\ \bibnamefont {Noh}}, \ and\
  \bibinfo {author} {\bibfnamefont {Y.~S.}\ \bibnamefont {Lee}},\ }\href
  {\doibase 10.1103/PhysRevB.80.195110} {\bibfield  {journal} {\bibinfo
  {journal} {Phys. Rev. B}\ }\textbf {\bibinfo {volume} {80}},\ \bibinfo
  {pages} {195110} (\bibinfo {year} {2009})}\BibitemShut {NoStop}%
\bibitem [{\citenamefont {Fujiyama}\ \emph
  {et~al.}(2012{\natexlab{b}})\citenamefont {Fujiyama}, \citenamefont {Ohashi},
  \citenamefont {Ohsumi}, \citenamefont {Sugimoto}, \citenamefont {Takayama},
  \citenamefont {Komesu}, \citenamefont {Takata}, \citenamefont {Arima},\ and\
  \citenamefont {Takagi}}]{PhysRevB.86.174414}%
  \BibitemOpen
  \bibfield  {author} {\bibinfo {author} {\bibfnamefont {S.}~\bibnamefont
  {Fujiyama}}, \bibinfo {author} {\bibfnamefont {K.}~\bibnamefont {Ohashi}},
  \bibinfo {author} {\bibfnamefont {H.}~\bibnamefont {Ohsumi}}, \bibinfo
  {author} {\bibfnamefont {K.}~\bibnamefont {Sugimoto}}, \bibinfo {author}
  {\bibfnamefont {T.}~\bibnamefont {Takayama}}, \bibinfo {author}
  {\bibfnamefont {T.}~\bibnamefont {Komesu}}, \bibinfo {author} {\bibfnamefont
  {M.}~\bibnamefont {Takata}}, \bibinfo {author} {\bibfnamefont
  {T.}~\bibnamefont {Arima}}, \ and\ \bibinfo {author} {\bibfnamefont
  {H.}~\bibnamefont {Takagi}},\ }\href {\doibase 10.1103/PhysRevB.86.174414}
  {\bibfield  {journal} {\bibinfo  {journal} {Phys. Rev. B}\ }\textbf {\bibinfo
  {volume} {86}},\ \bibinfo {pages} {174414} (\bibinfo {year}
  {2012}{\natexlab{b}})}\BibitemShut {NoStop}%
\bibitem [{\citenamefont {Park}\ \emph {et~al.}(2014)\citenamefont {Park},
  \citenamefont {Sohn}, \citenamefont {Jeong}, \citenamefont {Cao},
  \citenamefont {Kim}, \citenamefont {Moon}, \citenamefont {Jin}, \citenamefont
  {Cho},\ and\ \citenamefont {Noh}}]{Park2014}%
  \BibitemOpen
  \bibfield  {author} {\bibinfo {author} {\bibfnamefont {H.~J.}\ \bibnamefont
  {Park}}, \bibinfo {author} {\bibfnamefont {C.~H.}\ \bibnamefont {Sohn}},
  \bibinfo {author} {\bibfnamefont {D.~W.}\ \bibnamefont {Jeong}}, \bibinfo
  {author} {\bibfnamefont {G.}~\bibnamefont {Cao}}, \bibinfo {author}
  {\bibfnamefont {K.~W.}\ \bibnamefont {Kim}}, \bibinfo {author} {\bibfnamefont
  {S.~J.}\ \bibnamefont {Moon}}, \bibinfo {author} {\bibfnamefont
  {H.}~\bibnamefont {Jin}}, \bibinfo {author} {\bibfnamefont {D.-Y.}\
  \bibnamefont {Cho}}, \ and\ \bibinfo {author} {\bibfnamefont {T.~W.}\
  \bibnamefont {Noh}},\ }\href {\doibase 10.1103/PhysRevB.89.155115} {\bibfield
   {journal} {\bibinfo  {journal} {Phys. Rev. B}\ }\textbf {\bibinfo {volume}
  {89}},\ \bibinfo {pages} {155115} (\bibinfo {year} {2014})}\BibitemShut
  {NoStop}%
\bibitem [{\citenamefont {Boseggia}\ \emph {et~al.}(2012)\citenamefont
  {Boseggia}, \citenamefont {Springell}, \citenamefont {Walker}, \citenamefont
  {Boothroyd}, \citenamefont {Prabhakaran}, \citenamefont {Collins},\ and\
  \citenamefont {McMorrow}}]{Boseggia2012}%
  \BibitemOpen
  \bibfield  {author} {\bibinfo {author} {\bibfnamefont {S.}~\bibnamefont
  {Boseggia}}, \bibinfo {author} {\bibfnamefont {R.}~\bibnamefont {Springell}},
  \bibinfo {author} {\bibfnamefont {H.~C.}\ \bibnamefont {Walker}}, \bibinfo
  {author} {\bibfnamefont {A.~T.}\ \bibnamefont {Boothroyd}}, \bibinfo {author}
  {\bibfnamefont {D.}~\bibnamefont {Prabhakaran}}, \bibinfo {author}
  {\bibfnamefont {S.~P.}\ \bibnamefont {Collins}}, \ and\ \bibinfo {author}
  {\bibfnamefont {D.~F.}\ \bibnamefont {McMorrow}},\ }\href
  {http://stacks.iop.org/0953-8984/24/i=31/a=312202} {\bibfield  {journal}
  {\bibinfo  {journal} {Journal of Physics: Condensed Matter}\ }\textbf
  {\bibinfo {volume} {24}},\ \bibinfo {pages} {312202} (\bibinfo {year}
  {2012})}\BibitemShut {NoStop}%
\bibitem [{\citenamefont {Kim}\ \emph {et~al.}(2015{\natexlab{b}})\citenamefont
  {Kim}, \citenamefont {Chen},\ and\ \citenamefont {Kee}}]{Heung-Sik2015}%
  \BibitemOpen
  \bibfield  {author} {\bibinfo {author} {\bibfnamefont {H.-S.}\ \bibnamefont
  {Kim}}, \bibinfo {author} {\bibfnamefont {Y.}~\bibnamefont {Chen}}, \ and\
  \bibinfo {author} {\bibfnamefont {H.-Y.}\ \bibnamefont {Kee}},\ }\href
  {\doibase 10.1103/PhysRevB.91.235103} {\bibfield  {journal} {\bibinfo
  {journal} {Phys. Rev. B}\ }\textbf {\bibinfo {volume} {91}},\ \bibinfo
  {pages} {235103} (\bibinfo {year} {2015}{\natexlab{b}})}\BibitemShut
  {NoStop}%
\bibitem [{\citenamefont {Chen}\ \emph {et~al.}(2015)\citenamefont {Chen},
  \citenamefont {Lu},\ and\ \citenamefont {Kee}}]{Chen2015}%
  \BibitemOpen
  \bibfield  {author} {\bibinfo {author} {\bibfnamefont {Y.}~\bibnamefont
  {Chen}}, \bibinfo {author} {\bibfnamefont {Y.-M.}\ \bibnamefont {Lu}}, \ and\
  \bibinfo {author} {\bibfnamefont {H.-Y.}\ \bibnamefont {Kee}},\ }\href
  {https://www-nature-com.uaccess.univie.ac.at/articles/ncomms7593} {\bibfield
  {journal} {\bibinfo  {journal} {Nat Commun}\ }\textbf {\bibinfo {volume}
  {6}},\ \bibinfo {pages} {6593} (\bibinfo {year} {2015})}\BibitemShut
  {NoStop}%
\bibitem [{\citenamefont {Fujioka}\ \emph {et~al.}(2017)\citenamefont
  {Fujioka}, \citenamefont {Okawa}, \citenamefont {Yamamoto},\ and\
  \citenamefont {Tokura}}]{Fujioka2017}%
  \BibitemOpen
  \bibfield  {author} {\bibinfo {author} {\bibfnamefont {J.}~\bibnamefont
  {Fujioka}}, \bibinfo {author} {\bibfnamefont {T.}~\bibnamefont {Okawa}},
  \bibinfo {author} {\bibfnamefont {A.}~\bibnamefont {Yamamoto}}, \ and\
  \bibinfo {author} {\bibfnamefont {Y.}~\bibnamefont {Tokura}},\ }\href
  {\doibase 10.1103/PhysRevB.95.121102} {\bibfield  {journal} {\bibinfo
  {journal} {Phys. Rev. B}\ }\textbf {\bibinfo {volume} {95}},\ \bibinfo
  {pages} {121102} (\bibinfo {year} {2017})}\BibitemShut {NoStop}%
\bibitem [{\citenamefont {Zhang}\ \emph {et~al.}(2017)\citenamefont {Zhang},
  \citenamefont {Pang}, \citenamefont {Chen},\ and\ \citenamefont
  {Chen}}]{Lunyong2017_review113}%
  \BibitemOpen
  \bibfield  {author} {\bibinfo {author} {\bibfnamefont {L.}~\bibnamefont
  {Zhang}}, \bibinfo {author} {\bibfnamefont {B.}~\bibnamefont {Pang}},
  \bibinfo {author} {\bibfnamefont {Y.~B.}\ \bibnamefont {Chen}}, \ and\
  \bibinfo {author} {\bibfnamefont {Y.}~\bibnamefont {Chen}},\ }\href {\doibase
  10.1080/10408436.2017.1358147} {\bibfield  {journal} {\bibinfo  {journal}
  {Critical Reviews in Solid State and Materials Sciences}\ }\textbf {\bibinfo
  {volume} {0}},\ \bibinfo {pages} {1} (\bibinfo {year} {2017})}\BibitemShut
  {NoStop}%
\bibitem [{\citenamefont {Lee}\ \emph {et~al.}(2012)\citenamefont {Lee},
  \citenamefont {Krockenberger}, \citenamefont {Takahashi}, \citenamefont
  {Kawasaki},\ and\ \citenamefont {Tokura}}]{Lee2012_optics}%
  \BibitemOpen
  \bibfield  {author} {\bibinfo {author} {\bibfnamefont {J.~S.}\ \bibnamefont
  {Lee}}, \bibinfo {author} {\bibfnamefont {Y.}~\bibnamefont {Krockenberger}},
  \bibinfo {author} {\bibfnamefont {K.~S.}\ \bibnamefont {Takahashi}}, \bibinfo
  {author} {\bibfnamefont {M.}~\bibnamefont {Kawasaki}}, \ and\ \bibinfo
  {author} {\bibfnamefont {Y.}~\bibnamefont {Tokura}},\ }\href {\doibase
  10.1103/PhysRevB.85.035101} {\bibfield  {journal} {\bibinfo  {journal} {Phys.
  Rev. B}\ }\textbf {\bibinfo {volume} {85}},\ \bibinfo {pages} {035101}
  (\bibinfo {year} {2012})}\BibitemShut {NoStop}%
\bibitem [{\citenamefont {Kim}\ \emph {et~al.}(2012{\natexlab{c}})\citenamefont
  {Kim}, \citenamefont {Khaliullin},\ and\ \citenamefont
  {Min}}]{Kim2012_PRLoptics}%
  \BibitemOpen
  \bibfield  {author} {\bibinfo {author} {\bibfnamefont {B.~H.}\ \bibnamefont
  {Kim}}, \bibinfo {author} {\bibfnamefont {G.}~\bibnamefont {Khaliullin}}, \
  and\ \bibinfo {author} {\bibfnamefont {B.~I.}\ \bibnamefont {Min}},\ }\href
  {\doibase 10.1103/PhysRevLett.109.167205} {\bibfield  {journal} {\bibinfo
  {journal} {Phys. Rev. Lett.}\ }\textbf {\bibinfo {volume} {109}},\ \bibinfo
  {pages} {167205} (\bibinfo {year} {2012}{\natexlab{c}})}\BibitemShut
  {NoStop}%
\bibitem [{\citenamefont {Nichols}\ \emph
  {et~al.}(2013{\natexlab{a}})\citenamefont {Nichols}, \citenamefont {Korneta},
  \citenamefont {Terzic}, \citenamefont {Long}, \citenamefont {Cao},
  \citenamefont {Brill},\ and\ \citenamefont {Seo}}]{Nichols2013_APL}%
  \BibitemOpen
  \bibfield  {author} {\bibinfo {author} {\bibfnamefont {J.}~\bibnamefont
  {Nichols}}, \bibinfo {author} {\bibfnamefont {O.~B.}\ \bibnamefont
  {Korneta}}, \bibinfo {author} {\bibfnamefont {J.}~\bibnamefont {Terzic}},
  \bibinfo {author} {\bibfnamefont {L.~E.~D.}\ \bibnamefont {Long}}, \bibinfo
  {author} {\bibfnamefont {G.}~\bibnamefont {Cao}}, \bibinfo {author}
  {\bibfnamefont {J.~W.}\ \bibnamefont {Brill}}, \ and\ \bibinfo {author}
  {\bibfnamefont {S.~S.~A.}\ \bibnamefont {Seo}},\ }\href {\doibase
  10.1063/1.4822334} {\bibfield  {journal} {\bibinfo  {journal} {Applied
  Physics Letters}\ }\textbf {\bibinfo {volume} {103}},\ \bibinfo {pages}
  {131910} (\bibinfo {year} {2013}{\natexlab{a}})}\BibitemShut {NoStop}%
\bibitem [{\citenamefont {Nichols}\ \emph
  {et~al.}(2013{\natexlab{b}})\citenamefont {Nichols}, \citenamefont {Terzic},
  \citenamefont {Bittle}, \citenamefont {Korneta}, \citenamefont {Long},
  \citenamefont {Brill}, \citenamefont {Cao},\ and\ \citenamefont
  {Seo}}]{Nichols2013_optics}%
  \BibitemOpen
  \bibfield  {author} {\bibinfo {author} {\bibfnamefont {J.}~\bibnamefont
  {Nichols}}, \bibinfo {author} {\bibfnamefont {J.}~\bibnamefont {Terzic}},
  \bibinfo {author} {\bibfnamefont {E.~G.}\ \bibnamefont {Bittle}}, \bibinfo
  {author} {\bibfnamefont {O.~B.}\ \bibnamefont {Korneta}}, \bibinfo {author}
  {\bibfnamefont {L.~E.~D.}\ \bibnamefont {Long}}, \bibinfo {author}
  {\bibfnamefont {J.~W.}\ \bibnamefont {Brill}}, \bibinfo {author}
  {\bibfnamefont {G.}~\bibnamefont {Cao}}, \ and\ \bibinfo {author}
  {\bibfnamefont {S.~S.~A.}\ \bibnamefont {Seo}},\ }\href {\doibase
  10.1063/1.4801877} {\bibfield  {journal} {\bibinfo  {journal} {Applied
  Physics Letters}\ }\textbf {\bibinfo {volume} {102}},\ \bibinfo {pages}
  {141908} (\bibinfo {year} {2013}{\natexlab{b}})}\BibitemShut {NoStop}%
\bibitem [{\citenamefont {Liu}\ \emph {et~al.}(2013)\citenamefont {Liu},
  \citenamefont {Chu}, \citenamefont {Serrao}, \citenamefont {D.~Yi},
  \citenamefont {Nelson}, \citenamefont {Frontera}, \citenamefont {Kriegner},
  \citenamefont {Horak}, \citenamefont {Arenholz}, \citenamefont {Orenstein},
  \citenamefont {Vishwanath}, \citenamefont {Marti},\ and\ \citenamefont
  {Ramesh}}]{Liu2013}%
  \BibitemOpen
  \bibfield  {author} {\bibinfo {author} {\bibfnamefont {J.}~\bibnamefont
  {Liu}}, \bibinfo {author} {\bibfnamefont {J.-H.}\ \bibnamefont {Chu}},
  \bibinfo {author} {\bibfnamefont {C.~R.}\ \bibnamefont {Serrao}}, \bibinfo
  {author} {\bibfnamefont {J.~K.}\ \bibnamefont {D.~Yi}}, \bibinfo {author}
  {\bibfnamefont {C.}~\bibnamefont {Nelson}}, \bibinfo {author} {\bibfnamefont
  {C.}~\bibnamefont {Frontera}}, \bibinfo {author} {\bibfnamefont
  {D.}~\bibnamefont {Kriegner}}, \bibinfo {author} {\bibfnamefont
  {L.}~\bibnamefont {Horak}}, \bibinfo {author} {\bibfnamefont
  {E.}~\bibnamefont {Arenholz}}, \bibinfo {author} {\bibfnamefont
  {J.}~\bibnamefont {Orenstein}}, \bibinfo {author} {\bibfnamefont
  {A.}~\bibnamefont {Vishwanath}}, \bibinfo {author} {\bibfnamefont
  {X.}~\bibnamefont {Marti}}, \ and\ \bibinfo {author} {\bibfnamefont
  {R.}~\bibnamefont {Ramesh}},\ }\href {https://arxiv.org/abs/1305.1732}
  {\bibfield  {journal} {\bibinfo  {journal} {arXiv:1305.1732}\ } (\bibinfo
  {year} {2013})}\BibitemShut {NoStop}%
\bibitem [{\citenamefont {Kim}\ \emph {et~al.}(2016{\natexlab{a}})\citenamefont
  {Kim}, \citenamefont {Kim}, \citenamefont {Sandilands}, \citenamefont {Sohn},
  \citenamefont {Matsuno}, \citenamefont {Takagi}, \citenamefont {Kim},
  \citenamefont {Lee}, \citenamefont {Moon},\ and\ \citenamefont
  {Noh}}]{Kim2016_SL}%
  \BibitemOpen
  \bibfield  {author} {\bibinfo {author} {\bibfnamefont {S.~Y.}\ \bibnamefont
  {Kim}}, \bibinfo {author} {\bibfnamefont {C.~H.}\ \bibnamefont {Kim}},
  \bibinfo {author} {\bibfnamefont {L.~J.}\ \bibnamefont {Sandilands}},
  \bibinfo {author} {\bibfnamefont {C.~H.}\ \bibnamefont {Sohn}}, \bibinfo
  {author} {\bibfnamefont {J.}~\bibnamefont {Matsuno}}, \bibinfo {author}
  {\bibfnamefont {H.}~\bibnamefont {Takagi}}, \bibinfo {author} {\bibfnamefont
  {K.~W.}\ \bibnamefont {Kim}}, \bibinfo {author} {\bibfnamefont {Y.~S.}\
  \bibnamefont {Lee}}, \bibinfo {author} {\bibfnamefont {S.~J.}\ \bibnamefont
  {Moon}}, \ and\ \bibinfo {author} {\bibfnamefont {T.~W.}\ \bibnamefont
  {Noh}},\ }\href {\doibase 10.1103/PhysRevB.94.245113} {\bibfield  {journal}
  {\bibinfo  {journal} {Phys. Rev. B}\ }\textbf {\bibinfo {volume} {94}},\
  \bibinfo {pages} {245113} (\bibinfo {year} {2016}{\natexlab{a}})}\BibitemShut
  {NoStop}%
\bibitem [{\citenamefont {Sohn}\ \emph {et~al.}(2014)\citenamefont {Sohn},
  \citenamefont {Lee}, \citenamefont {Park}, \citenamefont {Noh}, \citenamefont
  {Yoo}, \citenamefont {Moon}, \citenamefont {Kim}, \citenamefont {Qi},
  \citenamefont {Cao}, \citenamefont {Cho},\ and\ \citenamefont
  {Noh}}]{Sohn2014_optics}%
  \BibitemOpen
  \bibfield  {author} {\bibinfo {author} {\bibfnamefont {C.~H.}\ \bibnamefont
  {Sohn}}, \bibinfo {author} {\bibfnamefont {M.-C.}\ \bibnamefont {Lee}},
  \bibinfo {author} {\bibfnamefont {H.~J.}\ \bibnamefont {Park}}, \bibinfo
  {author} {\bibfnamefont {K.~J.}\ \bibnamefont {Noh}}, \bibinfo {author}
  {\bibfnamefont {H.~K.}\ \bibnamefont {Yoo}}, \bibinfo {author} {\bibfnamefont
  {S.~J.}\ \bibnamefont {Moon}}, \bibinfo {author} {\bibfnamefont {K.~W.}\
  \bibnamefont {Kim}}, \bibinfo {author} {\bibfnamefont {T.~F.}\ \bibnamefont
  {Qi}}, \bibinfo {author} {\bibfnamefont {G.}~\bibnamefont {Cao}}, \bibinfo
  {author} {\bibfnamefont {D.-Y.}\ \bibnamefont {Cho}}, \ and\ \bibinfo
  {author} {\bibfnamefont {T.~W.}\ \bibnamefont {Noh}},\ }\href {\doibase
  10.1103/PhysRevB.90.041105} {\bibfield  {journal} {\bibinfo  {journal} {Phys.
  Rev. B}\ }\textbf {\bibinfo {volume} {90}},\ \bibinfo {pages} {041105}
  (\bibinfo {year} {2014})}\BibitemShut {NoStop}%
\bibitem [{\citenamefont {Pr\"opper}\ \emph {et~al.}(2016)\citenamefont
  {Pr\"opper}, \citenamefont {Yaresko}, \citenamefont {H\"oppner},
  \citenamefont {Matiks}, \citenamefont {Mathis}, \citenamefont {Takayama},
  \citenamefont {Matsumoto}, \citenamefont {Takagi}, \citenamefont {Keimer},\
  and\ \citenamefont {Boris}}]{Propper2016_optics}%
  \BibitemOpen
  \bibfield  {author} {\bibinfo {author} {\bibfnamefont {D.}~\bibnamefont
  {Pr\"opper}}, \bibinfo {author} {\bibfnamefont {A.~N.}\ \bibnamefont
  {Yaresko}}, \bibinfo {author} {\bibfnamefont {M.}~\bibnamefont {H\"oppner}},
  \bibinfo {author} {\bibfnamefont {Y.}~\bibnamefont {Matiks}}, \bibinfo
  {author} {\bibfnamefont {Y.-L.}\ \bibnamefont {Mathis}}, \bibinfo {author}
  {\bibfnamefont {T.}~\bibnamefont {Takayama}}, \bibinfo {author}
  {\bibfnamefont {A.}~\bibnamefont {Matsumoto}}, \bibinfo {author}
  {\bibfnamefont {H.}~\bibnamefont {Takagi}}, \bibinfo {author} {\bibfnamefont
  {B.}~\bibnamefont {Keimer}}, \ and\ \bibinfo {author} {\bibfnamefont {A.~V.}\
  \bibnamefont {Boris}},\ }\href {\doibase 10.1103/PhysRevB.94.035158}
  {\bibfield  {journal} {\bibinfo  {journal} {Phys. Rev. B}\ }\textbf {\bibinfo
  {volume} {94}},\ \bibinfo {pages} {035158} (\bibinfo {year}
  {2016})}\BibitemShut {NoStop}%
\bibitem [{\citenamefont {Souri}\ \emph {et~al.}(2017)\citenamefont {Souri},
  \citenamefont {Kim}, \citenamefont {Gruenewald}, \citenamefont {Connell},
  \citenamefont {Thompson}, \citenamefont {Nichols}, \citenamefont {Terzic},
  \citenamefont {Min}, \citenamefont {Cao}, \citenamefont {Brill},\ and\
  \citenamefont {Seo}}]{Souri2017_optics}%
  \BibitemOpen
  \bibfield  {author} {\bibinfo {author} {\bibfnamefont {M.}~\bibnamefont
  {Souri}}, \bibinfo {author} {\bibfnamefont {B.~H.}\ \bibnamefont {Kim}},
  \bibinfo {author} {\bibfnamefont {J.~H.}\ \bibnamefont {Gruenewald}},
  \bibinfo {author} {\bibfnamefont {J.~G.}\ \bibnamefont {Connell}}, \bibinfo
  {author} {\bibfnamefont {J.}~\bibnamefont {Thompson}}, \bibinfo {author}
  {\bibfnamefont {J.}~\bibnamefont {Nichols}}, \bibinfo {author} {\bibfnamefont
  {J.}~\bibnamefont {Terzic}}, \bibinfo {author} {\bibfnamefont {B.~I.}\
  \bibnamefont {Min}}, \bibinfo {author} {\bibfnamefont {G.}~\bibnamefont
  {Cao}}, \bibinfo {author} {\bibfnamefont {J.~W.}\ \bibnamefont {Brill}}, \
  and\ \bibinfo {author} {\bibfnamefont {A.}~\bibnamefont {Seo}},\ }\href
  {\doibase 10.1103/PhysRevB.95.235125} {\bibfield  {journal} {\bibinfo
  {journal} {Phys. Rev. B}\ }\textbf {\bibinfo {volume} {95}},\ \bibinfo
  {pages} {235125} (\bibinfo {year} {2017})}\BibitemShut {NoStop}%
\bibitem [{\citenamefont {Basov}\ \emph {et~al.}(2011)\citenamefont {Basov},
  \citenamefont {Averitt}, \citenamefont {van~der Marel}, \citenamefont
  {Dressel},\ and\ \citenamefont {Haule}}]{RevModPhys.83.471}%
  \BibitemOpen
  \bibfield  {author} {\bibinfo {author} {\bibfnamefont {D.~N.}\ \bibnamefont
  {Basov}}, \bibinfo {author} {\bibfnamefont {R.~D.}\ \bibnamefont {Averitt}},
  \bibinfo {author} {\bibfnamefont {D.}~\bibnamefont {van~der Marel}}, \bibinfo
  {author} {\bibfnamefont {M.}~\bibnamefont {Dressel}}, \ and\ \bibinfo
  {author} {\bibfnamefont {K.}~\bibnamefont {Haule}},\ }\href {\doibase
  10.1103/RevModPhys.83.471} {\bibfield  {journal} {\bibinfo  {journal} {Rev.
  Mod. Phys.}\ }\textbf {\bibinfo {volume} {83}},\ \bibinfo {pages} {471}
  (\bibinfo {year} {2011})}\BibitemShut {NoStop}%
\bibitem [{\citenamefont {Haskel}\ \emph {et~al.}(2012)\citenamefont {Haskel},
  \citenamefont {Fabbris}, \citenamefont {Zhernenkov}, \citenamefont {Kong},
  \citenamefont {Jin}, \citenamefont {Cao},\ and\ \citenamefont {van
  Veenendaal}}]{Haskel2012}%
  \BibitemOpen
  \bibfield  {author} {\bibinfo {author} {\bibfnamefont {D.}~\bibnamefont
  {Haskel}}, \bibinfo {author} {\bibfnamefont {G.}~\bibnamefont {Fabbris}},
  \bibinfo {author} {\bibfnamefont {M.}~\bibnamefont {Zhernenkov}}, \bibinfo
  {author} {\bibfnamefont {P.~P.}\ \bibnamefont {Kong}}, \bibinfo {author}
  {\bibfnamefont {C.~Q.}\ \bibnamefont {Jin}}, \bibinfo {author} {\bibfnamefont
  {G.}~\bibnamefont {Cao}}, \ and\ \bibinfo {author} {\bibfnamefont
  {M.}~\bibnamefont {van Veenendaal}},\ }\href {\doibase
  10.1103/PhysRevLett.109.027204} {\bibfield  {journal} {\bibinfo  {journal}
  {Phys. Rev. Lett.}\ }\textbf {\bibinfo {volume} {109}},\ \bibinfo {pages}
  {027204} (\bibinfo {year} {2012})}\BibitemShut {NoStop}%
\bibitem [{\citenamefont {Zhang}\ \emph {et~al.}(2013)\citenamefont {Zhang},
  \citenamefont {Haule},\ and\ \citenamefont {Vanderbilt}}]{Zhang2013_DMFT}%
  \BibitemOpen
  \bibfield  {author} {\bibinfo {author} {\bibfnamefont {H.}~\bibnamefont
  {Zhang}}, \bibinfo {author} {\bibfnamefont {K.}~\bibnamefont {Haule}}, \ and\
  \bibinfo {author} {\bibfnamefont {D.}~\bibnamefont {Vanderbilt}},\ }\href
  {\doibase 10.1103/PhysRevLett.111.246402} {\bibfield  {journal} {\bibinfo
  {journal} {Phys. Rev. Lett.}\ }\textbf {\bibinfo {volume} {111}},\ \bibinfo
  {pages} {246402} (\bibinfo {year} {2013})}\BibitemShut {NoStop}%
\bibitem [{\citenamefont {Kim}\ \emph {et~al.}(2016{\natexlab{b}})\citenamefont
  {Kim}, \citenamefont {Kim}, \citenamefont {Kim},\ and\ \citenamefont
  {Min}}]{Kim2016_optics}%
  \BibitemOpen
  \bibfield  {author} {\bibinfo {author} {\bibfnamefont {B.}~\bibnamefont
  {Kim}}, \bibinfo {author} {\bibfnamefont {B.~H.}\ \bibnamefont {Kim}},
  \bibinfo {author} {\bibfnamefont {K.}~\bibnamefont {Kim}}, \ and\ \bibinfo
  {author} {\bibfnamefont {B.~I.}\ \bibnamefont {Min}},\ }\href
  {https://www.nature.com/articles/srep27095} {\bibfield  {journal} {\bibinfo
  {journal} {Sci Rep}\ }\textbf {\bibinfo {volume} {6}},\ \bibinfo {pages}
  {27095} (\bibinfo {year} {2016}{\natexlab{b}})}\BibitemShut {NoStop}%
\bibitem [{\citenamefont {He}\ and\ \citenamefont
  {Franchini}(2012)}]{PhysRevB.86.235117}%
  \BibitemOpen
  \bibfield  {author} {\bibinfo {author} {\bibfnamefont {J.}~\bibnamefont
  {He}}\ and\ \bibinfo {author} {\bibfnamefont {C.}~\bibnamefont {Franchini}},\
  }\href {\doibase 10.1103/PhysRevB.86.235117} {\bibfield  {journal} {\bibinfo
  {journal} {Phys. Rev. B}\ }\textbf {\bibinfo {volume} {86}},\ \bibinfo
  {pages} {235117} (\bibinfo {year} {2012})}\BibitemShut {NoStop}%
\bibitem [{\citenamefont {Georges}\ \emph {et~al.}(1996)\citenamefont
  {Georges}, \citenamefont {Kotliar}, \citenamefont {Krauth},\ and\
  \citenamefont {Rozenberg}}]{RevModPhys.68.13}%
  \BibitemOpen
  \bibfield  {author} {\bibinfo {author} {\bibfnamefont {A.}~\bibnamefont
  {Georges}}, \bibinfo {author} {\bibfnamefont {G.}~\bibnamefont {Kotliar}},
  \bibinfo {author} {\bibfnamefont {W.}~\bibnamefont {Krauth}}, \ and\ \bibinfo
  {author} {\bibfnamefont {M.~J.}\ \bibnamefont {Rozenberg}},\ }\href {\doibase
  10.1103/RevModPhys.68.13} {\bibfield  {journal} {\bibinfo  {journal} {Rev.
  Mod. Phys.}\ }\textbf {\bibinfo {volume} {68}},\ \bibinfo {pages} {13}
  (\bibinfo {year} {1996})}\BibitemShut {NoStop}%
\bibitem [{\citenamefont {Hanke}\ and\ \citenamefont {Sham}(1980)}]{Hanke1980}%
  \BibitemOpen
  \bibfield  {author} {\bibinfo {author} {\bibfnamefont {W.}~\bibnamefont
  {Hanke}}\ and\ \bibinfo {author} {\bibfnamefont {L.~J.}\ \bibnamefont
  {Sham}},\ }\href {\doibase 10.1103/PhysRevB.21.4656} {\bibfield  {journal}
  {\bibinfo  {journal} {Phys. Rev. B}\ }\textbf {\bibinfo {volume} {21}},\
  \bibinfo {pages} {4656} (\bibinfo {year} {1980})}\BibitemShut {NoStop}%
\bibitem [{\citenamefont {Onida}\ \emph {et~al.}(2002)\citenamefont {Onida},
  \citenamefont {Reining},\ and\ \citenamefont {Rubio}}]{RevModPhys.74.601}%
  \BibitemOpen
  \bibfield  {author} {\bibinfo {author} {\bibfnamefont {G.}~\bibnamefont
  {Onida}}, \bibinfo {author} {\bibfnamefont {L.}~\bibnamefont {Reining}}, \
  and\ \bibinfo {author} {\bibfnamefont {A.}~\bibnamefont {Rubio}},\ }\href
  {\doibase 10.1103/RevModPhys.74.601} {\bibfield  {journal} {\bibinfo
  {journal} {Rev. Mod. Phys.}\ }\textbf {\bibinfo {volume} {74}},\ \bibinfo
  {pages} {601} (\bibinfo {year} {2002})}\BibitemShut {NoStop}%
\bibitem [{\citenamefont {van Schilfgaarde}\ \emph {et~al.}(2006)\citenamefont
  {van Schilfgaarde}, \citenamefont {Kotani},\ and\ \citenamefont
  {Faleev}}]{Schilfgaarde2006}%
  \BibitemOpen
  \bibfield  {author} {\bibinfo {author} {\bibfnamefont {M.}~\bibnamefont {van
  Schilfgaarde}}, \bibinfo {author} {\bibfnamefont {T.}~\bibnamefont {Kotani}},
  \ and\ \bibinfo {author} {\bibfnamefont {S.}~\bibnamefont {Faleev}},\ }\href
  {\doibase 10.1103/PhysRevLett.96.226402} {\bibfield  {journal} {\bibinfo
  {journal} {Phys. Rev. Lett.}\ }\textbf {\bibinfo {volume} {96}},\ \bibinfo
  {pages} {226402} (\bibinfo {year} {2006})}\BibitemShut {NoStop}%
\bibitem [{\citenamefont {Hedin}(1965)}]{Hedin1965}%
  \BibitemOpen
  \bibfield  {author} {\bibinfo {author} {\bibfnamefont {L.}~\bibnamefont
  {Hedin}},\ }\href {\doibase 10.1103/PhysRev.139.A796} {\bibfield  {journal}
  {\bibinfo  {journal} {Phys. Rev.}\ }\textbf {\bibinfo {volume} {139}},\
  \bibinfo {pages} {A796} (\bibinfo {year} {1965})}\BibitemShut {NoStop}%
\bibitem [{\citenamefont {Strinati}\ \emph {et~al.}(1982)\citenamefont
  {Strinati}, \citenamefont {Mattausch},\ and\ \citenamefont
  {Hanke}}]{Strinati1982}%
  \BibitemOpen
  \bibfield  {author} {\bibinfo {author} {\bibfnamefont {G.}~\bibnamefont
  {Strinati}}, \bibinfo {author} {\bibfnamefont {H.~J.}\ \bibnamefont
  {Mattausch}}, \ and\ \bibinfo {author} {\bibfnamefont {W.}~\bibnamefont
  {Hanke}},\ }\href {\doibase 10.1103/PhysRevB.25.2867} {\bibfield  {journal}
  {\bibinfo  {journal} {Phys. Rev. B}\ }\textbf {\bibinfo {volume} {25}},\
  \bibinfo {pages} {2867} (\bibinfo {year} {1982})}\BibitemShut {NoStop}%
\bibitem [{\citenamefont {Hybertsen}\ and\ \citenamefont
  {Louie}(1985)}]{Louie1985}%
  \BibitemOpen
  \bibfield  {author} {\bibinfo {author} {\bibfnamefont {M.~S.}\ \bibnamefont
  {Hybertsen}}\ and\ \bibinfo {author} {\bibfnamefont {S.~G.}\ \bibnamefont
  {Louie}},\ }\href {\doibase 10.1103/PhysRevLett.55.1418} {\bibfield
  {journal} {\bibinfo  {journal} {Phys. Rev. Lett.}\ }\textbf {\bibinfo
  {volume} {55}},\ \bibinfo {pages} {1418} (\bibinfo {year}
  {1985})}\BibitemShut {NoStop}%
\bibitem [{\citenamefont {Klime\v{s}}\ \emph {et~al.}(2014)\citenamefont
  {Klime\v{s}}, \citenamefont {Kaltak},\ and\ \citenamefont
  {Kresse}}]{Klime2014}%
  \BibitemOpen
  \bibfield  {author} {\bibinfo {author} {\bibfnamefont {J.}~\bibnamefont
  {Klime\v{s}}}, \bibinfo {author} {\bibfnamefont {M.}~\bibnamefont {Kaltak}},
  \ and\ \bibinfo {author} {\bibfnamefont {G.}~\bibnamefont {Kresse}},\ }\href
  {\doibase 10.1103/PhysRevB.90.075125} {\bibfield  {journal} {\bibinfo
  {journal} {Phys. Rev. B}\ }\textbf {\bibinfo {volume} {90}},\ \bibinfo
  {pages} {075125} (\bibinfo {year} {2014})}\BibitemShut {NoStop}%
\bibitem [{\citenamefont {Liu}\ \emph {et~al.}(2016{\natexlab{b}})\citenamefont
  {Liu}, \citenamefont {Kaltak}, \citenamefont {Klime\v{s}},\ and\
  \citenamefont {Kresse}}]{PhysRevB.94.165109}%
  \BibitemOpen
  \bibfield  {author} {\bibinfo {author} {\bibfnamefont {P.}~\bibnamefont
  {Liu}}, \bibinfo {author} {\bibfnamefont {M.}~\bibnamefont {Kaltak}},
  \bibinfo {author} {\bibfnamefont {J.}~\bibnamefont {Klime\v{s}}}, \ and\
  \bibinfo {author} {\bibfnamefont {G.}~\bibnamefont {Kresse}},\ }\href
  {\doibase 10.1103/PhysRevB.94.165109} {\bibfield  {journal} {\bibinfo
  {journal} {Phys. Rev. B}\ }\textbf {\bibinfo {volume} {94}},\ \bibinfo
  {pages} {165109} (\bibinfo {year} {2016}{\natexlab{b}})}\BibitemShut
  {NoStop}%
\bibitem [{\citenamefont {Erg\"onenc}\ \emph {et~al.}(2018)\citenamefont
  {Erg\"onenc}, \citenamefont {Kim}, \citenamefont {Liu}, \citenamefont
  {Kresse},\ and\ \citenamefont {Franchini}}]{Zeynep2018}%
  \BibitemOpen
  \bibfield  {author} {\bibinfo {author} {\bibfnamefont {Z.}~\bibnamefont
  {Erg\"onenc}}, \bibinfo {author} {\bibfnamefont {B.}~\bibnamefont {Kim}},
  \bibinfo {author} {\bibfnamefont {P.}~\bibnamefont {Liu}}, \bibinfo {author}
  {\bibfnamefont {G.}~\bibnamefont {Kresse}}, \ and\ \bibinfo {author}
  {\bibfnamefont {C.}~\bibnamefont {Franchini}},\ }\href {\doibase
  10.1103/PhysRevMaterials.2.024601} {\bibfield  {journal} {\bibinfo  {journal}
  {Phys. Rev. Materials}\ }\textbf {\bibinfo {volume} {2}},\ \bibinfo {pages}
  {024601} (\bibinfo {year} {2018})}\BibitemShut {NoStop}%
\bibitem [{\citenamefont {Rohlfing}\ and\ \citenamefont
  {Louie}(2000)}]{Rohlfing2000}%
  \BibitemOpen
  \bibfield  {author} {\bibinfo {author} {\bibfnamefont {M.}~\bibnamefont
  {Rohlfing}}\ and\ \bibinfo {author} {\bibfnamefont {S.~G.}\ \bibnamefont
  {Louie}},\ }\href {\doibase 10.1103/PhysRevB.62.4927} {\bibfield  {journal}
  {\bibinfo  {journal} {Phys. Rev. B}\ }\textbf {\bibinfo {volume} {62}},\
  \bibinfo {pages} {4927} (\bibinfo {year} {2000})}\BibitemShut {NoStop}%
\bibitem [{\citenamefont {Tiago}\ and\ \citenamefont
  {Chelikowsky}(2006)}]{Tiago2006}%
  \BibitemOpen
  \bibfield  {author} {\bibinfo {author} {\bibfnamefont {M.~L.}\ \bibnamefont
  {Tiago}}\ and\ \bibinfo {author} {\bibfnamefont {J.~R.}\ \bibnamefont
  {Chelikowsky}},\ }\href {\doibase 10.1103/PhysRevB.73.205334} {\bibfield
  {journal} {\bibinfo  {journal} {Phys. Rev. B}\ }\textbf {\bibinfo {volume}
  {73}},\ \bibinfo {pages} {205334} (\bibinfo {year} {2006})}\BibitemShut
  {NoStop}%
\bibitem [{\citenamefont {K\"{o}rbel}\ \emph {et~al.}(2014)\citenamefont
  {K\"{o}rbel}, \citenamefont {Boulanger}, \citenamefont {Duchemin},
  \citenamefont {Blase}, \citenamefont {Marques},\ and\ \citenamefont
  {Botti}}]{Sabine2014}%
  \BibitemOpen
  \bibfield  {author} {\bibinfo {author} {\bibfnamefont {S.}~\bibnamefont
  {K\"{o}rbel}}, \bibinfo {author} {\bibfnamefont {P.}~\bibnamefont
  {Boulanger}}, \bibinfo {author} {\bibfnamefont {I.}~\bibnamefont {Duchemin}},
  \bibinfo {author} {\bibfnamefont {X.}~\bibnamefont {Blase}}, \bibinfo
  {author} {\bibfnamefont {M.~A.~L.}\ \bibnamefont {Marques}}, \ and\ \bibinfo
  {author} {\bibfnamefont {S.}~\bibnamefont {Botti}},\ }\href {\doibase
  10.1021/ct5003658} {\bibfield  {journal} {\bibinfo  {journal} {Journal of
  Chemical Theory and Computation}\ }\textbf {\bibinfo {volume} {10}},\
  \bibinfo {pages} {3934} (\bibinfo {year} {2014})}\BibitemShut {NoStop}%
\bibitem [{\citenamefont {He}\ \emph {et~al.}(2014)\citenamefont {He},
  \citenamefont {Hummer},\ and\ \citenamefont {Franchini}}]{Jiangang2014}%
  \BibitemOpen
  \bibfield  {author} {\bibinfo {author} {\bibfnamefont {J.}~\bibnamefont
  {He}}, \bibinfo {author} {\bibfnamefont {K.}~\bibnamefont {Hummer}}, \ and\
  \bibinfo {author} {\bibfnamefont {C.}~\bibnamefont {Franchini}},\ }\href
  {\doibase 10.1103/PhysRevB.89.075409} {\bibfield  {journal} {\bibinfo
  {journal} {Phys. Rev. B}\ }\textbf {\bibinfo {volume} {89}},\ \bibinfo
  {pages} {075409} (\bibinfo {year} {2014})}\BibitemShut {NoStop}%
\bibitem [{\citenamefont {Cunningham}\ \emph {et~al.}(2018)\citenamefont
  {Cunningham}, \citenamefont {Gr\"uning}, \citenamefont {Azarhoosh},
  \citenamefont {Pashov},\ and\ \citenamefont {van
  Schilfgaarde}}]{PhysRevMaterials.2.034603}%
  \BibitemOpen
  \bibfield  {author} {\bibinfo {author} {\bibfnamefont {B.}~\bibnamefont
  {Cunningham}}, \bibinfo {author} {\bibfnamefont {M.}~\bibnamefont
  {Gr\"uning}}, \bibinfo {author} {\bibfnamefont {P.}~\bibnamefont
  {Azarhoosh}}, \bibinfo {author} {\bibfnamefont {D.}~\bibnamefont {Pashov}}, \
  and\ \bibinfo {author} {\bibfnamefont {M.}~\bibnamefont {van Schilfgaarde}},\
  }\href {\doibase 10.1103/PhysRevMaterials.2.034603} {\bibfield  {journal}
  {\bibinfo  {journal} {Phys. Rev. Materials}\ }\textbf {\bibinfo {volume}
  {2}},\ \bibinfo {pages} {034603} (\bibinfo {year} {2018})}\BibitemShut
  {NoStop}%
\bibitem [{\citenamefont {Franchini}\ \emph {et~al.}(2010)\citenamefont
  {Franchini}, \citenamefont {Sanna}, \citenamefont {Marsman},\ and\
  \citenamefont {Kresse}}]{PhysRevB.81.085213}%
  \BibitemOpen
  \bibfield  {author} {\bibinfo {author} {\bibfnamefont {C.}~\bibnamefont
  {Franchini}}, \bibinfo {author} {\bibfnamefont {A.}~\bibnamefont {Sanna}},
  \bibinfo {author} {\bibfnamefont {M.}~\bibnamefont {Marsman}}, \ and\
  \bibinfo {author} {\bibfnamefont {G.}~\bibnamefont {Kresse}},\ }\href
  {\doibase 10.1103/PhysRevB.81.085213} {\bibfield  {journal} {\bibinfo
  {journal} {Phys. Rev. B}\ }\textbf {\bibinfo {volume} {81}},\ \bibinfo
  {pages} {085213} (\bibinfo {year} {2010})}\BibitemShut {NoStop}%
\bibitem [{\citenamefont {Sponza}\ \emph {et~al.}(2013)\citenamefont {Sponza},
  \citenamefont {V\'eniard}, \citenamefont {Sottile}, \citenamefont
  {Giorgetti},\ and\ \citenamefont {Reining}}]{PhysRevB.87.235102}%
  \BibitemOpen
  \bibfield  {author} {\bibinfo {author} {\bibfnamefont {L.}~\bibnamefont
  {Sponza}}, \bibinfo {author} {\bibfnamefont {V.}~\bibnamefont {V\'eniard}},
  \bibinfo {author} {\bibfnamefont {F.}~\bibnamefont {Sottile}}, \bibinfo
  {author} {\bibfnamefont {C.}~\bibnamefont {Giorgetti}}, \ and\ \bibinfo
  {author} {\bibfnamefont {L.}~\bibnamefont {Reining}},\ }\href {\doibase
  10.1103/PhysRevB.87.235102} {\bibfield  {journal} {\bibinfo  {journal} {Phys.
  Rev. B}\ }\textbf {\bibinfo {volume} {87}},\ \bibinfo {pages} {235102}
  (\bibinfo {year} {2013})}\BibitemShut {NoStop}%
\bibitem [{\citenamefont {He}\ and\ \citenamefont
  {Franchini}(2014)}]{PhysRevB.89.045104}%
  \BibitemOpen
  \bibfield  {author} {\bibinfo {author} {\bibfnamefont {J.}~\bibnamefont
  {He}}\ and\ \bibinfo {author} {\bibfnamefont {C.}~\bibnamefont {Franchini}},\
  }\href {\doibase 10.1103/PhysRevB.89.045104} {\bibfield  {journal} {\bibinfo
  {journal} {Phys. Rev. B}\ }\textbf {\bibinfo {volume} {89}},\ \bibinfo
  {pages} {045104} (\bibinfo {year} {2014})}\BibitemShut {NoStop}%
\bibitem [{\citenamefont {Gatti}\ \emph {et~al.}(2015)\citenamefont {Gatti},
  \citenamefont {Sottile},\ and\ \citenamefont {Reining}}]{PhysRevB.91.195137}%
  \BibitemOpen
  \bibfield  {author} {\bibinfo {author} {\bibfnamefont {M.}~\bibnamefont
  {Gatti}}, \bibinfo {author} {\bibfnamefont {F.}~\bibnamefont {Sottile}}, \
  and\ \bibinfo {author} {\bibfnamefont {L.}~\bibnamefont {Reining}},\ }\href
  {\doibase 10.1103/PhysRevB.91.195137} {\bibfield  {journal} {\bibinfo
  {journal} {Phys. Rev. B}\ }\textbf {\bibinfo {volume} {91}},\ \bibinfo
  {pages} {195137} (\bibinfo {year} {2015})}\BibitemShut {NoStop}%
\bibitem [{\citenamefont {Bl\"ochl}(1994)}]{PhysRevB.50.17953}%
  \BibitemOpen
  \bibfield  {author} {\bibinfo {author} {\bibfnamefont {P.~E.}\ \bibnamefont
  {Bl\"ochl}},\ }\href {\doibase 10.1103/PhysRevB.50.17953} {\bibfield
  {journal} {\bibinfo  {journal} {Phys. Rev. B}\ }\textbf {\bibinfo {volume}
  {50}},\ \bibinfo {pages} {17953} (\bibinfo {year} {1994})}\BibitemShut
  {NoStop}%
\bibitem [{\citenamefont {Kresse}\ and\ \citenamefont
  {Hafner}(1993)}]{PhysRevB.47.558}%
  \BibitemOpen
  \bibfield  {author} {\bibinfo {author} {\bibfnamefont {G.}~\bibnamefont
  {Kresse}}\ and\ \bibinfo {author} {\bibfnamefont {J.}~\bibnamefont
  {Hafner}},\ }\href {\doibase 10.1103/PhysRevB.47.558} {\bibfield  {journal}
  {\bibinfo  {journal} {Phys. Rev. B}\ }\textbf {\bibinfo {volume} {47}},\
  \bibinfo {pages} {558} (\bibinfo {year} {1993})}\BibitemShut {NoStop}%
\bibitem [{\citenamefont {Kresse}\ and\ \citenamefont
  {Furthm\"uller}(1996)}]{PhysRevB.54.11169}%
  \BibitemOpen
  \bibfield  {author} {\bibinfo {author} {\bibfnamefont {G.}~\bibnamefont
  {Kresse}}\ and\ \bibinfo {author} {\bibfnamefont {J.}~\bibnamefont
  {Furthm\"uller}},\ }\href {\doibase 10.1103/PhysRevB.54.11169} {\bibfield
  {journal} {\bibinfo  {journal} {Phys. Rev. B}\ }\textbf {\bibinfo {volume}
  {54}},\ \bibinfo {pages} {11169} (\bibinfo {year} {1996})}\BibitemShut
  {NoStop}%
\bibitem [{\citenamefont {Crawford}\ \emph {et~al.}(1994)\citenamefont
  {Crawford}, \citenamefont {Subramanian}, \citenamefont {Harlow},
  \citenamefont {Fernandez-Baca}, \citenamefont {Wang},\ and\ \citenamefont
  {Johnston}}]{PhysRevB.49.9198}%
  \BibitemOpen
  \bibfield  {author} {\bibinfo {author} {\bibfnamefont {M.~K.}\ \bibnamefont
  {Crawford}}, \bibinfo {author} {\bibfnamefont {M.~A.}\ \bibnamefont
  {Subramanian}}, \bibinfo {author} {\bibfnamefont {R.~L.}\ \bibnamefont
  {Harlow}}, \bibinfo {author} {\bibfnamefont {J.~A.}\ \bibnamefont
  {Fernandez-Baca}}, \bibinfo {author} {\bibfnamefont {Z.~R.}\ \bibnamefont
  {Wang}}, \ and\ \bibinfo {author} {\bibfnamefont {D.~C.}\ \bibnamefont
  {Johnston}},\ }\href {\doibase 10.1103/PhysRevB.49.9198} {\bibfield
  {journal} {\bibinfo  {journal} {Phys. Rev. B}\ }\textbf {\bibinfo {volume}
  {49}},\ \bibinfo {pages} {9198} (\bibinfo {year} {1994})}\BibitemShut
  {NoStop}%
\bibitem [{\citenamefont {Subramanian}\ \emph {et~al.}(1994)\citenamefont
  {Subramanian}, \citenamefont {Crawford},\ and\ \citenamefont
  {Harlow}}]{SUBRAMANIAN1994645}%
  \BibitemOpen
  \bibfield  {author} {\bibinfo {author} {\bibfnamefont {M.}~\bibnamefont
  {Subramanian}}, \bibinfo {author} {\bibfnamefont {M.}~\bibnamefont
  {Crawford}}, \ and\ \bibinfo {author} {\bibfnamefont {R.}~\bibnamefont
  {Harlow}},\ }\href {\doibase https://doi.org/10.1016/0025-5408(94)90120-1}
  {\bibfield  {journal} {\bibinfo  {journal} {Materials Research Bulletin}\
  }\textbf {\bibinfo {volume} {29}},\ \bibinfo {pages} {645 } (\bibinfo {year}
  {1994})}\BibitemShut {NoStop}%
\bibitem [{\citenamefont {Zhao}\ \emph {et~al.}(2008)\citenamefont {Zhao},
  \citenamefont {Yang}, \citenamefont {Yu}, \citenamefont {Li}, \citenamefont
  {Yu}, \citenamefont {Fang}, \citenamefont {Chen},\ and\ \citenamefont
  {Jin}}]{Zhao2008}%
  \BibitemOpen
  \bibfield  {author} {\bibinfo {author} {\bibfnamefont {J.~G.}\ \bibnamefont
  {Zhao}}, \bibinfo {author} {\bibfnamefont {L.~X.}\ \bibnamefont {Yang}},
  \bibinfo {author} {\bibfnamefont {Y.}~\bibnamefont {Yu}}, \bibinfo {author}
  {\bibfnamefont {F.~Y.}\ \bibnamefont {Li}}, \bibinfo {author} {\bibfnamefont
  {R.~C.}\ \bibnamefont {Yu}}, \bibinfo {author} {\bibfnamefont
  {Z.}~\bibnamefont {Fang}}, \bibinfo {author} {\bibfnamefont {L.~C.}\
  \bibnamefont {Chen}}, \ and\ \bibinfo {author} {\bibfnamefont {C.~Q.}\
  \bibnamefont {Jin}},\ }\href {\doibase 10.1063/1.2908879} {\bibfield
  {journal} {\bibinfo  {journal} {Journal of Applied Physics}\ }\textbf
  {\bibinfo {volume} {103}},\ \bibinfo {pages} {103706} (\bibinfo {year}
  {2008})}\BibitemShut {NoStop}%
\bibitem [{\citenamefont {Puggioni}\ and\ \citenamefont
  {Rondinelli}(2016)}]{Puggioni2016}%
  \BibitemOpen
  \bibfield  {author} {\bibinfo {author} {\bibfnamefont {D.}~\bibnamefont
  {Puggioni}}\ and\ \bibinfo {author} {\bibfnamefont {J.~M.}\ \bibnamefont
  {Rondinelli}},\ }\href {\doibase 10.1063/1.4942651} {\bibfield  {journal}
  {\bibinfo  {journal} {Journal of Applied Physics}\ }\textbf {\bibinfo
  {volume} {119}},\ \bibinfo {pages} {086102} (\bibinfo {year}
  {2016})}\BibitemShut {NoStop}%
\bibitem [{\citenamefont {Hobbs}\ \emph {et~al.}(2000)\citenamefont {Hobbs},
  \citenamefont {Kresse},\ and\ \citenamefont {Hafner}}]{Hobbs2000_ncl_VASP}%
  \BibitemOpen
  \bibfield  {author} {\bibinfo {author} {\bibfnamefont {D.}~\bibnamefont
  {Hobbs}}, \bibinfo {author} {\bibfnamefont {G.}~\bibnamefont {Kresse}}, \
  and\ \bibinfo {author} {\bibfnamefont {J.}~\bibnamefont {Hafner}},\ }\href
  {\doibase 10.1103/PhysRevB.62.11556} {\bibfield  {journal} {\bibinfo
  {journal} {Phys. Rev. B}\ }\textbf {\bibinfo {volume} {62}},\ \bibinfo
  {pages} {11556} (\bibinfo {year} {2000})}\BibitemShut {NoStop}%
\bibitem [{\citenamefont {Aryasetiawan}\ and\ \citenamefont
  {Biermann}(2008)}]{Biermann2008_GW_SOC}%
  \BibitemOpen
  \bibfield  {author} {\bibinfo {author} {\bibfnamefont {F.}~\bibnamefont
  {Aryasetiawan}}\ and\ \bibinfo {author} {\bibfnamefont {S.}~\bibnamefont
  {Biermann}},\ }\href {\doibase 10.1103/PhysRevLett.100.116402} {\bibfield
  {journal} {\bibinfo  {journal} {Phys. Rev. Lett.}\ }\textbf {\bibinfo
  {volume} {100}},\ \bibinfo {pages} {116402} (\bibinfo {year}
  {2008})}\BibitemShut {NoStop}%
\bibitem [{\citenamefont {Steiner}\ \emph {et~al.}(2016)\citenamefont
  {Steiner}, \citenamefont {Khmelevskyi}, \citenamefont {Marsmann},\ and\
  \citenamefont {Kresse}}]{PhysRevB.93.224425}%
  \BibitemOpen
  \bibfield  {author} {\bibinfo {author} {\bibfnamefont {S.}~\bibnamefont
  {Steiner}}, \bibinfo {author} {\bibfnamefont {S.}~\bibnamefont
  {Khmelevskyi}}, \bibinfo {author} {\bibfnamefont {M.}~\bibnamefont
  {Marsmann}}, \ and\ \bibinfo {author} {\bibfnamefont {G.}~\bibnamefont
  {Kresse}},\ }\href {\doibase 10.1103/PhysRevB.93.224425} {\bibfield
  {journal} {\bibinfo  {journal} {Phys. Rev. B}\ }\textbf {\bibinfo {volume}
  {93}},\ \bibinfo {pages} {224425} (\bibinfo {year} {2016})}\BibitemShut
  {NoStop}%
\bibitem [{\citenamefont {Fuchs}\ \emph {et~al.}(2007)\citenamefont {Fuchs},
  \citenamefont {Furthm\"uller}, \citenamefont {Bechstedt}, \citenamefont
  {Shishkin},\ and\ \citenamefont {Kresse}}]{Fuchs2007}%
  \BibitemOpen
  \bibfield  {author} {\bibinfo {author} {\bibfnamefont {F.}~\bibnamefont
  {Fuchs}}, \bibinfo {author} {\bibfnamefont {J.}~\bibnamefont
  {Furthm\"uller}}, \bibinfo {author} {\bibfnamefont {F.}~\bibnamefont
  {Bechstedt}}, \bibinfo {author} {\bibfnamefont {M.}~\bibnamefont {Shishkin}},
  \ and\ \bibinfo {author} {\bibfnamefont {G.}~\bibnamefont {Kresse}},\ }\href
  {\doibase 10.1103/PhysRevB.76.115109} {\bibfield  {journal} {\bibinfo
  {journal} {Phys. Rev. B}\ }\textbf {\bibinfo {volume} {76}},\ \bibinfo
  {pages} {115109} (\bibinfo {year} {2007})}\BibitemShut {NoStop}%
\bibitem [{\citenamefont {Shishkin}\ \emph {et~al.}(2007)\citenamefont
  {Shishkin}, \citenamefont {Marsman},\ and\ \citenamefont
  {Kresse}}]{Shishkin2007_PRL}%
  \BibitemOpen
  \bibfield  {author} {\bibinfo {author} {\bibfnamefont {M.}~\bibnamefont
  {Shishkin}}, \bibinfo {author} {\bibfnamefont {M.}~\bibnamefont {Marsman}}, \
  and\ \bibinfo {author} {\bibfnamefont {G.}~\bibnamefont {Kresse}},\ }\href
  {\doibase 10.1103/PhysRevLett.99.246403} {\bibfield  {journal} {\bibinfo
  {journal} {Phys. Rev. Lett.}\ }\textbf {\bibinfo {volume} {99}},\ \bibinfo
  {pages} {246403} (\bibinfo {year} {2007})}\BibitemShut {NoStop}%
\bibitem [{\citenamefont {Jiang}\ \emph {et~al.}(2009)\citenamefont {Jiang},
  \citenamefont {Gomez-Abal}, \citenamefont {Rinke},\ and\ \citenamefont
  {Scheffler}}]{JiangHong_2009PRL}%
  \BibitemOpen
  \bibfield  {author} {\bibinfo {author} {\bibfnamefont {H.}~\bibnamefont
  {Jiang}}, \bibinfo {author} {\bibfnamefont {R.~I.}\ \bibnamefont
  {Gomez-Abal}}, \bibinfo {author} {\bibfnamefont {P.}~\bibnamefont {Rinke}}, \
  and\ \bibinfo {author} {\bibfnamefont {M.}~\bibnamefont {Scheffler}},\ }\href
  {\doibase 10.1103/PhysRevLett.102.126403} {\bibfield  {journal} {\bibinfo
  {journal} {Phys. Rev. Lett.}\ }\textbf {\bibinfo {volume} {102}},\ \bibinfo
  {pages} {126403} (\bibinfo {year} {2009})}\BibitemShut {NoStop}%
\bibitem [{\citenamefont {Jiang}\ \emph {et~al.}(2010)\citenamefont {Jiang},
  \citenamefont {Gomez-Abal}, \citenamefont {Rinke},\ and\ \citenamefont
  {Scheffler}}]{JiangHong_2010PRB}%
  \BibitemOpen
  \bibfield  {author} {\bibinfo {author} {\bibfnamefont {H.}~\bibnamefont
  {Jiang}}, \bibinfo {author} {\bibfnamefont {R.~I.}\ \bibnamefont
  {Gomez-Abal}}, \bibinfo {author} {\bibfnamefont {P.}~\bibnamefont {Rinke}}, \
  and\ \bibinfo {author} {\bibfnamefont {M.}~\bibnamefont {Scheffler}},\ }\href
  {\doibase 10.1103/PhysRevB.82.045108} {\bibfield  {journal} {\bibinfo
  {journal} {Phys. Rev. B}\ }\textbf {\bibinfo {volume} {82}},\ \bibinfo
  {pages} {045108} (\bibinfo {year} {2010})}\BibitemShut {NoStop}%
\bibitem [{\citenamefont {Dudarev}\ \emph {et~al.}(1998)\citenamefont
  {Dudarev}, \citenamefont {Botton}, \citenamefont {Savrasov}, \citenamefont
  {Humphreys},\ and\ \citenamefont {Sutton}}]{PhysRevB.57.1505}%
  \BibitemOpen
  \bibfield  {author} {\bibinfo {author} {\bibfnamefont {S.~L.}\ \bibnamefont
  {Dudarev}}, \bibinfo {author} {\bibfnamefont {G.~A.}\ \bibnamefont {Botton}},
  \bibinfo {author} {\bibfnamefont {S.~Y.}\ \bibnamefont {Savrasov}}, \bibinfo
  {author} {\bibfnamefont {C.~J.}\ \bibnamefont {Humphreys}}, \ and\ \bibinfo
  {author} {\bibfnamefont {A.~P.}\ \bibnamefont {Sutton}},\ }\href {\doibase
  10.1103/PhysRevB.57.1505} {\bibfield  {journal} {\bibinfo  {journal} {Phys.
  Rev. B}\ }\textbf {\bibinfo {volume} {57}},\ \bibinfo {pages} {1505}
  (\bibinfo {year} {1998})}\BibitemShut {NoStop}%
\bibitem [{\citenamefont {Aryasetiawan}\ \emph {et~al.}(2006)\citenamefont
  {Aryasetiawan}, \citenamefont {Karlsson}, \citenamefont {Jepsen},\ and\
  \citenamefont {Sch\"onberger}}]{PhysRevB.74.125106}%
  \BibitemOpen
  \bibfield  {author} {\bibinfo {author} {\bibfnamefont {F.}~\bibnamefont
  {Aryasetiawan}}, \bibinfo {author} {\bibfnamefont {K.}~\bibnamefont
  {Karlsson}}, \bibinfo {author} {\bibfnamefont {O.}~\bibnamefont {Jepsen}}, \
  and\ \bibinfo {author} {\bibfnamefont {U.}~\bibnamefont {Sch\"onberger}},\
  }\href {\doibase 10.1103/PhysRevB.74.125106} {\bibfield  {journal} {\bibinfo
  {journal} {Phys. Rev. B}\ }\textbf {\bibinfo {volume} {74}},\ \bibinfo
  {pages} {125106} (\bibinfo {year} {2006})}\BibitemShut {NoStop}%
\bibitem [{\citenamefont {Kaltak}(2015)}]{Merzuk2015}%
  \BibitemOpen
  \bibfield  {author} {\bibinfo {author} {\bibfnamefont {M.}~\bibnamefont
  {Kaltak}},\ }\emph {\bibinfo {title} {Merging GW with DMFT}},\ \href
  {http://othes.univie.ac.at/38099/} {Ph.D. thesis},\ \bibinfo  {school}
  {University of Vienna} (\bibinfo {year} {2015})\BibitemShut {NoStop}%
\bibitem [{\citenamefont {Marzari}\ and\ \citenamefont
  {Vanderbilt}(1997)}]{PhysRevB.56.12847}%
  \BibitemOpen
  \bibfield  {author} {\bibinfo {author} {\bibfnamefont {N.}~\bibnamefont
  {Marzari}}\ and\ \bibinfo {author} {\bibfnamefont {D.}~\bibnamefont
  {Vanderbilt}},\ }\href {\doibase 10.1103/PhysRevB.56.12847} {\bibfield
  {journal} {\bibinfo  {journal} {Phys. Rev. B}\ }\textbf {\bibinfo {volume}
  {56}},\ \bibinfo {pages} {12847} (\bibinfo {year} {1997})}\BibitemShut
  {NoStop}%
\bibitem [{\citenamefont {Mostofi}\ \emph {et~al.}(2008)\citenamefont
  {Mostofi}, \citenamefont {Yates}, \citenamefont {Lee}, \citenamefont {Souza},
  \citenamefont {Vanderbilt},\ and\ \citenamefont {Marzari}}]{MOSTOFI2008685}%
  \BibitemOpen
  \bibfield  {author} {\bibinfo {author} {\bibfnamefont {A.~A.}\ \bibnamefont
  {Mostofi}}, \bibinfo {author} {\bibfnamefont {J.~R.}\ \bibnamefont {Yates}},
  \bibinfo {author} {\bibfnamefont {Y.-S.}\ \bibnamefont {Lee}}, \bibinfo
  {author} {\bibfnamefont {I.}~\bibnamefont {Souza}}, \bibinfo {author}
  {\bibfnamefont {D.}~\bibnamefont {Vanderbilt}}, \ and\ \bibinfo {author}
  {\bibfnamefont {N.}~\bibnamefont {Marzari}},\ }\href {\doibase
  https://doi.org/10.1016/j.cpc.2007.11.016} {\bibfield  {journal} {\bibinfo
  {journal} {Computer Physics Communications}\ }\textbf {\bibinfo {volume}
  {178}},\ \bibinfo {pages} {685 } (\bibinfo {year} {2008})}\BibitemShut
  {NoStop}%
\bibitem [{\citenamefont {Franchini}\ \emph {et~al.}(2012)\citenamefont
  {Franchini}, \citenamefont {Kovcik}, \citenamefont {Marsman}, \citenamefont
  {Murthy}, \citenamefont {He}, \citenamefont {Ederer},\ and\ \citenamefont
  {Kresse}}]{Franchini2012}%
  \BibitemOpen
  \bibfield  {author} {\bibinfo {author} {\bibfnamefont {C.}~\bibnamefont
  {Franchini}}, \bibinfo {author} {\bibfnamefont {R.}~\bibnamefont {Kovcik}},
  \bibinfo {author} {\bibfnamefont {M.}~\bibnamefont {Marsman}}, \bibinfo
  {author} {\bibfnamefont {S.~S.}\ \bibnamefont {Murthy}}, \bibinfo {author}
  {\bibfnamefont {J.}~\bibnamefont {He}}, \bibinfo {author} {\bibfnamefont
  {C.}~\bibnamefont {Ederer}}, \ and\ \bibinfo {author} {\bibfnamefont
  {G.}~\bibnamefont {Kresse}},\ }\href
  {http://stacks.iop.org/0953-8984/24/i=23/a=235602} {\bibfield  {journal}
  {\bibinfo  {journal} {Journal of Physics: Condensed Matter}\ }\textbf
  {\bibinfo {volume} {24}},\ \bibinfo {pages} {235602} (\bibinfo {year}
  {2012})}\BibitemShut {NoStop}%
\bibitem [{\citenamefont {Arita}\ \emph {et~al.}(2012)\citenamefont {Arita},
  \citenamefont {Kune\ifmmode~\check{s}\else \v{s}\fi{}}, \citenamefont
  {Kozhevnikov}, \citenamefont {Eguiluz},\ and\ \citenamefont
  {Imada}}]{PhysRevLett.108.086403}%
  \BibitemOpen
  \bibfield  {author} {\bibinfo {author} {\bibfnamefont {R.}~\bibnamefont
  {Arita}}, \bibinfo {author} {\bibfnamefont {J.}~\bibnamefont
  {Kune\ifmmode~\check{s}\else \v{s}\fi{}}}, \bibinfo {author} {\bibfnamefont
  {A.~V.}\ \bibnamefont {Kozhevnikov}}, \bibinfo {author} {\bibfnamefont
  {A.~G.}\ \bibnamefont {Eguiluz}}, \ and\ \bibinfo {author} {\bibfnamefont
  {M.}~\bibnamefont {Imada}},\ }\href {\doibase 10.1103/PhysRevLett.108.086403}
  {\bibfield  {journal} {\bibinfo  {journal} {Phys. Rev. Lett.}\ }\textbf
  {\bibinfo {volume} {108}},\ \bibinfo {pages} {086403} (\bibinfo {year}
  {2012})}\BibitemShut {NoStop}%
\bibitem [{\citenamefont {Sander}\ \emph {et~al.}(2015)\citenamefont {Sander},
  \citenamefont {Maggio},\ and\ \citenamefont {Kresse}}]{Tobias2015}%
  \BibitemOpen
  \bibfield  {author} {\bibinfo {author} {\bibfnamefont {T.}~\bibnamefont
  {Sander}}, \bibinfo {author} {\bibfnamefont {E.}~\bibnamefont {Maggio}}, \
  and\ \bibinfo {author} {\bibfnamefont {G.}~\bibnamefont {Kresse}},\ }\href
  {\doibase 10.1103/PhysRevB.92.045209} {\bibfield  {journal} {\bibinfo
  {journal} {Phys. Rev. B}\ }\textbf {\bibinfo {volume} {92}},\ \bibinfo
  {pages} {045209} (\bibinfo {year} {2015})}\BibitemShut {NoStop}%
\bibitem [{\citenamefont {Ambrosch-Draxl}\ and\ \citenamefont
  {Sofo}(2006)}]{Ambrosch-Draxl2004}%
  \BibitemOpen
  \bibfield  {author} {\bibinfo {author} {\bibfnamefont {C.}~\bibnamefont
  {Ambrosch-Draxl}}\ and\ \bibinfo {author} {\bibfnamefont {J.~O.}\
  \bibnamefont {Sofo}},\ }\href {\doibase
  https://doi.org/10.1016/j.cpc.2006.03.005} {\bibfield  {journal} {\bibinfo
  {journal} {Computer Physics Communications}\ }\textbf {\bibinfo {volume}
  {175}},\ \bibinfo {pages} {1 } (\bibinfo {year} {2006})}\BibitemShut
  {NoStop}%
\bibitem [{\citenamefont {Bechstedt}\ \emph {et~al.}(1992)\citenamefont
  {Bechstedt}, \citenamefont {Sole}, \citenamefont {Cappellini},\ and\
  \citenamefont {Reining}}]{BECHSTEDT1992765}%
  \BibitemOpen
  \bibfield  {author} {\bibinfo {author} {\bibfnamefont {F.}~\bibnamefont
  {Bechstedt}}, \bibinfo {author} {\bibfnamefont {R.~D.}\ \bibnamefont {Sole}},
  \bibinfo {author} {\bibfnamefont {G.}~\bibnamefont {Cappellini}}, \ and\
  \bibinfo {author} {\bibfnamefont {L.}~\bibnamefont {Reining}},\ }\href
  {\doibase https://doi.org/10.1016/0038-1098(92)90476-P} {\bibfield  {journal}
  {\bibinfo  {journal} {Solid State Communications}\ }\textbf {\bibinfo
  {volume} {84}},\ \bibinfo {pages} {765 } (\bibinfo {year}
  {1992})}\BibitemShut {NoStop}%
\bibitem [{\citenamefont {Fuchs}\ \emph {et~al.}(2008)\citenamefont {Fuchs},
  \citenamefont {R\"odl}, \citenamefont {Schleife},\ and\ \citenamefont
  {Bechstedt}}]{PhysRevB.78.085103}%
  \BibitemOpen
  \bibfield  {author} {\bibinfo {author} {\bibfnamefont {F.}~\bibnamefont
  {Fuchs}}, \bibinfo {author} {\bibfnamefont {C.}~\bibnamefont {R\"odl}},
  \bibinfo {author} {\bibfnamefont {A.}~\bibnamefont {Schleife}}, \ and\
  \bibinfo {author} {\bibfnamefont {F.}~\bibnamefont {Bechstedt}},\ }\href
  {\doibase 10.1103/PhysRevB.78.085103} {\bibfield  {journal} {\bibinfo
  {journal} {Phys. Rev. B}\ }\textbf {\bibinfo {volume} {78}},\ \bibinfo
  {pages} {085103} (\bibinfo {year} {2008})}\BibitemShut {NoStop}%
\bibitem [{\citenamefont {Bokdam}\ \emph {et~al.}(2016)\citenamefont {Bokdam},
  \citenamefont {Sander}, \citenamefont {Stroppa}, \citenamefont {Picozzi},
  \citenamefont {Sarma}, \citenamefont {Franchini},\ and\ \citenamefont
  {Kresse}}]{Bokdam2016}%
  \BibitemOpen
  \bibfield  {author} {\bibinfo {author} {\bibfnamefont {M.}~\bibnamefont
  {Bokdam}}, \bibinfo {author} {\bibfnamefont {T.}~\bibnamefont {Sander}},
  \bibinfo {author} {\bibfnamefont {A.}~\bibnamefont {Stroppa}}, \bibinfo
  {author} {\bibfnamefont {S.}~\bibnamefont {Picozzi}}, \bibinfo {author}
  {\bibfnamefont {D.~D.}\ \bibnamefont {Sarma}}, \bibinfo {author}
  {\bibfnamefont {C.}~\bibnamefont {Franchini}}, \ and\ \bibinfo {author}
  {\bibfnamefont {G.}~\bibnamefont {Kresse}},\ }\href
  {https://www.nature.com/articles/srep28618} {\bibfield  {journal} {\bibinfo
  {journal} {Sci Rep}\ }\textbf {\bibinfo {volume} {6}},\ \bibinfo {pages}
  {28618} (\bibinfo {year} {2016})}\BibitemShut {NoStop}%
\bibitem [{\citenamefont {Paier}\ \emph {et~al.}(2008)\citenamefont {Paier},
  \citenamefont {Marsman},\ and\ \citenamefont {Kresse}}]{PhysRevB.78.121201}%
  \BibitemOpen
  \bibfield  {author} {\bibinfo {author} {\bibfnamefont {J.}~\bibnamefont
  {Paier}}, \bibinfo {author} {\bibfnamefont {M.}~\bibnamefont {Marsman}}, \
  and\ \bibinfo {author} {\bibfnamefont {G.}~\bibnamefont {Kresse}},\ }\href
  {\doibase 10.1103/PhysRevB.78.121201} {\bibfield  {journal} {\bibinfo
  {journal} {Phys. Rev. B}\ }\textbf {\bibinfo {volume} {78}},\ \bibinfo
  {pages} {121201} (\bibinfo {year} {2008})}\BibitemShut {NoStop}%
\bibitem [{\citenamefont {Beznosikov}\ and\ \citenamefont
  {Aleksandrov}(2000)}]{Beznosikov2000}%
  \BibitemOpen
  \bibfield  {author} {\bibinfo {author} {\bibfnamefont {B.~V.}\ \bibnamefont
  {Beznosikov}}\ and\ \bibinfo {author} {\bibfnamefont {K.~S.}\ \bibnamefont
  {Aleksandrov}},\ }\href {https://link.springer.com/article/10.1134/1.1312923}
  {\bibfield  {journal} {\bibinfo  {journal} {Crystallography Reports}\
  }\textbf {\bibinfo {volume} {45}},\ \bibinfo {pages} {792} (\bibinfo {year}
  {2000})}\BibitemShut {NoStop}%
\bibitem [{\citenamefont {Longo}\ \emph {et~al.}(1971)\citenamefont {Longo},
  \citenamefont {Kafalas},\ and\ \citenamefont {Arnott}}]{LONGO1971174}%
  \BibitemOpen
  \bibfield  {author} {\bibinfo {author} {\bibfnamefont {J.}~\bibnamefont
  {Longo}}, \bibinfo {author} {\bibfnamefont {J.}~\bibnamefont {Kafalas}}, \
  and\ \bibinfo {author} {\bibfnamefont {R.}~\bibnamefont {Arnott}},\ }\href
  {\doibase https://doi.org/10.1016/0022-4596(71)90022-3} {\bibfield  {journal}
  {\bibinfo  {journal} {Journal of Solid State Chemistry}\ }\textbf {\bibinfo
  {volume} {3}},\ \bibinfo {pages} {174 } (\bibinfo {year} {1971})}\BibitemShut
  {NoStop}%
\bibitem [{\citenamefont {Nie}\ \emph {et~al.}(2015)\citenamefont {Nie},
  \citenamefont {King}, \citenamefont {Kim}, \citenamefont {Uchida},
  \citenamefont {Wei}, \citenamefont {Faeth}, \citenamefont {Ruf},
  \citenamefont {Ruff}, \citenamefont {Xie}, \citenamefont {Pan}, \citenamefont
  {Fennie}, \citenamefont {Schlom},\ and\ \citenamefont
  {Shen}}]{PhysRevLett.114.016401}%
  \BibitemOpen
  \bibfield  {author} {\bibinfo {author} {\bibfnamefont {Y.~F.}\ \bibnamefont
  {Nie}}, \bibinfo {author} {\bibfnamefont {P.~D.~C.}\ \bibnamefont {King}},
  \bibinfo {author} {\bibfnamefont {C.~H.}\ \bibnamefont {Kim}}, \bibinfo
  {author} {\bibfnamefont {M.}~\bibnamefont {Uchida}}, \bibinfo {author}
  {\bibfnamefont {H.~I.}\ \bibnamefont {Wei}}, \bibinfo {author} {\bibfnamefont
  {B.~D.}\ \bibnamefont {Faeth}}, \bibinfo {author} {\bibfnamefont {J.~P.}\
  \bibnamefont {Ruf}}, \bibinfo {author} {\bibfnamefont {J.~P.~C.}\
  \bibnamefont {Ruff}}, \bibinfo {author} {\bibfnamefont {L.}~\bibnamefont
  {Xie}}, \bibinfo {author} {\bibfnamefont {X.}~\bibnamefont {Pan}}, \bibinfo
  {author} {\bibfnamefont {C.~J.}\ \bibnamefont {Fennie}}, \bibinfo {author}
  {\bibfnamefont {D.~G.}\ \bibnamefont {Schlom}}, \ and\ \bibinfo {author}
  {\bibfnamefont {K.~M.}\ \bibnamefont {Shen}},\ }\href {\doibase
  10.1103/PhysRevLett.114.016401} {\bibfield  {journal} {\bibinfo  {journal}
  {Phys. Rev. Lett.}\ }\textbf {\bibinfo {volume} {114}},\ \bibinfo {pages}
  {016401} (\bibinfo {year} {2015})}\BibitemShut {NoStop}%
\bibitem [{\citenamefont {de~la Torre}\ \emph {et~al.}(2015)\citenamefont
  {de~la Torre}, \citenamefont {McKeown~Walker}, \citenamefont {Bruno},
  \citenamefont {Ricc\'o}, \citenamefont {Wang}, \citenamefont
  {Gutierrez~Lezama}, \citenamefont {Scheerer}, \citenamefont {Giriat},
  \citenamefont {Jaccard}, \citenamefont {Berthod}, \citenamefont {Kim},
  \citenamefont {Hoesch}, \citenamefont {Hunter}, \citenamefont {Perry},
  \citenamefont {Tamai},\ and\ \citenamefont
  {Baumberger}}]{PhysRevLett.115.176402}%
  \BibitemOpen
  \bibfield  {author} {\bibinfo {author} {\bibfnamefont {A.}~\bibnamefont
  {de~la Torre}}, \bibinfo {author} {\bibfnamefont {S.}~\bibnamefont
  {McKeown~Walker}}, \bibinfo {author} {\bibfnamefont {F.~Y.}\ \bibnamefont
  {Bruno}}, \bibinfo {author} {\bibfnamefont {S.}~\bibnamefont {Ricc\'o}},
  \bibinfo {author} {\bibfnamefont {Z.}~\bibnamefont {Wang}}, \bibinfo {author}
  {\bibfnamefont {I.}~\bibnamefont {Gutierrez~Lezama}}, \bibinfo {author}
  {\bibfnamefont {G.}~\bibnamefont {Scheerer}}, \bibinfo {author}
  {\bibfnamefont {G.}~\bibnamefont {Giriat}}, \bibinfo {author} {\bibfnamefont
  {D.}~\bibnamefont {Jaccard}}, \bibinfo {author} {\bibfnamefont
  {C.}~\bibnamefont {Berthod}}, \bibinfo {author} {\bibfnamefont {T.~K.}\
  \bibnamefont {Kim}}, \bibinfo {author} {\bibfnamefont {M.}~\bibnamefont
  {Hoesch}}, \bibinfo {author} {\bibfnamefont {E.~C.}\ \bibnamefont {Hunter}},
  \bibinfo {author} {\bibfnamefont {R.~S.}\ \bibnamefont {Perry}}, \bibinfo
  {author} {\bibfnamefont {A.}~\bibnamefont {Tamai}}, \ and\ \bibinfo {author}
  {\bibfnamefont {F.}~\bibnamefont {Baumberger}},\ }\href {\doibase
  10.1103/PhysRevLett.115.176402} {\bibfield  {journal} {\bibinfo  {journal}
  {Phys. Rev. Lett.}\ }\textbf {\bibinfo {volume} {115}},\ \bibinfo {pages}
  {176402} (\bibinfo {year} {2015})}\BibitemShut {NoStop}%
\bibitem [{\citenamefont {Zeb}\ and\ \citenamefont {Kee}(2012)}]{Kee2012}%
  \BibitemOpen
  \bibfield  {author} {\bibinfo {author} {\bibfnamefont {M.~A.}\ \bibnamefont
  {Zeb}}\ and\ \bibinfo {author} {\bibfnamefont {H.-Y.}\ \bibnamefont {Kee}},\
  }\href {\doibase 10.1103/PhysRevB.86.085149} {\bibfield  {journal} {\bibinfo
  {journal} {Phys. Rev. B}\ }\textbf {\bibinfo {volume} {86}},\ \bibinfo
  {pages} {085149} (\bibinfo {year} {2012})}\BibitemShut {NoStop}%
\bibitem [{\citenamefont {Carter}\ \emph {et~al.}(2012)\citenamefont {Carter},
  \citenamefont {Shankar}, \citenamefont {Zeb},\ and\ \citenamefont
  {Kee}}]{Carter2012}%
  \BibitemOpen
  \bibfield  {author} {\bibinfo {author} {\bibfnamefont {J.-M.}\ \bibnamefont
  {Carter}}, \bibinfo {author} {\bibfnamefont {V.~V.}\ \bibnamefont {Shankar}},
  \bibinfo {author} {\bibfnamefont {M.~A.}\ \bibnamefont {Zeb}}, \ and\
  \bibinfo {author} {\bibfnamefont {H.-Y.}\ \bibnamefont {Kee}},\ }\href
  {\doibase 10.1103/PhysRevB.85.115105} {\bibfield  {journal} {\bibinfo
  {journal} {Phys. Rev. B}\ }\textbf {\bibinfo {volume} {85}},\ \bibinfo
  {pages} {115105} (\bibinfo {year} {2012})}\BibitemShut {NoStop}%
\bibitem [{\citenamefont {Liu}\ \emph {et~al.}(2005)\citenamefont {Liu},
  \citenamefont {Masumoto},\ and\ \citenamefont {Goto}}]{Liu2005}%
  \BibitemOpen
  \bibfield  {author} {\bibinfo {author} {\bibfnamefont {Y.}~\bibnamefont
  {Liu}}, \bibinfo {author} {\bibfnamefont {H.}~\bibnamefont {Masumoto}}, \
  and\ \bibinfo {author} {\bibfnamefont {T.}~\bibnamefont {Goto}},\ }\href
  {https://www.jstage.jst.go.jp/article/matertrans/46/1/46_1_100/_article}
  {\bibfield  {journal} {\bibinfo  {journal} {Materials Transactions}\ }\textbf
  {\bibinfo {volume} {46}},\ \bibinfo {pages} {100} (\bibinfo {year}
  {2005})}\BibitemShut {NoStop}%
\bibitem [{\citenamefont {Groenendijk}\ \emph {et~al.}(2016)\citenamefont
  {Groenendijk}, \citenamefont {Manca}, \citenamefont {Mattoni}, \citenamefont
  {Kootstra}, \citenamefont {Gariglio}, \citenamefont {Huang}, \citenamefont
  {van Heumen},\ and\ \citenamefont {Caviglia}}]{doi:10.1063/1.4960101}%
  \BibitemOpen
  \bibfield  {author} {\bibinfo {author} {\bibfnamefont {D.~J.}\ \bibnamefont
  {Groenendijk}}, \bibinfo {author} {\bibfnamefont {N.}~\bibnamefont {Manca}},
  \bibinfo {author} {\bibfnamefont {G.}~\bibnamefont {Mattoni}}, \bibinfo
  {author} {\bibfnamefont {L.}~\bibnamefont {Kootstra}}, \bibinfo {author}
  {\bibfnamefont {S.}~\bibnamefont {Gariglio}}, \bibinfo {author}
  {\bibfnamefont {Y.}~\bibnamefont {Huang}}, \bibinfo {author} {\bibfnamefont
  {E.}~\bibnamefont {van Heumen}}, \ and\ \bibinfo {author} {\bibfnamefont
  {A.~D.}\ \bibnamefont {Caviglia}},\ }\href {\doibase 10.1063/1.4960101}
  {\bibfield  {journal} {\bibinfo  {journal} {Applied Physics Letters}\
  }\textbf {\bibinfo {volume} {109}},\ \bibinfo {pages} {041906} (\bibinfo
  {year} {2016})}\BibitemShut {NoStop}%
\bibitem [{\citenamefont {Nishio}\ \emph {et~al.}(2016)\citenamefont {Nishio},
  \citenamefont {Hwang},\ and\ \citenamefont {Hikita}}]{Kazunori2016}%
  \BibitemOpen
  \bibfield  {author} {\bibinfo {author} {\bibfnamefont {K.}~\bibnamefont
  {Nishio}}, \bibinfo {author} {\bibfnamefont {H.~Y.}\ \bibnamefont {Hwang}}, \
  and\ \bibinfo {author} {\bibfnamefont {Y.}~\bibnamefont {Hikita}},\ }\href
  {\doibase 10.1063/1.4943519} {\bibfield  {journal} {\bibinfo  {journal} {APL
  Materials}\ }\textbf {\bibinfo {volume} {4}},\ \bibinfo {pages} {036102}
  (\bibinfo {year} {2016})}\BibitemShut {NoStop}%
\bibitem [{\citenamefont {Sung}\ \emph {et~al.}(2016)\citenamefont {Sung},
  \citenamefont {Gretarsson}, \citenamefont {Proepper}, \citenamefont {Porras},
  \citenamefont {Tacon}, \citenamefont {Boris}, \citenamefont {Keimer},\ and\
  \citenamefont {Kim}}]{doi:10.1080/14786435.2015.1134835}%
  \BibitemOpen
  \bibfield  {author} {\bibinfo {author} {\bibfnamefont {N.~H.}\ \bibnamefont
  {Sung}}, \bibinfo {author} {\bibfnamefont {H.}~\bibnamefont {Gretarsson}},
  \bibinfo {author} {\bibfnamefont {D.}~\bibnamefont {Proepper}}, \bibinfo
  {author} {\bibfnamefont {J.}~\bibnamefont {Porras}}, \bibinfo {author}
  {\bibfnamefont {M.~L.}\ \bibnamefont {Tacon}}, \bibinfo {author}
  {\bibfnamefont {A.~V.}\ \bibnamefont {Boris}}, \bibinfo {author}
  {\bibfnamefont {B.}~\bibnamefont {Keimer}}, \ and\ \bibinfo {author}
  {\bibfnamefont {B.~J.}\ \bibnamefont {Kim}},\ }\href {\doibase
  10.1080/14786435.2015.1134835} {\bibfield  {journal} {\bibinfo  {journal}
  {Philosophical Magazine}\ }\textbf {\bibinfo {volume} {96}},\ \bibinfo
  {pages} {413} (\bibinfo {year} {2016})}\BibitemShut {NoStop}%
\end{thebibliography}%

\end{document}